\documentclass[a4paper,11pt]{article}
\usepackage{jheppubMod}
\usepackage{amsmath,amsfonts,amssymb,array,stackrel}
\usepackage{bm,bbm,bbm,bbold}
 \usepackage{cancel}
\usepackage{float}
\usepackage[T1]{fontenc}
\usepackage{dsfont}
\usepackage{enumerate}
\usepackage{graphicx}
\usepackage{lipsum}
\usepackage[displaymath, mathlines,running]{lineno}
\usepackage{multirow,multicol}
\usepackage{subfigure,slashed}
\usepackage{tabu}
\usepackage{datetime}
\usepackage{ulem}
\usepackage{booktabs}
\usepackage{rotating}
\usepackage[utf8]{inputenc}
\usepackage[dvipsnames]{xcolor}
\usepackage{hyperref}
\hypersetup{bookmarks=true,unicode=true,pdftoolbar=true,pdfmenubar=true,
  pdffitwindow=false,pdfstartview={FitH},
  pdftitle={Vector boson fusion at multi-TeV muon colliders},pdfauthor={Costantini, et al},
  pdfsubject={Subject},pdfcreator={Creator},pdfproducer={Producer},
  pdfkeywords={Lepton Colliders}{Muon Collider}{Vector Boson Fusion}{Beyond the Standard Model},
  pdfnewwindow=true,colorlinks=true
}

\renewcommand{\arraystretch}{1.2}

\providecommand{\tabularnewline}{\\}

\renewcommand{\arraystretch}{1.2}
\newcolumntype{x}[1]{
{\centering\hspace{0pt}}p{#1}}

\newcommand{\GeV}{{\rm ~GeV}}
\newcommand{\TeV}{{\rm ~TeV}}

\newcommand{\ab}{{\rm ~ab}}
\newcommand{\invfb}{{\rm ~fb^{-1}}}
\newcommand{\invab}{{\rm ~ab^{-1}}}

\newcommand{\mgFull}{{\sc MadGraph5\_aMC@NLO}}
\newcommand{\mgamc}{{\sc mg5amc}}

\newcommand{\mpm}{\mu^\pm}
\newcommand{\mpmm}{\mu^+\mu^-}
\newcommand\sss{\scriptscriptstyle}
\newcommand{\Op}[1]{\OO_{\sss #1}}

\newcommand{\OO}{\ensuremath{\mathcal{O}}}
\newcommand{\pdp}{\ensuremath{\phi^\dagger\phi}}
\renewcommand{\phi}{\ensuremath{\varphi}}

\newcommand{\bpm}{\begin{pmatrix}}      
\newcommand{\epm}{\end{pmatrix}} 
 \def\lra#1{\overset{\text{\scriptsize$\leftrightarrow$}}{#1}}

\newcommand{\confirm}[1]{{\color{black}#1}}

\title{Vector boson fusion at multi-TeV muon colliders}

\author[a]{Antonio Costantini,}
\author[b]{Federico De Lillo,}
\author[b,c]{Fabio Maltoni,}
\author[b,d]{Luca Mantani,}
\author[b]{Olivier Mattelaer,}
\author[b]{Richard Ruiz}
\author[b]{and Xiaoran Zhao}

\affiliation[a]{INFN, Sezione di Bologna, via Irnerio 46, I-40126 Bologna, Italy}
\affiliation[b]{Centre for Cosmology, Particle Physics and Phenomenology {\rm (CP3)},\\
Universit\'e Catholique de Louvain, Chemin du Cyclotron, B-1348 Louvain la Neuve, Belgium}
\affiliation[c]{Dipartimento di Fisica e Astronomia, Universit\`a di Bologna e INFN, Sezione di Bologna, via Irnerio 46, I-40126 Bologna, Italy}
\affiliation[d]{Institut f\"ur Theoretische Physik, Universit\"at Heidelberg, Germany}
 
\emailAdd{antonio.costantini@bo.infn.it}
\emailAdd{federico.delillo@uclouvain.be}
\emailAdd{fabio.maltoni@uclouvain.be}
\emailAdd{luca.mantani@uclouvain.be}
\emailAdd{olivier.mattelaer@uclouvain.be}
\emailAdd{richard.ruiz@uclouvain.be}
\emailAdd{xiaoran.zhao@uclouvain.be}

\abstract{High-energy lepton colliders with a centre-of-mass energy in the multi-TeV range are currently considered among the most challenging and far-reaching future accelerator projects. 
Studies performed so far have mostly focused on the reach for new phenomena in lepton-antilepton annihilation channels. 
In this work we observe that starting from collider energies of a few TeV, electroweak (EW) vector boson fusion/scattering (VBF) 
at lepton colliders becomes the dominant production mode for all Standard Model processes relevant to studying the EW sector.
In many cases we find that this also holds for new physics.
We  quantify the size and the growth of VBF cross sections with collider energy for a number of SM and new physics processes. 
By considering luminosity scenarios achievable at a muon collider,  we conclude that such a machine would effectively be a ``high-luminosity weak boson collider,''
and subsequently offer a wide range of opportunities to precisely measure EW and Higgs couplings as well as  discover new particles.
}

\preprint{CP3-20-20, MCNET-20-12, VBSCAN-PUB-03-20}
\begin{document}
\maketitle
\setcounter{page}{2}
\flushbottom

\section{Introduction}\label{sec:intro}

Standing out among the important results that the Large Hadron Collider (LHC) has thus far delivered  are the discovery of the Higgs boson $(H)$ and the measurements of its properties. 
On the other hand, long-awaited evidence of new physics based on theoretical arguments, such as the stabilization of the electroweak (EW) scale, 
or on experimental grounds, such as dark matter and neutrino masses, have evaded our scrutiny.  
Despite the fact that the LHC's physics program is far from over, and with Run III and the upgrade to the 
high luminosity LHC (HL-LHC) already lined up, 
the time has come for the high-energy community to assess what could be next in exploring the energy frontier. 
Such a question, which has been the main theme of the activities built around the European Strategy Update for Particle Physics~\cite{Strategy:2019vxc,EuropeanStrategyGroup:2020pow}, is not an easy one: 
Current physics and technology challenges are formidable.
The fact that we have no clear indication where the scale of new physics might reside hampers the definition of a clear target.
And depending on the properties of the new phenomena,  either ``low-energy'' precision measurements or searching for new states in ``high-energy'' direct production may be the most sensitive and informative strategy to follow.  
In any case, exploration of the energy frontier will require building a new collider. 

So far, two main options have actively been discussed by the community: 
a very energetic hadron collider with a center-of-mass (c.m.) energy of $\sqrt{s}=100$ TeV,  and an $e^+e^-$ collider, at either high energy (up to a few TeV) or ultra high luminosity. 
These two classes have very different characteristics.
The former has a much higher discovery reach of new states,
while the latter is feasible on a shorter time scale and allows a precision-measurement campaign of the Higgs/EW sector. 
Both avenues entail incredible investments, an intense research and development program, and formidable engineering capabilities. 
However, as construction of such collider experiments will not start for at least another 15-20 years from now and then require up to 20-40 more years of operation to achieve tentative physics targets,
the community has started  to seriously consider other avenues.
This includes scenarios once thought too audacious or just impossible with even foreseeable technology. 

In this context, both linear $e^+e^-$ and circular $\mu^+\mu^-$ machines  running at energies of several-to-many TeV have recently experienced a boost of interest within the community.  
 In the former case, novel techniques based on plasma acceleration could potentially deliver up to several GeV/$m$ acceleration gradients, thereby reaching TeV scales on the km range~\cite{Gschwendtner:2015rni}.  
 An outstanding challenge in this case, however, is delivering the instantaneous luminosity needed to meet physics goals.
  Accelerating muons, on the other hand, would allow one to merge the best of both hadron and $e^+e^-$ colliders~\cite{Palmer:1996gs,Ankenbrandt:1999cta}, i.e., a high energy reach on one side and a ``clean'' environment on the other.
Such a  facility could possibly reach luminosities in the range of $L=10^{35}$  cm$^{-2}$ s$^{-1}$  (or 100 nb$^{-1}$ Hz)~\cite{Delahaye:2019omf} and, importantly,  be hosted at preexisting laboratory sites and tunnel complexes.  
These dream-like features are counterbalanced by a number of outstanding challenges, all of which eventually originate from a simple fact:  muons are unstable and decay weakly into electrons and neutrinos. 

Conceptual studies of muon colliders started decades ago and recently resulted in the Muon Accelerator Program (MAP) project \cite{Palmer:2014nza}. 
In the MAP proposal, muons are produced as tertiary particles in the decays of pions, 
which  themselves are created by an intense proton beam impinging on a heavy-material target, as already achievable at accelerator-based muon and neutrino facilities. 
The challenge is that muons from pion decays have relatively low energy but large transverse and longitudinal emittance. Hence, they must be ``cooled'' in order to achieve high beam luminosities. 
More recently,  a different approach  to muon production has been proposed: 
in the Low Emission Muon Accelerator (LEMMA) muons are produced in $e^+ e^-$ annihilation near the threshold for $\mu^+ \mu^-$ pair creation \cite{Antonelli:2013mmk,Antonelli:2015nla}. 
A novelty is that  muon beams do not require cooling to reach high instantaneous luminosities. 
This is because when a high-energy positron beam annihilates with electrons from a target the resulting muons are highly collimated and possess very small emittance.
Muons are then already highly boosted with $\gamma\sim 200$ and reach a lifetime of $\tau\sim 500\,\mu$s \cite{Antonelli:2013mmk}. 
The low emittance of the muons may further allow high beam luminosities with a smaller number of muons per bunch.
This results in a lower level of expected beam-induced background, alleviating also potential radiation hazards due to neutrinos \cite{Bartosik:2019dzq}.   

Given the recent interest and fast progress on how to overcome   technological challenges, the most  urgent mission becomes to clearly identify the  reach and physics possibilities that such machines could offer. 
Available studies at  the CLIC $e^+ e^-$ linear collider at 3 TeV have been used to gauge  the potential of a muon collider in the multi-TeV range. 
Earlier, dedicated studies focusing mostly on processes arising from $\mu^+ \mu^-$ annihilation are 
available~\cite{Barger:1995hr,Palmer:1996gs,Ankenbrandt:1999cta,Han:2012rb,Chakrabarty:2014pja,Greco:2016izi,Buttazzo:2018qqp,Delahaye:2019omf,Ruhdorfer:2019utl}, 
and indicate promising potential for finding  new physics from direct searches as well as from indirect searches with precision measurements of EW physics.

The  work here is motivated by the observation that at sufficiently high energies  we expect EW vector boson fusion and scattering (collectedly denoted as VBF) 
to become the dominant production mode at a multi-TeV lepton collider.
While well-established for (heavy) Higgs production~\cite{Dawson:1984ta,Hikasa:1985ee,Altarelli:1987ue,Kilian:1995tr,Gunion:1998jc} and
more recently for the production of heavy singlet scalars~\cite{Buttazzo:2018qqp}, 
we anticipate that this behavior  holds  more broadly for all Standard Model (SM) final states relevant to studying  the EW sector and/or the direct search of (not too heavy) new physics. 
To this aim, we present  a systematic exploration of SM processes  featuring  $W$, $Z$, $H$ bosons and top quarks $t$ in the final state.
We investigate and compare $s$-channel annihilation and VBF cross sections in  high-energy, lepton-lepton  collisions, quantifying the size and the growth of the latter with collider energy. 
We consider the potential utility of precision SM measurements, focusing on a few representative examples, namely in the context of the SM effective field theory (SMEFT)~\cite{Grzadkowski:2010es,Aebischer:2017ugx,Brivio:2017btx}.
Finally, we consider the direct and indirect production of new, heavy states in a number of benchmark, beyond the SM (BSM) scenarios.
Having in mind the luminosity scenarios envisaged for a muon collider~\cite{Delahaye:2019omf}, 
we conclude that a multi-TeV lepton collider could offer a wide range of precision measurements of EW and Higgs couplings as well as sensitivity to new resonances beyond present experiments. 
For example: a $\sqrt{s}=10$ TeV muon collider with an integrated luminosity of $\mathcal{L}=10\, \invab$ would produce about $ 8\cdot 10^6$ Higgs bosons, 
with about $ 3\cdot 10^4$ from pair production alone.
This provides direct access to the trilinear coupling of the Higgs~\cite{Delahaye:2019omf} and gives an excellent perspective on the quartic coupling~\cite{Chiesa:2020awd}.

This study is organized in the following manner: 
In section~\ref{sec:setup} we briefly summarize our computational setup and SM inputs for reproducibility.
We then set the stage in section~\ref{sec:ppvsmuon}  by presenting and critically evaluating simple methods to  
estimate and compare the discovery potential of a hadron collider with that of a high-energy lepton collider. 
In section~\ref{sec:sm} we present production cross sections for SM  processes involving the Higgs bosons, top quark pairs, and EW gauge bosons in $\mpmm$ collisions.  
In particular, we report the total c.m.~energies at which cross sections for VBF processes, which grow as $\log {s}$, 
overcome the corresponding $s$-channel, annihilation ones, which instead decrease as $1/{s}$.
In section~\ref{sec:eft} we  consider the potential  of a multi-TeV muon collider to facilitate precision measurements of EW processes.
We do this by exploring in detail limits that can be obtained in the context of the  SMEFT by measuring $HH$ and $HHH$ production as well as final states involving the top quarks and weak bosons. 
Section~\ref{sec:bsm} presents an overview on the possibilities for direct searches for new resonances at a  multi-TeV muon collider, 
comparing the reach with those attainable at  hadron colliders.
We further investigate and compare the relative importance of VBF production in BSM searches in section~\ref{sec:bsm_vbf}.
 We summarize our work in section~\ref{sec:conclusions}.

\section{Computational setup}\label{sec:setup}
We briefly summarize here  our computational setup.
Throughout this work the evaluation of leading order matrix elements and phase space integration are handled numerically using the general purpose event generator \mgFull (\mgamc) v2.6.5~\cite{Alwall:2014hca}. 
For SM interactions we use the default setup, which assumes the following EW inputs:
\begin{eqnarray}
G_F &= 1.166390 \cdot 10^{-5} \textrm{ GeV}^{-2},  \quad \alpha_{EW}(M_Z) = 1/132.5
\nonumber\\
M_Z &= 91.188\GeV, \quad  M_t = 173\GeV, \quad M_H = 125\GeV.
\end{eqnarray}
For relevant computations, we employ the NNPDF 3.0 LO parton distribution functions (PDFs) with $\alpha_s(M_Z)=0.118$~\cite{Ball:2014uwa},
as maintained using the LHAPDF 6 libraries~\cite{Buckley:2014ana}.
To gain confidence in our results, especially at very high energies where we find that phase space integration converges much more slowly,
we employ \mgamc~v2.7.2, which includes a ``hard survey'' option for improved phase space sampling.
In addition, some SM results have been cross-checked with {\sc Whizard}~\cite{Kilian:2007gr} and in-house MC computations using matrix elements provided by {\sc Recola2} \cite{Denner:2017wsf}.

\section{Comparing proton colliders and muon colliders}\label{sec:ppvsmuon}

In trying to assess and compare qualitatively different colliders, it is constructive to first define a translatable measure of reach.
Therefore, in this section, we propose a simple methodology for comparing the reach of a hypothetical muon collider to what is attainable at a proton collider.
The obvious difference between the two classes of colliders is that  protons are composite particles while muons are not.
This means  that proton collisions involve the scattering of (primarily) QCD partons that carry only a fraction of the total available energy,
whereas muon collisions, up to radiative corrections, involve the scattering of particles carrying the total available energy.
Concretely, we investigate three process categories:
(\ref{sec:ppvsmuon_2to1}) the annihilation of initial-state particles (partons in the $pp$ case) into a single-particle
final state at a fixed final-state invariant mass $(\sqrt{\hat{s}})$;
(\ref{sec:ppvsmuon_2to2}) the two-particle final state analogue of this; and
(\ref{sec:ppvsmuon_vbf}) the scattering of weak gauge bosons.

\subsection{$2\to1$ annihilations}\label{sec:ppvsmuon_2to1}
Despite obvious differences,  our aim is to compare the reach of  muon and hadron colliders in  (as much as possible) a model independent manner.
 In all cases, we find it useful to formulate the comparison in terms of  ``generalized parton luminosities''~\cite{Quigg:2009gg}, 
 where a parton can be any particle in the initial state, be it a lepton, a QCD parton, or an EW boson.
In this language, the total, inclusive cross section for a given process  in $pp$ collisions is
\begin{equation}
    \sigma(p p \to X + \text{anything}) = \int_{\tau_0}^1 d\tau \sum_{ij} \Phi_{ij}(\tau, \mu_f) \, \hat{\sigma}(i j \to X) \, .
\end{equation}
Here $X$ is a generic final state with invariant mass $m_X=\sqrt{\hat{s}}=\sqrt{\tau s}$;
the parton-level cross section is given by $\hat{\sigma}(i j \to X)$ and is kinematically forbidden for $\tau<\tau_0 \equiv \min(m_X^2)/s$;
 and for a c.m.~hadron collider energy $\sqrt{s}$, $\Phi_{ij}$ is the $ij$ parton luminosity, defined as 
\begin{equation}\label{eq:parton_lumi_def}
    \Phi_{ij}(\tau, \mu_f) \equiv \frac{1}{1 + \delta_{ij}} \int_{\tau}^1 \frac{d\xi}{\xi} \left[f_{i/p}(\xi, \mu_f) \, f_{j/p}\left(\frac{\tau}{\xi}, \mu_f\right)
    + (i \leftrightarrow j)
    \right] \, .
\end{equation}
The  $f_{i/p}(\xi,\mu_f)$  are the collinear PDFs for parton $i$ carrying a longitudinal momentum $p_z^i = \xi E_p$,
out of a hadron $p$ with  momentum $p_z^p =  E_p= \sqrt{s}/2$, when renormalization group (RG)-evolved to a collinear factorization scale $\mu_f$.
Where applicable, we set $\mu_f$ to half the partonic c.m.~energy, i.e., set $\mu_f = \sqrt{\hat{s}}/2$.
The Kronecker $\delta_{ij}$ removes double counting of identical initial states in $i \leftrightarrow j$ beam symmetrization.

Given equation~\ref{eq:parton_lumi_def}, then for a  muon collider process $\mpmm\to Y$ and its cross section $\sigma_\mu$ at a fixed muon collider energy $\sqrt{s_\mu}$,
 we define the ``equivalent proton collider energy'' as the  corresponding $pp$ collider energy $\sqrt{s_p}$ 
 such that the analogous hadron-collider process $pp\to Y$ has the same hadronic cross section $\sigma_p$.
 That is,  $\sqrt{s_\mu}$ and $\sqrt{s_p}$ such that  $\sigma_p = \sigma_\mu$.
 
 Now, for the case of a $1$-body final state $Y$ with mass $M=\sqrt{\hat{s}}$, we have
\begin{align}
    \sigma_p(s_p) 		&= \int_{\tau_0}^1 d\tau ~ \sum_{ij} \Phi_{ij}(\tau, \mu_f) \, ~ [\hat{\sigma}_{ij}]_p \, \delta\left(\tau - \frac{M^2}{s_p}\right) \, , \\
    \sigma_\mu(s_\mu) 	&= [\hat{\sigma}]_\mu \, ,
\end{align}
where $[\hat{\sigma}_{ij}]_p$ and $[\hat{\sigma}]_\mu$ are the characteristic partonic cross sections of the two collider processes.
For the $pp$  case, we make explicit the Dirac $\delta$ function arising from the $1$-body phase space measure.
For the $\mpmm$ case, we assume that this is absorbed through, for example, the use of the narrow width approximation
and rescaling by a suitably chosen branching rate.
By construction, $s_\mu = \hat{s}=M^2$, since production can only happen on threshold. 
Requiring that $\sigma_p = \sigma_\mu$ and evaluating the beam-threshold integral, we obtain
\begin{eqnarray}
[\hat{\sigma}]_\mu 	&=& \sigma_\mu(s_\mu)  = \sigma_p(s_p)  \\
				&=&   \sum_{ij} \Phi_{ij}\left(\frac{s_\mu}{s_p}, \mu_f\right) \times [\hat{\sigma}_{ij}]_p \approx  
				[\hat{\sigma}]_p \times \sum_{ij} \Phi_{ij}\left(\frac{s_\mu}{s_p}, \frac{\sqrt{s_\mu}}{2}\right) . 
\label{eq:single_prod}
\end{eqnarray}
In the last step we assume that the $ij$-specific partonic cross section can be approximated by a universal, 
 $ij$-independent cross section $[\hat{\sigma}]_p$.
Crucially, in the luminosity function $\Phi(\tau,\mu_f)$, we identify the kinematic threshold as $\tau = s_\mu / s_p$,
and likewise the factorization scale as $\mu_f = \sqrt{s_\mu}/2$.
If one further assumes a relationship between  the  partonic cross sections, 
then this identification allows us to write equation~\ref{eq:single_prod} as
\begin{equation}
\sum_{ij} \Phi_{ij}\left(\frac{s_\mu}{s_p}, \frac{\sqrt{s_\mu}}{2}\right)    = \cfrac{[\hat{\sigma}]_\mu}{[\hat{\sigma}]_p } \equiv \frac{1}{\beta}.
\label{eq:single_prod_beta}
\end{equation}
which can be  solved\footnote{Explicitly, we use the \texttt{scipy} function \texttt{fsolve} to carry out a brute force computation of this transcendental equation. 
We report a reasonable computation time on a {2-core personal laptop}.
\label{foot:scipy}} 
numerically for  $s_p$ as a function of $s_\mu$ and $\beta$.

\begin{figure}[t]
\begin{center}
\subfigure[]{\includegraphics[width=.48\textwidth]{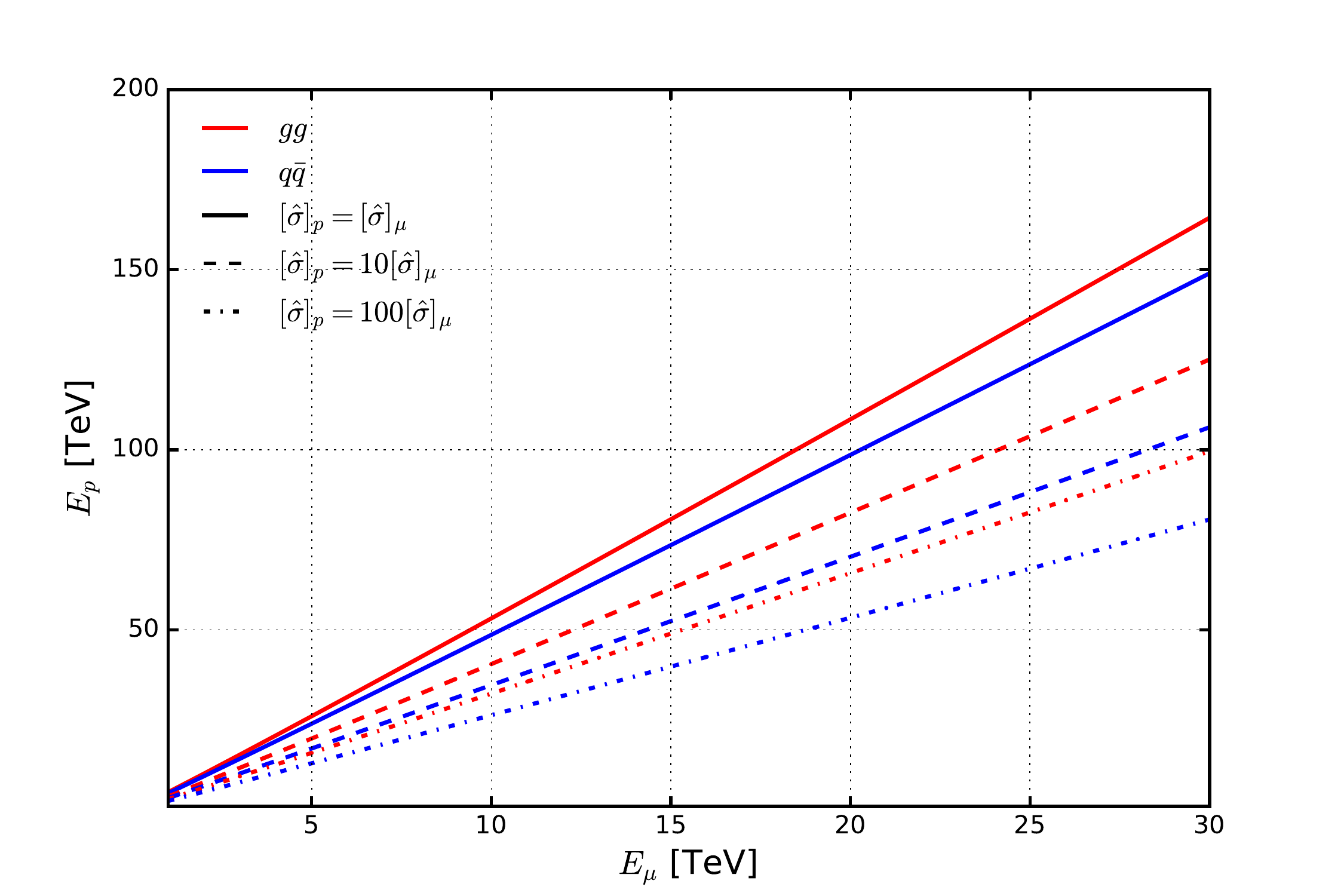}	\label{fig:p_vs_muon_1body}}
\subfigure[]{\includegraphics[width=.48\textwidth]{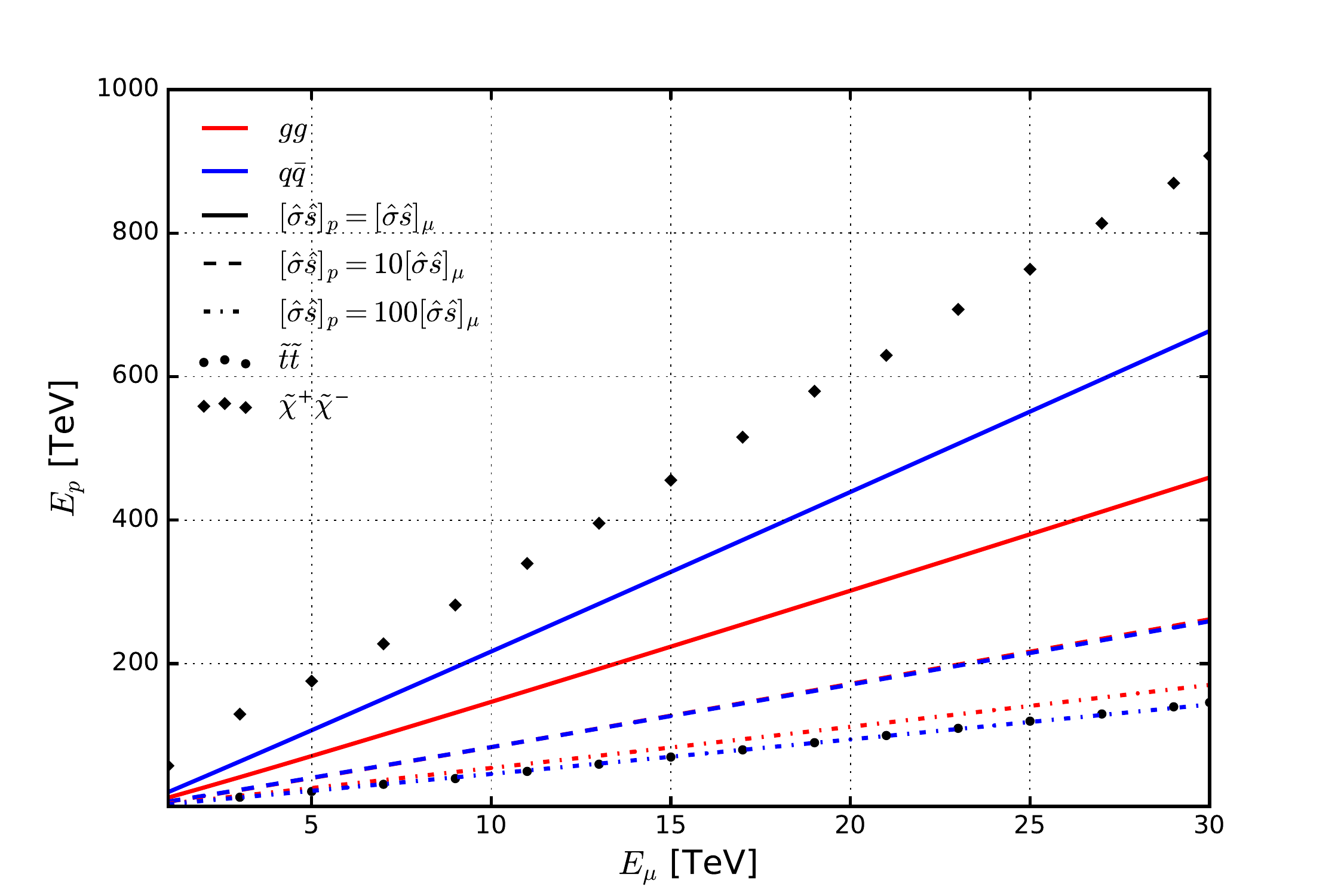}	\label{fig:p_vs_muon_2body}}
\end{center}
\caption{
The equivalent proton collider energy $\sqrt{s_p}$ [TeV]
required to reach the same beam-level cross section as a $\mpmm$ collider with energy $\sqrt{s_\mu}$ [TeV]
for (a) $2\to1$ and (b) $2\to2$ parton-level process, 
for benchmark scaling relationships between the parton-level cross sections $[\hat{\sigma}]_p$ and $[\hat{\sigma}]_\mu$
as well as for pair production of $\tilde{t}\overline{\tilde{t}}$ and $\tilde{\chi}^+\tilde{\chi}^-$ through their leading $2\to2$ production modes.
\label{fig:p_vs_muon}}
\end{figure}

For various benchmark assumptions $(\beta)$ on the  partonic cross sections $[\hat{\sigma}]_p$ and $[\hat{\sigma}]_\mu$, 
and for the parton luminosity configurations $ij=gg$ (red) and $ij=q\overline{q}$ (blue), where $q\in\{u,c,d,s\}$ is any light quark,
we plot in figure~\ref{fig:p_vs_muon_1body} the equivalent proton collider energy $\sqrt{s_p}$ as a function of $\sqrt{s_\mu}$, for a generic $2\to1$, neutral current process.
In particular, for each partonic configuration, 
we consider the case where the  $ij$ and $\mpmm$ partonic rates are the same, i.e.,
when $\beta=1$ (solid line) in equation~\ref{eq:single_prod_beta},  as well as when $\beta=10$ (dash) and $\beta=100$ (dash-dot).
The purpose of these benchmarks is to cover various coupling regimes, 
such as when $ij\to Y$ and $\mpmm\to Y$ are governed by the same physics $(\beta=1)$
or when  $ij\to Y$ is governed by, say, QCD  but $\mpmm\to Y$ by QED $(\beta=10)$.

Overall, we find several notable features.
First is the general expectation that a larger $pp$ collider energy is needed to achieve the same partonic cross section as a $\mpmm$ collider.
This follows from the fact that $pp$ beam energies are distributed among many partons whereas $\mpmm$ collider energies are effectively held by just two incoming partons.
Interestingly, we find a surprisingly simple \textit{linear} scaling between the two colliders for all $ij$ and $\beta$ combinations.
For the $ij=q\overline{q}$ configuration and equal partonic coupling strength, i.e., $\beta=1$, we report a scaling relationship of \confirm{$\sqrt{s_p}\sim 5\times \sqrt{s_\mu}$}.
Under the above assumptions, one would need a muon collider energy  of $\sqrt{s_\mu} \sim 10~(20)~[30]$ to match the reach of a hadron collider with $\sqrt{s_p}\sim 50~(100)~[150]$\TeV.
Specifically for the $\sqrt{s_p}=14\TeV$ LHC and its potential upgrade to \confirm{$\sqrt{s_p}=28\TeV$, one needs $\sqrt{s_\mu} \sim 3\TeV$ and $5.5\TeV$,} respectively.
For the realistic case where the $\mpmm$ dynamics is ultra weakly coupled but $pp$ dynamics is strong, i.e., $\beta=100$, 
and proceeds through the $ij=gg$ partonic channel, we report a milder scaling of \confirm{$\sqrt{s_p}\sim 3.3\times \sqrt{s_\mu}$.}
This translates to needing a higher $ \sqrt{s_\mu}$ to achieve the same reach at a fixed  $\sqrt{s_p}$.
For example: for \confirm{$\sqrt{s_p}=14~(28)\TeV$, one requires instead  $ \sqrt{s_\mu}\sim4.25~(8.5)\TeV$}.
As a cautionary note, we stress that the precise numerical values of scaling ratios reported here are somewhat accidental and can 
shift were one to assume alternative PDF sets or $\mu_f$.
The qualitative behavior, however, should remain.

\subsection{$2\to2$ annihilations}\label{sec:ppvsmuon_2to2}
Instead of comparing the two colliders' equivalent reach for  $2\to1$ processes,
another possibility is to compare the reach for the pair production of heavy states.
Doing so accounts for the nontrivial opening  of new phase space configurations and kinematic thresholds.
In the $2\to2$ case, we assume that the muon collider is optimally configured, i.e., 
that $\sqrt{s_\mu}$ is chosen slightly above threshold, where the production cross section is at its maximum.
For $pp$ collisions, the situation differs from the previous consideration in that
pair production cross sections do not occur at fixed $\hat s$ and, in general, are suppressed by $[\hat{\sigma}_{ij} ]_p\sim1/\hat{s}$, once far above threshold.
 Hence, we make the approximation that the quantity $[\hat{\sigma}_{ij} \hat{s}]_p$  does not depend on $\sqrt{\hat{s}}$,
 and  recast beam-level cross sections in the following way:
\begin{align}
    \sigma_p(s_p) &= \frac{1}{s_p}\int_{\tau_0}^1 d\tau \frac{1}{\tau} ~ \sum_{ij} \Phi_{ij}(\tau, \mu_f) ~ \, [\hat{\sigma}_{ij} \hat{s}]_p \, , \\
    \sigma_\mu(s_\mu) &= \frac{1}{s_\mu}[\hat{\sigma} \hat{s}]_\mu\,.
\end{align}
Assuming again that $[\hat{\sigma}_{ij} ]_p$ can be approximated by the $ij$-independent $[\hat{\sigma}]_p$,
and making analogous identifications as in equation~\ref{eq:single_prod}, then after equating $ \sigma_\mu(s_\mu)=\sigma_p(s_p)$, we obtain
\begin{equation}\label{eq:pair_prod}
    \frac{s_\mu}{s_p} ~ \int_{\frac{s_\mu}{s_p}}^1 d\tau ~ \frac{1}{\tau} ~ \sum_{ij} \Phi_{ij}\left(\tau, \frac{\sqrt{s_\mu}}{2}\right) ~=~ \frac{[\hat{\sigma} \hat{s}]_\mu}{[\hat{\sigma} \hat{s}]_p} ~\equiv~  \frac{1}{\beta} \, .
\end{equation}
Here, the parton luminosity $ij$ runs over the same configurations as in the $2\to1$ case and $\beta$ similarly 
models the relationship between the  (weighted) characteristic, partonic cross sections $[\hat{\sigma}\hat{s}]_p$ and $[\hat{\sigma}\hat{s}]_\mu$.
As in the previous case, we can solve equation~\ref{eq:pair_prod}
numerically (see footnote \ref{foot:scipy}) for the equivalent $pp$ collider energy $\sqrt{s_p}$ as a function of ${s_\mu}$ and $\beta$.

For the same benchmark assumptions on parton luminosities and partonic cross sections $[\hat{\sigma}]_p$ and $[\hat{\sigma}]_\mu$
as considered in figure~\ref{fig:p_vs_muon_1body}, we plot in figure~\ref{fig:p_vs_muon_2body}
 the equivalent proton collider energy $\sqrt{s_p}$ [TeV] as a function of $\sqrt{s_\mu}$ [TeV], for a generic, $2\to2$, neutral current process.
 For concreteness, we also  consider the LO production of 
 \confirm{top squark pairs $\tilde{t}\overline{\tilde{t}}$ through QCD currents in $pp$ collisions but EW currents in $\mpmm$ collisions,
 as well as of chargino pairs $\tilde{\chi}^+\tilde{\chi}^-$ through EW currents.}
 For these cases, we fixed particle masses $M$ such that $2M$ constitutes 90\% of the total muon collider energy, i.e.,  $M=0.9\times\sqrt{s_\mu}/2$.
 
 As in the previous case, we again observe that a much higher energy $pp$ collider exhibits the same reach as lower energy $\mpmm$ colliders.
However, we find the scaling to be more drastic, with higher equivalent proton collider energies being reached for the same muon collider energies.
 We attribute this difference to the fact that, while a spectrum of $\sqrt{\hat{s}}$ is sampled in $pp$ collisions, 
pair production beyond threshold is kinematically suppressed; this is unlike $\mpmm$ collisions where $\sqrt{\hat{s}}$ is fixed.
Remarkably, we also find that the scaling relationship between $\sqrt{s_p}$ and $\sqrt{s_\mu}$ for $2\to2$ processes retains its
 linear behavior for all our representative cases.
 In this measure of comparing colliders, we report a scaling relationship of \confirm{$\sqrt{s_p}\sim 22\times \sqrt{s_\mu}$}
 for the $ij=q\overline{q}$ configuration and equal partonic coupling strength, i.e., $\beta=1$.
 This indicates that a muon collider of \confirm{$\sqrt{s_\mu}\sim10~(20)~[30]\TeV$ has roughly the same reach 
 as a proton collider at $\sqrt{s_p}\sim220~(440)~[660]\TeV$.}
 For the physics scenario where pair production is governed by weak (strong) dynamics in muon (proton), i.e., $\beta=100$,
 we find very similar behavior for both the $ij=gg$ and $q\overline{q}$ parton configurations.
 As in the $2\to1$ case, we report a smaller linear scaling of about \confirm{$\sqrt{s_p}\sim 5.5\times \sqrt{s_\mu}$},
 indicating that the reach of a hypothetical muon collider of \confirm{$\sqrt{s}=2.5~(5)~[14]\TeV$ can only exceed or match the reach of 
 proton colliders up to $\sqrt{s_p}=14~(28)~[80]\TeV$.}
 
 For the concrete cases of stop (dot) and chargino (diamond) pair production, we observe several additional trends in figure~\ref{fig:p_vs_muon_2body}.
 Starkly, we see that the $\tilde{t}\overline{\tilde{t}}$ scaling is in good agreement with 
 the scenario where production is governed by ultra weak (strong) dynamics in muon (proton), i.e., $\beta=100$, for the $ij=q\overline{q}$  configuration. 
 The preferred agreement for $ij=q\overline{q}$ over $ij=gg$ follows from the production of high-mass states in $pp$ collisions being typically driven by $q\overline{q}$ annihilation,
 where $q\in\{u,d\}$ is a valence quark.
 For $\tilde{\chi}^+\tilde{\chi}^-$, we find poorer agreement with na\"ive scaling, with \confirm{$\sqrt{s_p}\sim 30\times \sqrt{s_\mu}$}.
 This is about $\sim1.5\times$ the estimation of the $ij=q\overline{q}$ configuration with equal partonic coupling strength $(\beta=1)$.
 We attribute this difference to the individual EW charges carried by elementary particles:
 as the $\mu\mu Z$ coupling is suppressed, $\mpmm\to\tilde{\chi}^+\tilde{\chi}^-$ is dominated by the $\gamma^*$ subchannel.
The $uuZ$ and $ddZ$ couplings in $qq\to \tilde{\chi}^+\tilde{\chi}^-$, on the other hand, are more sizable, and interfere destructively with the $\gamma^*$ subchannel,
which itself is suppressed due to quarks' fractional electric charge.
This is unlike stop pair production since QCD and QED processes are less flavor dependent.
The disagreement is hence tied to a breakdown of the assumption that  $[\hat{\sigma}_{ij} ]_p$ are  $ij$-independent.
 Nevertheless, we importantly find that our scaling relationships, as derived from equations~\ref{eq:single_prod_beta} and \ref{eq:pair_prod}, 
 provide reliable, if not conservative,  estimates for the equivalent $pp$ collider energy for a given $\mpmm$ collider energy.

\subsection{Weak boson fusion}\label{sec:ppvsmuon_vbf}

We conclude this section by comparing the potential for EW VBF at  high-energy  $\mpmm$ and $pp$ collider facilities.
 As we will analyze in the following sections, one of the main features of a multi-TeV  lepton collider is the increased relevance of VBF  
 over $s$-channel scattering as the total collider energy increases. 
From this perspective, a muon collider could effectively be considered  a ``weak boson collider''. 
It is therefore interesting to compare the potential for  VBF at a muon collider to that at a $pp$  collider.

To make this comparison, we find it useful to continue in the language of parton luminosities and employ the Effective $W$ Approximation (EWA)~\cite{Dawson:1984gx,Kane:1984bb},
which allows us to treat weak bosons on the same footing as QCD partons.
That is to say, enabling us to consistently define $V_\lambda V'_{\lambda'}$ parton luminosities in both $pp$ and $\mpmm$ collisions.
The validity of the EWA as an extension of standard collinear factorization in non-Abelian gauge theories~\cite{Collins:2011zzd} has long been 
discussed in literature~\cite{Cahn:1984tx,Willenbrock:1986cr,Dawson:1986tc,Altarelli:1987ue,Kunszt:1987tk}.
More recent investigations have focused on 
reformulations that make power counting  manifest~\cite{Borel:2012by,Wulzer:2013mza,Chen:2019dkx}
and matching prescriptions between the broken and unbroken phases of the EW theory~\cite{Chen:2016wkt,Chen:2017ekt,Cuomo:2019siu}. 

Under the EWA, splitting functions are used to describe the likelihood of forward emissions of weak bosons off light, initial-state fermions. 
In the notation of equation~\ref{eq:parton_lumi_def},
the helicity-dependent PDFs that describe the radiation 
of a weak vector boson $V$ in helicity state $\lambda$ and carrying a longitudinal energy fraction $\xi$ from a fermion $a$ are~\cite{Dawson:1984gx,Kane:1984bb,Kunszt:1987tk}:
\begin{align}
f_{V_\lambda/a}(\xi, \mu_f, \lambda=\pm1) &= \frac{C}{16\pi^2} \frac{(g_V^a \mp g_A^a)^2 +(g_V^a \pm g_A^a)^2(1-\xi)^2}{\xi}\log{\left(\frac{\mu_f^2}{M_V^2}\right)}, 
\label{eq:ewa_VT}
\\
f_{V_0/a}(\xi, \mu_f) &= \frac{C}{4\pi^2} (g_V^{a~2} + g_A^{a~2})\left( \frac{1-\xi}{\xi}\right)\,.
\label{eq:ewa_V0}
\end{align}
Here $C,~g_V^a,$ and $g_A^a$ represent the appropriate weak gauge couplings of $a$, given by
\begin{eqnarray}
\text{for}~V=W		:&	C=\frac{g^2}{8}, \qquad 			& g_V^a=-g_A^a=1 \, ,
\\
\text{for}~V=Z		:& C=\frac{g^2}{\cos^2{\theta_W}}, \quad & g_V^a=\frac{1}{2}\left(T^3_L\right)^a- Q^a\sin^2{\theta_W},  \quad g_A^a = -\frac{1}{2} \left(T^3_L\right)^a .  
\end{eqnarray}
At this order, the PDFs do not describe QED charge inversion, i.e., $f_{W^\mp/\mu^\pm}=0+\mathcal{O}(g^2)$.
For simplicity, we further define the spin-averaged transverse parton distribution as
\begin{equation}
f_{V_T/a}(\xi, \mu_f) \equiv \cfrac{f_{V_{+1}/a}(\xi, \mu_f^2, \lambda=+1) + f_{V_{-1}/a}(\xi, \mu_f^2, \lambda=-1)}{2} \, .
\end{equation}
For a lepton collider, the $V_\lambda V'_{\lambda'}$ luminosities $\Phi_{V_\lambda V'_{\lambda'}}(\tau,\mu_f)$  are defined as in equation~\ref{eq:parton_lumi_def},
but with substituting the QCD parton PDFs $f_{i/p}$ with the weak boson PDFs $f_{V_\lambda/a}$.
 In particular, for $W^+_{\lambda_1} W^-_{\lambda_2}$ in $\mpmm$ collisions for $\sqrt{s \tau}>2M_W$, we have
\begin{equation}
 \Phi_{W^+_{\lambda_1}  W^-_{\lambda_2} }(\tau, \mu_f) = \int_{\tau}^1 \frac{d\xi}{\xi} ~ f_{W_{\lambda_1} /\mu}\left(\xi, \mu_f\right) \, f_{W_{\lambda_2} /\mu}\left(\frac{\tau}{\xi}, \mu_f\right)  \, .
\end{equation}
For the $pp$ case, the $V_\lambda V'_{\lambda'}$ luminosities are obtained after making the substitution
\begin{equation}
f_{i/p}(\xi,\mu_f) \to f_{V_\lambda/p} (\xi,\mu_f) = \sum_q \int^1_\xi \frac{dz}{z} f_{V_\lambda/q}(z,\mu_f) f_{q/p}\left(\frac{\xi}{z},\mu_f\right),
\end{equation}
which is essentially the  EW gauge boson PDF of the proton. 
The additional convolution accounts for the fact that $q$ in $p$ carries a variable  momentum.
(For simplicity, we keep all $\mu_f$ the same as in equation~\ref{eq:parton_lumi_def}.)
The $V_\lambda V'_{\lambda'}$ luminosity at a scale $\tau$ in $pp$ collisions is then,
\begin{eqnarray}
 \Phi_{V_\lambda V'_{\lambda'}}(\tau, \mu_f) &=& \frac{1}{1+\delta_{V_\lambda V'_{\lambda'}}} \int_\tau^1 \frac{d\xi}{\xi}\int_{\tau/\xi}^1 \frac{dz_1}{z_1}\int_{\tau/(\xi z_1)}^1 \frac{dz_2}{z_2} \sum_{q, q'} 
\\
& &
f_{V_{\lambda}/q}(z_2)f_{V'_{\lambda'}/q'}(z_1)
\left[
f_{q/p}(\xi)f_{q'/p}\left(\frac{\tau}{\xi z_1 z_2}\right) 
+ 
f_{q/p}\left(\frac{\tau}{\xi z_1 z_2}\right)f_{q'/p}(\xi)\right] \, .
\nonumber
\end{eqnarray}

\begin{figure}[t]
\begin{center}
\subfigure[]{\includegraphics[width=.49\textwidth]{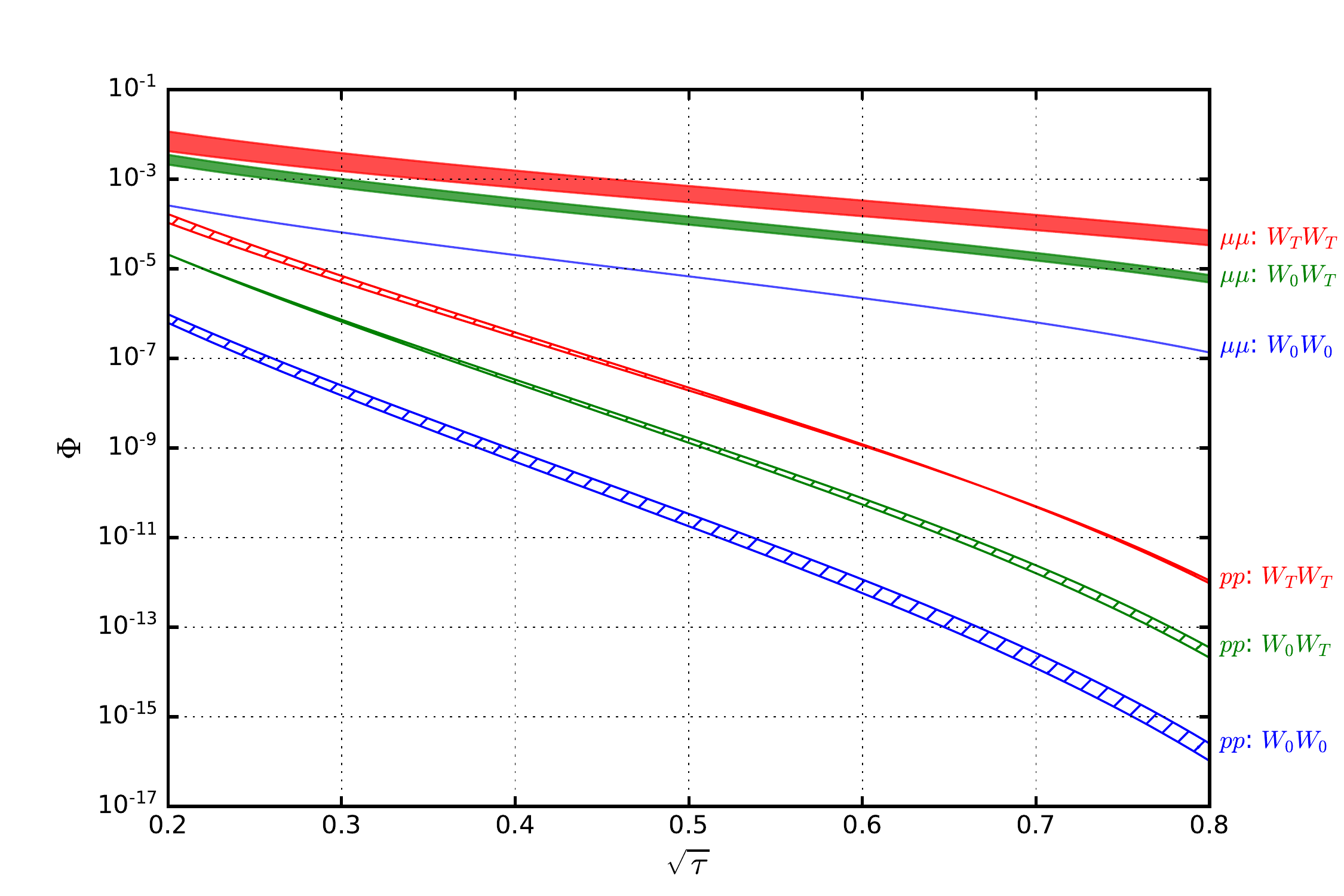}\label{fig:p_vs_muon_VBF_comparison}}
\subfigure[]{\includegraphics[width=.49\textwidth]{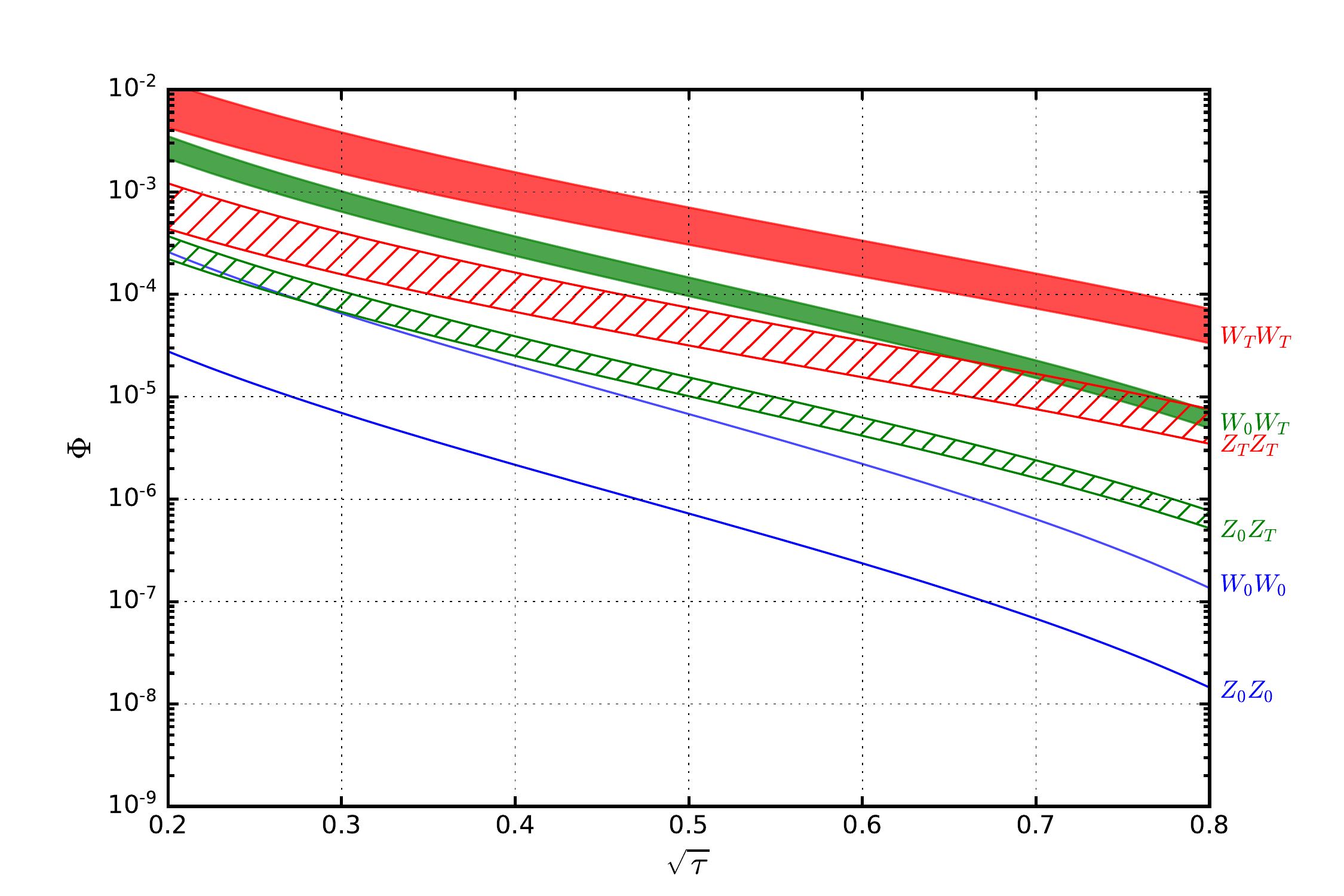}\label{fig:p_vs_muon_VBF_wz}}
\end{center}
\caption{
(a)
As a function of fractional scattering scale $\sqrt{\tau}=M_{VV'}/\sqrt{s}$,
 the (dimensionless) parton luminosities $\Phi$ for 
$W_T^+ W_T^-$ (red), 
$W_T^\pm W_0^\mp$ (green),
$W_0^+W_0^-$ (blue)
in both $pp$ (hatched shading) and $\mpmm$ (solid shading) collisions.
(b) The same but for
$W_\lambda^+ W_{\lambda'}^-$ (solid shading) and 
$Z_\lambda Z_{\lambda'}$ (hatched shading)
in $\mpmm$  collisions with
$(\lambda,\lambda')= (T,T$) (red), $(0,T)+(T,0)$ (green), and $(0,0)$ (blue).
Band thickness corresponds to the $\mu_f$ dependency  as quantified in the text.
}
\label{fig:p_vs_muon_VBF}
\end{figure}

As a function of fractional scattering scale $\sqrt{\tau}=M_{VV'}/\sqrt{s}$, where $\sqrt{s}$ is the total collider energy and $M_{VV'}$ is the $VV'$-system invariant mass, 
we plot in figure~\ref{fig:p_vs_muon_VBF_comparison} the parton luminosities for 
$W_T^+ W_T^-$ (red), 
$W_T^\pm W_0^\mp$ (green),
$W_0^+W_0^-$ (blue)
in both $pp$ (hatched shading) and $\mpmm$ (solid shading) collisions.
Due to our choice to set the collinear factorization scale $\mu_f$ to half the partonic c.m.~energy (see below equation \ref{eq:parton_lumi_def} for details),
the curves possess a (logarithmic) dependence on the collider energy.
To take this ambiguity/dependency into account, we plot the envelopes 
for each parton luminosity spanned by varying $\sqrt{s}=14\TeV-200\TeV~(3\TeV-30\TeV)$ for the proton (muon) case.
The precise ranges of $\sqrt{s}$ and $\sqrt{\tau}$ that we consider help ensure that 
 the partonic fraction of energy  is neither too small nor too big, and hence that the EWA remains reliable~ \cite{Borel:2012by}.
We report that this ``uncertainty'' has little impact on our qualitative and quantitative assessments.

In figure~\ref{fig:p_vs_muon_VBF_comparison}, we find that for each helicity polarization configuration,
the $W_\lambda W_{\lambda'}$ luminosity in $\mpmm$ collisions unambiguously exceeds the analogous luminosity in $pp$ collisions over the $\sqrt{\tau}$ considered.
At $\sqrt{\tau}=0.2~(0.5)~[0.8]$, we find that the $W_\lambda W_{\lambda'}$ luminosities at a muon collider  are roughly \confirm{$10^2-10^3~(10^4-10^6)~[10^8-10^9]$}
times  those of a proton collider.
Hence, for a fixed collider energy $\sqrt{s_\mu}=\sqrt{s_p}$, the likelihood of $WW$ scattering in $\mpmm$ collisions is much higher than for $pp$ collisions.
We  attribute this to several factors:
First is that the emerging EW PDFs in proton beams are a subdominant phenomenon in perturbation theory whereas in muon beams they arise at lowest order.
Relatedly, as muons are point particles, they carry more energy than typical partons in a proton beam with the same beam energy.
This enables EW PDFs in $\mpmm$ collisions to access smaller momentum fractions $\xi$, thereby accessing larger PDF enhancements at smaller $\xi$.

To further explore this hierarchy, we compare in figure~\ref{fig:p_vs_muon_VBF_wz} 
the $\mpmm$ collider luminosities for $W_\lambda^+W_{\lambda'}^-$ (solid shading) and $Z_{\lambda}Z_{\lambda'}$ (hatched shading) pairs,
for $(\lambda,\lambda')=(T,T)$ (red), $(0,T)+(T,0)$ (green), and $(0,0)$ (blue).
Globally, we see that the $WW$ and $ZZ$ luminosities exhibit a very similar shape dependence on $\sqrt{\tau}$,
which follows from the functional form of $f_{V/a}(\xi)$.
The normalization difference is due to the SU$(2)_L$ quantum number of muons, which results in the well-known suppression of $\mu\mu Z$ couplings in the SM.
Indeed, for  $(\lambda,\lambda')=(0,0)$, we find that the ratio of luminosities exhibits the constant relationship
\begin{equation}
\cfrac{\Phi_{W_0W_0}(\tau)}{\Phi_{Z_0Z_0}(\tau)}\Bigg\vert_{\rm fixed~\tau} = 
\left[\cfrac{\cos^2\theta_W}{\left(T^{3~\mu}_L- 2Q^\mu\sin^2{\theta_W}\right)^2 + \left(T^{3~\mu}_L\right)^2 }\right]^2
\approx
\cfrac{\cos^4\theta_W}{\left(T^{3~\mu}_L\right)^4 }
\approx 9.
\end{equation}
While not shown, we report that the $W_{\lambda}Z_{\lambda'}$ luminosities also have similar shapes and are located roughly at the geometric average of the $WW$ and $ZZ$ curves.
Furthermore, due to gauge coupling universality, we 
anticipate that the luminosity hierarchy observed between muon and proton colliders 
  extends to luminosities involving $\gamma$ and $Z$ bosons.

\section{Standard Model processes at muon colliders}\label{sec:sm}

In this section we investigate  and present cross sections for 
various EW boson and top quark final states of the form $X= n\, t \bar{t} + m\, V + k\, H $, 
where $n$, $m$ and $k$ are integers that respectively denote the number of top quark pairs, weak vector bosons $V$, and Higgs bosons $H$.
One of our goals of this survey is to systematically compare  $s$-channel annihilation processes with EW VBF production channels in $\mpmm$ collisions, 
and in particular identify the c.m.~energies at which VBF rates overtake $s$-channel ones.

Specifically, we consider VBF process $VV\to X$ as obtainable from a $\mu^+\mu^-$ initial state.
This consists of the sub-channels
$W^{+}W^{-}$ fusion (section~\ref{sec:sm_ww}), 
$ZZ/Z\gamma/\gamma\gamma$ fusion  (section~\ref{sec:sm_zz}),
and $W^{\pm}Z/W^\pm\gamma$ fusion (section~\ref{sec:sm_wz}):
\begin{center}
\begin{tabular}{ll}
$\mu^+\mu^-\to X\, {\nu}_{\mu}\overline{\nu}_{\mu}$ &($WW$\, \textrm{fusion}),\nonumber\\
$\mu^+\mu^-\to X\, \mu^+\mu^-$&($ZZ/Z\gamma/\gamma\gamma$\, \rm{fusion}).\nonumber\\
$\mu^+\mu^-\to X\, \mu^\pm \overset{(-)}{\nu_\mu}$& ($WZ$\, \rm{fusion}),\nonumber
\label{procs}
\end{tabular}
\end{center}
We also consider collisions with same-sign muon pairs, $\mu^+\mu^+$ (section~\ref{sec:sm_samesign}).
In this case, the $WZ$ and $ZZ$ modes give rise to the same final state $X$, up to charge multiplicities, at the same rate as $\mpmm$ collisions.
The $W^+ W^+$ mode, on the other hand, opens truly new kinds of signatures
while possessing the same luminosity as $W^{+}W^{-}$ fusion reported in section~\ref{sec:ppvsmuon}.
Before presenting our survey, we briefly comment in section~\ref{sec:sm_technicalities} on a few technical details
related to simulating many-particle final states in multi-TeV lepton collisions.

\subsection{Technical nuances at high energies}\label{sec:sm_technicalities}

An important issue in this study involves the fact that the final states above also receive contributions from non-VBF processes,
like associated production of $X$ and a $W$ or $Z$ boson.
That is to say, from an $s$-channel process but with an additional $V$-strahlung emission that then splits into a lepton pair.
In general, these contributions interfere with VBF topologies at the amplitude level and are not all separately gauge-invariant subprocesses. 
Therefore, in principle, they need to be considered together with VBF.
However, the $V$ boson decay contributions are dominated by regions of phase space where $V$ are on their  mass shells.
Especially, at a lepton collider, where the initial-state energy of the collision is known accurately, 
such resonant configurations can be experimentally distinguished from the non-resonant continuum. 
In fact, the relative contributions of those resonant topologies 
as well as 
of their interference with the gauge-invariant VBF contributions 
are small when far from the on-shell region, 
i.e., where most of the VBF cross section is populated. 

Therefore, in order to avoid double counting of results that we will present for $s$-channel processes, as well as to make computations more efficient, 
we remove contributions with instances of on-shell $Z\to\mpmm$, $Z \to \nu \bar{\nu}$, and $W^- \to \mu^- \bar{\nu}$ decays.
In general,  removing diagrams  breaks gauge invariance and so we refrain from doing this. 
A simple solution, adopted for instance in Ref.~\cite{Chiesa:2020awd}, is to impose a minimum on the invariant mass of final-state lepton pairs.  
In this work, we adopt an even simpler prescription by  simulating  an initial state possessing a non-zero, net muon and electron flavor, i.e. $\mu^- e^+$ collisions. 
In so doing, $s$-channel annihilations are forbidden and VBF channels are automatically retained. 
We have checked for a few processes that the numerical differences 
with scattering rates of the analogous $\mu^+ \mu^-\to X$ channels in the far off-shell region are small at high energies.

\begin{figure}[!t]
\centering\mbox{\subfigure[]{\includegraphics[width=0.45\textwidth]{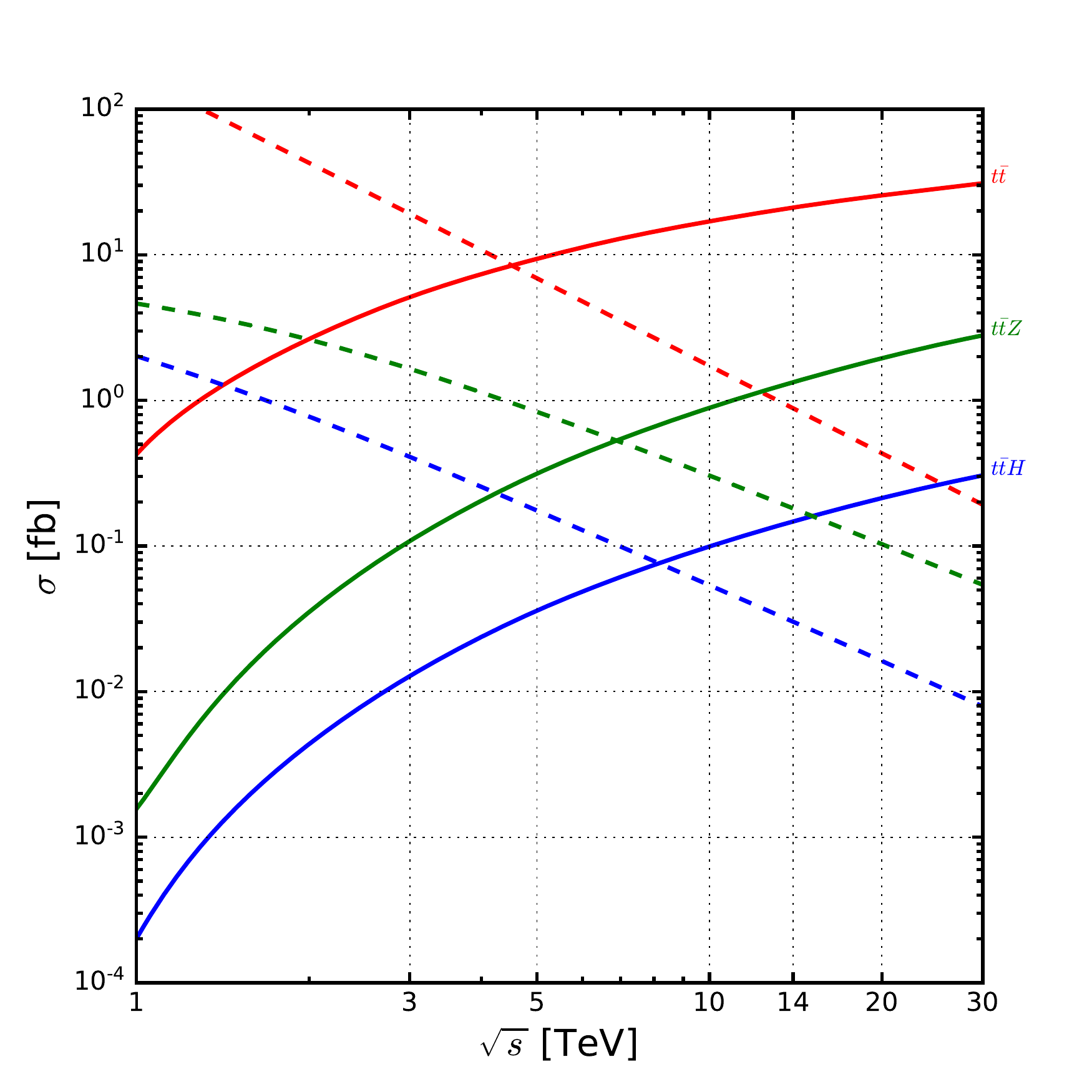}		\label{fig:SMt_ttVX}}}
\centering\mbox{\subfigure[]{\includegraphics[width=0.45\textwidth]{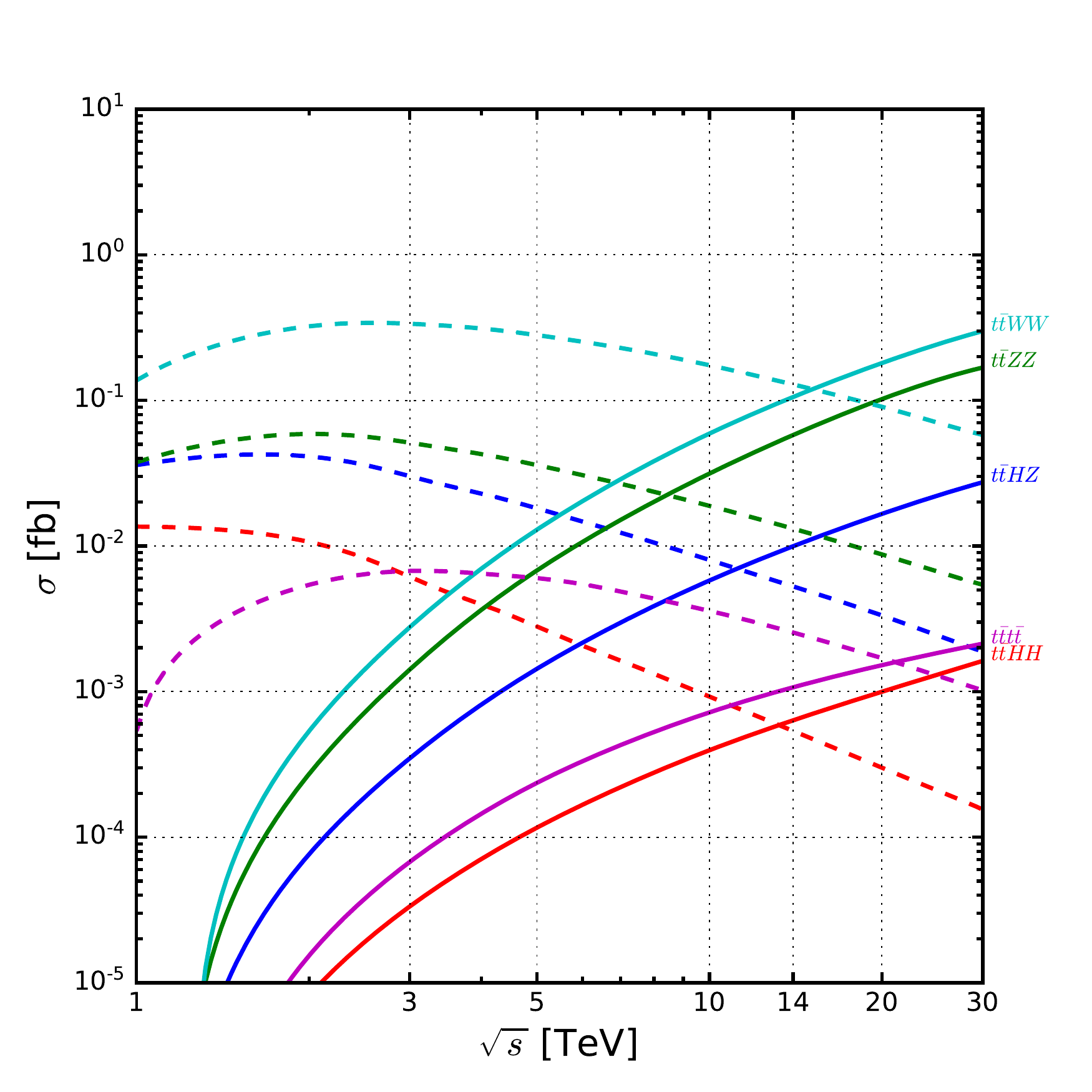}	\label{fig:SMt_ttVV}}}
\caption{$W^+W^-$ fusion (solid) and analogous $s$-channel annihilation (dashed) cross sections $\sigma$ [fb] for (a) $t\overline{t}X$ and (b) $t\overline{t}XX$ associated production as a function of collider energy $\sqrt{s}$ [TeV].}
\label{fig:SMt}
\end{figure}

\begin{figure}[!t]
\centering\mbox{\subfigure[]{\includegraphics[width=0.445\textwidth]{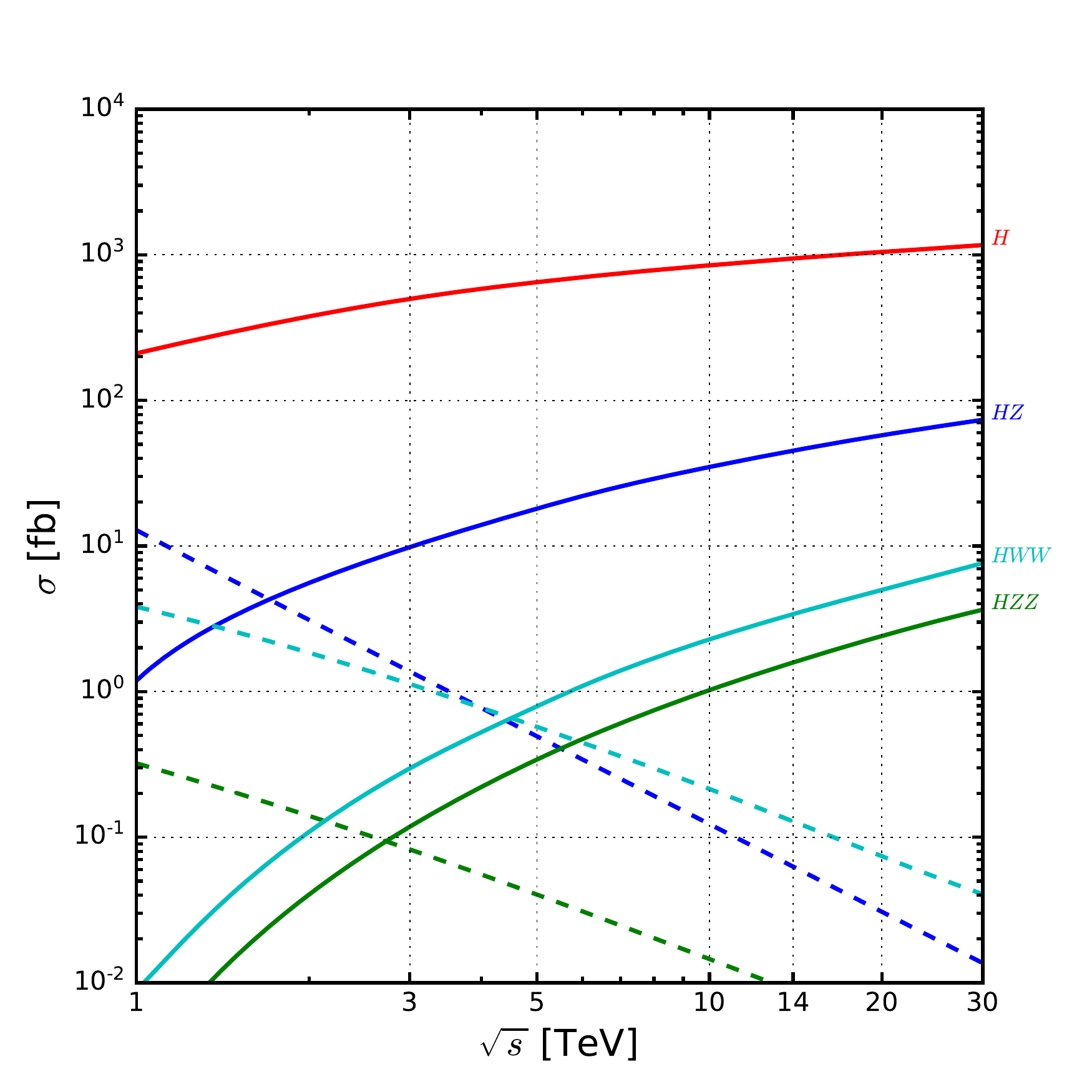}		\label{fig:SMh_hxxx}}}
\centering\mbox{\subfigure[]{\includegraphics[width=0.445\textwidth]{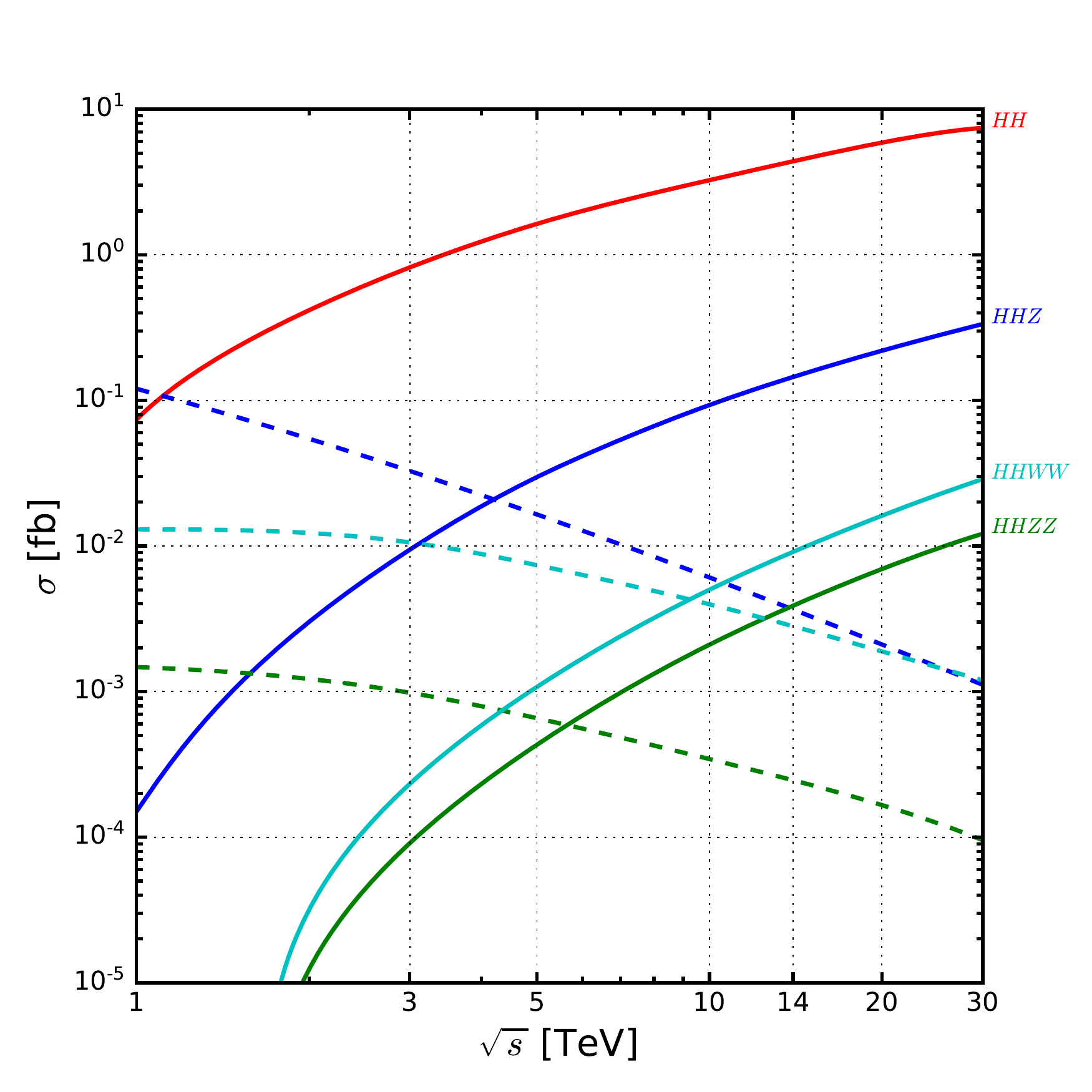}		\label{fig:SMh_hhxx}}}\\
\centering\mbox{\subfigure[]{\includegraphics[width=0.445\textwidth]{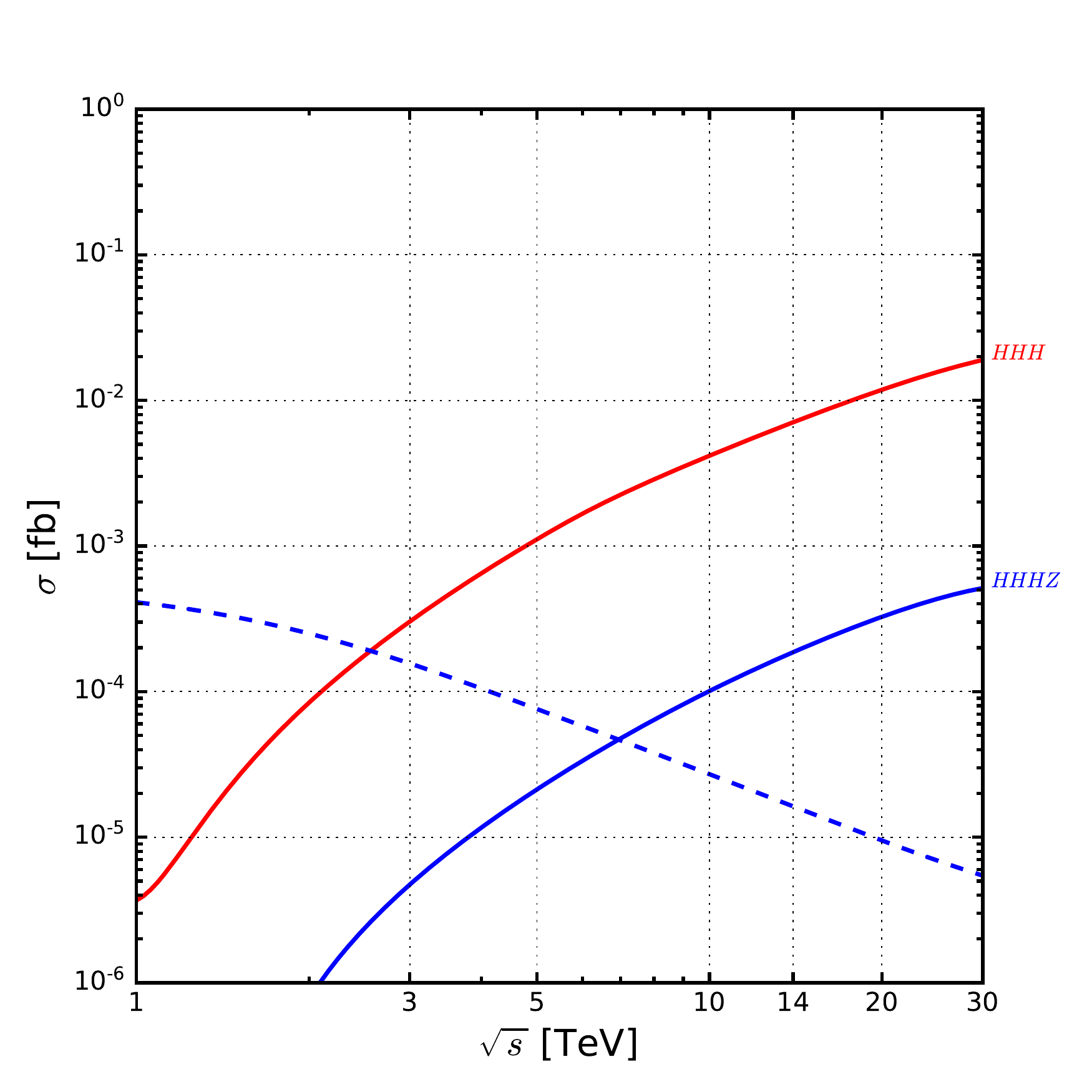}	\label{fig:SMh_hhhx}}}
\centering\mbox{\subfigure[]{\includegraphics[width=0.445\textwidth]{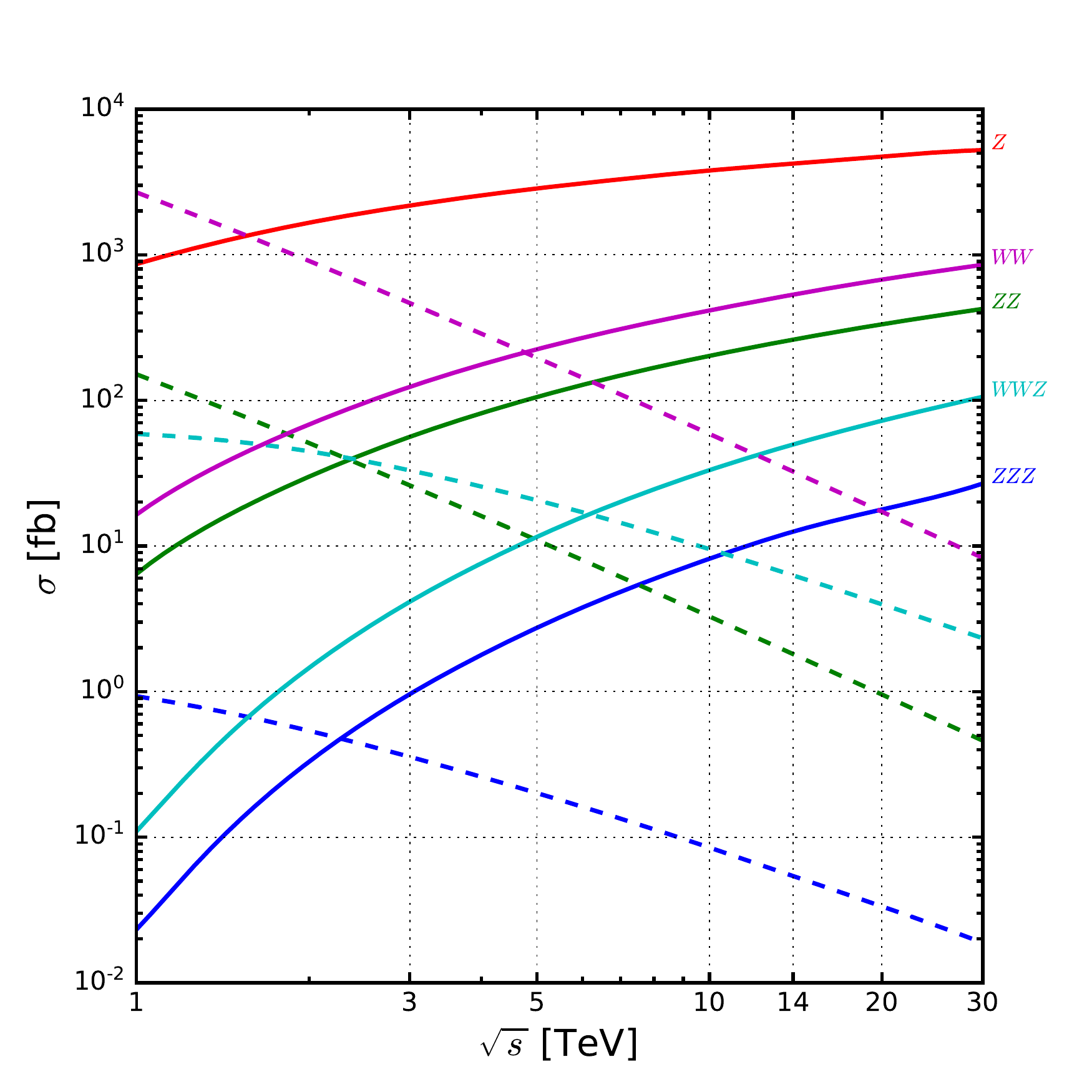}		\label{fig:SMh_vvvx}}}
\caption{
Same as figure~\ref{fig:SMt} but for  (a) $HX$, (b) $HHX$, and (c) $HHHX$ associated production as well as (d) multiboson  production.}
\label{fig:SMh}
\end{figure}

\begin{table}[!t]
\begin{center}
\resizebox{\textwidth}{!}{
\begin{tabular}{l||cc||cc||cc||cc}
\toprule
\toprule
\multirow{2}{*}{$\sigma$ [fb]} &        \multicolumn{2}{c}{$\sqrt s=$ 1 TeV} &    \multicolumn{2}{c}{$\sqrt s=$ 3 TeV} &    \multicolumn{2}{c}{$\sqrt s=$ 14 TeV} &      \multicolumn{2}{c}{$\sqrt s=$ 30 TeV}\\
		&VBF&s-ch.&VBF&s-ch.&VBF&s-ch.&VBF&s-ch\\
\midrule
\midrule
$t \bar{t}$           	&  4.3$\cdot 10^{-1}$ &1.7$\cdot 10^{2}$ &  5.1$\cdot 10^{0}$ &1.9$\cdot 10^{1}$ &  2.1$\cdot 10^{1}$ &8.8$\cdot 10^{-1}$ &  3.1$\cdot 10^{1}$ &1.9$\cdot 10^{-1}$ \\
\hline
$t \bar{t} Z$         	&  1.6$\cdot 10^{-3}$ &4.6$\cdot 10^{0}$ &  1.1$\cdot 10^{-1}$ &1.6$\cdot 10^{0}$ &  1.3$\cdot 10^{0}$ &1.8$\cdot 10^{-1}$ &  2.8$\cdot 10^{0}$ &5.4$\cdot 10^{-2}$ \\
$t \bar{t} H$         	&  2.0$\cdot 10^{-4}$ &2.0$\cdot 10^{0}$ &  1.3$\cdot 10^{-2}$ &4.1$\cdot 10^{-1}$ &  1.5$\cdot 10^{-1}$ &3.0$\cdot 10^{-2}$ &  3.1$\cdot 10^{-1}$ &7.9$\cdot 10^{-3}$ \\
\hline
$t \bar{t} W W$       	&  4.8$\cdot 10^{-6}$ &1.4$\cdot 10^{-1}$ &  2.8$\cdot 10^{-3}$ &3.4$\cdot 10^{-1}$ &  1.1$\cdot 10^{-1}$ &1.3$\cdot 10^{-1}$ &  3.0$\cdot 10^{-1}$ &5.8$\cdot 10^{-2}$ \\
$t \bar{t} Z Z$       	&  2.3$\cdot 10^{-6}$ &3.8$\cdot 10^{-2}$ &  1.4$\cdot 10^{-3}$ &5.1$\cdot 10^{-2}$ &  5.8$\cdot 10^{-2}$ &1.3$\cdot 10^{-2}$ &  1.7$\cdot 10^{-1}$ &5.4$\cdot 10^{-3}$ \\
$t \bar{t} H Z$       	&  7.1$\cdot 10^{-7}$ &3.6$\cdot 10^{-2}$ &  3.5$\cdot 10^{-4}$ &3.0$\cdot 10^{-2}$ &  1.0$\cdot 10^{-2}$ &5.3$\cdot 10^{-3}$ &  2.7$\cdot 10^{-2}$ &1.9$\cdot 10^{-3}$ \\
$t \bar{t} H H$       	&  7.2$\cdot 10^{-8}$ &1.4$\cdot 10^{-2}$ &  3.4$\cdot 10^{-5}$ &6.1$\cdot 10^{-3}$ &  6.4$\cdot 10^{-4}$ &5.4$\cdot 10^{-4}$ &  1.6$\cdot 10^{-3}$ &1.5$\cdot 10^{-4}$ \\
\hline
$t \bar{t} t \bar{t}\;\;(i)$ 	&  5.1$\cdot 10^{-8}$ &5.4$\cdot 10^{-4}$ &  6.8$\cdot 10^{-5}$ &6.7$\cdot 10^{-3}$ &  1.1$\cdot 10^{-3}$ &2.5$\cdot 10^{-3}$ &  2.1$\cdot 10^{-3}$ &1.0$\cdot 10^{-3}$ \\
$t \bar{t} t \bar{t}\;\;(ii)$ 	&  6.2$\cdot 10^{-9}$ &7.9$\cdot10^{-4}$ & 3.7$\cdot 10^{-5}$ &6.9$\cdot10^{-3}$&  1.7$\cdot 10^{-3}$ &2.3$\cdot10^{-3}$ &  4.7$\cdot 10^{-3}$ &9.0$\cdot10^{-4}$ \\
\midrule
\midrule
$H$                   	&        2.1$\cdot 10^{2}$ &- &        5.0$\cdot 10^{2}$ &- &        9.4$\cdot 10^{2}$ &- &        1.2$\cdot 10^{3}$ &- \\
$H H$                 	&        7.4$\cdot 10^{-2}$ &- &        8.2$\cdot 10^{-1}$ &- &        4.4$\cdot 10^{0}$ &- &        7.4$\cdot 10^{0}$ &- \\
$H H H$               	&        3.7$\cdot 10^{-6}$ &- &        3.0$\cdot 10^{-4}$ &- &        7.1$\cdot 10^{-3}$ &- &        1.9$\cdot 10^{-2}$ &- \\
\hline
$H Z$                 	&  1.2$\cdot 10^{0}$ &1.3$\cdot 10^{1}$ &  9.8$\cdot 10^{0}$ &1.4$\cdot 10^{0}$ &  4.5$\cdot 10^{1}$ &6.3$\cdot 10^{-2}$ &  7.4$\cdot 10^{1}$ &1.4$\cdot 10^{-2}$ \\
$H H Z$               	&  1.5$\cdot 10^{-4}$ &1.2$\cdot 10^{-1}$ &  9.4$\cdot 10^{-3}$ &3.3$\cdot 10^{-2}$ &  1.4$\cdot 10^{-1}$ &3.7$\cdot 10^{-3}$ &  3.3$\cdot 10^{-1}$ &1.1$\cdot 10^{-3}$ \\
$H H H Z$             	&  1.5$\cdot 10^{-8}$ &4.1$\cdot 10^{-4}$ &  4.7$\cdot 10^{-6}$ &1.6$\cdot 10^{-4}$ &  1.9$\cdot 10^{-4}$ &1.6$\cdot 10^{-5}$ &  5.1$\cdot 10^{-4}$ &5.4$\cdot 10^{-6}$ \\
\hline
$H W W$               	&  8.9$\cdot 10^{-3}$ &3.8$\cdot 10^{0}$ &  3.0$\cdot 10^{-1}$ &1.1$\cdot 10^{0}$ &  3.4$\cdot 10^{0}$ &1.3$\cdot 10^{-1}$ &  7.6$\cdot 10^{0}$ &4.1$\cdot 10^{-2}$ \\
$H H W W$             	&  7.2$\cdot 10^{-7}$ &1.3$\cdot 10^{-2}$ &  2.3$\cdot 10^{-4}$ &1.1$\cdot 10^{-2}$ &  9.1$\cdot 10^{-3}$ &2.8$\cdot 10^{-3}$ &  2.9$\cdot 10^{-2}$ &1.2$\cdot 10^{-3}$ \\
\hline
$H Z Z$               	&  2.7$\cdot 10^{-3}$ &3.2$\cdot 10^{-1}$ &  1.2$\cdot 10^{-1}$ &8.2$\cdot 10^{-2}$ &  1.6$\cdot 10^{0}$ &8.8$\cdot 10^{-3}$ &  3.7$\cdot 10^{0}$ &2.5$\cdot 10^{-3}$ \\
$H H Z Z$             	&  2.4$\cdot 10^{-7}$ &1.5$\cdot 10^{-3}$ &  9.1$\cdot 10^{-5}$ &9.8$\cdot 10^{-4}$ &  3.9$\cdot 10^{-3}$ &2.5$\cdot 10^{-4}$ &  1.2$\cdot 10^{-2}$ &9.5$\cdot 10^{-5}$ \\
\midrule
\midrule
$W W$		&1.6$\cdot 10^{1}$&2.7$\cdot 10^{3}$&1.2$\cdot 10^{2}$&4.7$\cdot 10^{2}$&5.3$\cdot 10^{2}$&3.2$\cdot 10^{1}$&8.5$\cdot 10^{2}$&8.3$\cdot 10^{0}$\\
$Z Z$		&6.4$\cdot 10^{0}$&1.5$\cdot 10^{2}$&5.6$\cdot 10^{1}$&2.6$\cdot 10^{1}$&2.6$\cdot 10^{2}$&1.8$\cdot 10^{0}$&4.2$\cdot 10^{2}$&4.6$\cdot 10^{-1}$\\
\hline
$W W Z$		&1.1$\cdot 10^{-1}$&5.9$\cdot 10^{1}$&4.1$\cdot 10^{0}$&3.3$\cdot 10^{1}$&5.0$\cdot 10^{1}$&6.3$\cdot 10^{0}$&1.0$\cdot 10^{2}$&2.3$\cdot 10^{0}$\\
$Z Z Z$		&2.3$\cdot 10^{-2}$&9.3$\cdot 10^{-1}$&9.6$\cdot 10^{-1}$&3.5$\cdot 10^{-1}$&1.2$\cdot 10^{1}$&5.4$\cdot 10^{-2}$&2.7$\cdot 10^{1}$&1.9$\cdot 10^{-2}$\\
\bottomrule
\bottomrule
\end{tabular}
}
\caption{Same as figures \ref{fig:SMt} and \ref{fig:SMh} but tabulated for representative collider energies. For the $t\overline{t}t\overline{t}$ processes, scenario $(i)$ considers mixed EW-QCD production and $(ii)$ considers pure EW production.}
\label{tab:neutralSM}
\end{center}
\end{table}

A second technical issue that requires care is the treatment of unstable particles, and in particular the inclusion of fixed widths $(\Gamma)$ in Breit-Wigner propagators. 
While formally suppressed by $\mathcal{O}\left(\Gamma/M\right)$ for resonances of mass $M$,
these terms can break gauge invariance
as well as spoil delicate unitarity cancellations at high energies. 
Indeed, we find that these disruptions can grow with energy for some processes  and spoil the correctness of our calculations.
 A well-known solution is to consider the complex mass scheme~\cite{Denner:1999gp,Denner:2005fg},  an option that is available in \mgamc~\cite{Alwall:2014hca}. 
 However, in this case, all unstable particles can  only appear as internal states, not as external ones. 
 This implies that  when modeling each particle in our final state $X$ we always must include a decay channel (or decay channel chain), complicating our work considerably.
Subsequently,  we have opted for the solution of simulating external, on-shell $W,Z,H,t$ with all widths set to zero.
In doing so, gauge invariance is automatically preserved.
Moreover, potential singularities in $W,Z,H,t$ propagators  are also automatically regulated due to their mass differences.

\begin{table}[h!]
\begin{center}
\renewcommand*{\arraystretch}{1}
\begin{tabular}{l |cc || l |cc}
\toprule
\toprule
\multicolumn{1}{l}{} &$\sigma$ [fb]&$\sqrt s$ [TeV]        &	\multicolumn{1}{l}{} &$\sigma$ [fb]&$\sqrt s$ [TeV]\\
\hline\hline
$t\bar t$&$8.4\cdot 10^0$		&$4.5$        	&$t\bar t ZZ$&$2.2\cdot 10^{-2}$		&$8.4$\\
$t\bar t Z$&$5.3\cdot 10^{-1}$		&$6.9$        	&$t\bar t HZ$&$7.0\cdot 10^{-3}$		&$11$\\
$t\bar t H$&$7.6\cdot 10^{-2}$		&$8.2$        	& $t\bar t HH$&$5.9\cdot 10^{-4}$		&$13$ \\
$t\bar t WW$&$1.2\cdot 10^{-1}$	&$15$     		& $t\bar t t\bar t$&$1.6\cdot10^{-3}$	&$22$\\
\midrule
$HZ$&$4.3\cdot 10^0$		&$1.7$	&$HHWW$&$4.3\cdot 10^{-3}$	&$9.2$\\
$HHZ$&$2.1\cdot 10^{-2}$	&$4.2$	&$HZZ$&$9.4\cdot 10^{-2}$		&$2.7$         \\
$HHHZ$&$4.7\cdot 10^{-5}$	&$6.9$	& $HHZZ$&$5.9\cdot 10^{-4}$	&$5.7$\\
$HWW$&$6.6\cdot 10^{-1}$	&$4.5$	&	&\\
\midrule
$WW$&$2.1\cdot 10^2$		&$4.8$	&$WWZ$		&$1.6\cdot 10^{1}$	&$6.2$\\
$ZZ$&$3.9\cdot 10^{1}$		&$2.4$	&$ZZZ$		&$4.8\cdot 10^{-1}$	&$2.3$\\
\bottomrule
\bottomrule
\end{tabular}
\caption{
The value of collider energy $\sqrt s$ [TeV] and the corresponding cross section $\sigma$ [fb] that satisfy $\sigma^{VBF}=\sigma^{s-ch.}$ for processes considered in  figures \ref{fig:SMt} and \ref{fig:SMh}.
}
\label{tab:sameXS}
\end{center}
\end{table}

\subsection{$W^+ W^-$ fusion}\label{sec:sm_ww}

We begin our survey by considering the production of up to four heavy particles from $W^+W^-$ fusion (solid lines) and $s$-channel, $\mpmm$ annihilation (dashed lines).
As a function of muon collider energy $(\sqrt{s})$ [TeV],
we plot cross sections $(\sigma)$ [fb]  
in figure~\ref{fig:SMt} for (a) $t\overline{t}X$ and (b) $t\overline{t}XX$ associated production.
In figure~\ref{fig:SMh} we plot the same for (a) $HX$, (b) $HHX$, and (c) $HHHX$ associated production, and (d) multiboson production.
We summarize our findings in table~\ref{tab:neutralSM} for representative collider energies and processes.

To summarize the global picture: as expected from the different production mechanism, 
$s$-channel annihilation rates categorically scale and decrease with collider energy at least as $\sigma\sim1/s$, when collider energies are far beyond kinematic threshold.
This is contrary to VBF processes where cross sections mildly increase with collider energy at least as a power of  $\log(s/M_W^2)$, in the high energy limit. 
 Consequentially, we find that  for all processes considered there is a $\sqrt{s}$ where  VBF production overcomes $s$-channel production. 
In table~\ref{tab:sameXS} we report this  $\sqrt s$ and the corresponding $\sigma$ at which the $s$-channel and VBF cross sections are the same.
 In general, the larger the final state multiplicity, the larger the value of $\sqrt s$ where  the cross section curves  cross.
A few more remarks are in order. 

First, for processes involving a top quark pair, as shown in figure~\ref{fig:SMt}, 
the $s$-channel cross sections at lower energies of $\mathcal{O}(1)$ TeV are comparable to if not larger than those from VBF at $\mathcal{O}(30)$ TeV, i.e., the highest energy that we consider.
In terms of statistics only, $s$-channel annihilations at lower energies serve as a larger source of $t \bar t $ events.
Hence, one may wonder if there is any gain  in going to higher $\sqrt{s}$.
This is addressed at length in section~\ref{sec:eft}.
Here it suffices to say that sensitivity to anomalous couplings  greatly improves with increasing $\sqrt{s}$, in particular for VBF processes. 
For example: 
At lowest order, $\mpmm\to \gamma^*/Z^* \to t\overline{t}$ is only sensitive to anomalous $ttZ/\gamma^*$ and $\mu\mu t t$ interactions;
the channel is insensitive, e.g., to unitarity cancellations in the Higgs sector.
This is unlike $W^+W^- \to  t\overline{t}$, which is also sensitive to anomalous $WWH, ttH$, and $tWb$ couplings, including relative CP phases.
In addition, the VBF channel features a strong, non-Abelian gauge cancellation, and therefore probes anomalous contributions that are enhanced by energy factors. 

A second interesting observation is the hierarchy of $t\overline{t}XX$ production from $W^+W^-$ fusion.
As seen in figure~\ref{fig:SMt_ttVV}, the rates for $t\overline{t}VV$ $(V=W,Z)$  between $\sqrt{s}=3-30\TeV$ systematically sit about an order of magnitude higher than $t\overline{t}HV$,
which in turn is another order of magnitude higher than $t\overline{t}HH$.
In fact, the $t\overline{t}HH$ rate sits just under the mixed EW-QCD $t\overline{t}t\overline{t}$ rate, despite being less phase space-suppressed.
We attribute the strong hierarchy to the relative minus signs among the top quark's Yukawa coupling, the Higgs boson's self-couplings, and the various weak gauge couplings,
which together lead to large destructive interference.

Third, for processes involving neutral bosons $H$ and/or $Z$ in the final state,  VBF cross sections are systematically larger than $s$-channel ones already at collider energies of a few TeV. 
This follows from the strong suppression of the $\ell\ell Z$ gauge coupling relative to the unsuppressed $\ell\nu W$ gauge interaction.
(Numerically, the further power of $\alpha_W$ in $WW$ fusion is still larger than 
the  vector and axial-vector couplings of electrically charged leptons  to $Z$ bosons.) 
Among the processes investigated multi-Higgs production in VBF stands out. 
For the specific cases of $HZ$ annd $HHZ$ production in figures~\ref{fig:SMh_hxxx} and  \ref{fig:SMh_hhxx},
we find that VBF already exceeds $s$-channel annihilation at $\sqrt{s}=2\TeV$ and $4\TeV$, respectively.
  
Lastly,  the energy growth of VBF scattering rates is in general steeper for final states with larger particle multiplicities than for lower ones.
This is due to many reasons.
The first is that the increase in energy crucially  opens phase space. 
For example: $t\overline{t}WW$ and $t\overline{t}HH$ have kinematic thresholds of $M_{\min}\approx 0.5\TeV$ and $0.6\TeV$,
indicating that their  VBF production rates are phase space-starved for $\sqrt{s}\lesssim2-3\TeV$.
The second relates to (collinear) logarithmic enhancements in processes with $t$-channel gauge bosons.
Final states with $m$ gauge bosons entail contributions from the  exchange of $m$ $t$-channel  gauge bosons.
At very high energies, such contributions become dominant and give rise to cross sections that  behave at least as $\sigma \sim \log^m (s/M_V^2)$.
Even though this largest log might not  always be dominant, 
we verify that the growth pattern as a function of final-state multiplicity corresponds to this expectation and is rather clearly visible in plotted curves.

\subsection{$ZZ$, $Z\gamma$, and $\gamma\gamma$  fusion}\label{sec:sm_zz}

\begin{table}[!t]
\begin{center}
\renewcommand*{\arraystretch}{0.95}
\begin{tabular}{l||cccc}
\toprule
\toprule
$\sigma$ [fb] &  $\sqrt s=$ 1 TeV &  $\sqrt s=$ 3 TeV &  $\sqrt s=$ 14 TeV &  $\sqrt s=$ 30 TeV\\
\midrule
\midrule
$t \bar{t}$     &  1.0 $\cdot 10^{-1}$ &   1.1 $\cdot 10^{0}$ &   4.3 $\cdot 10^{0}$ &   6.2 $\cdot 10^{0}$ \\
\hline
$t \bar{t} Z$   &  1.2 $\cdot 10^{-4}$ &  6.7 $\cdot 10^{-3}$ &  5.2 $\cdot 10^{-2}$ &  8.5 $\cdot 10^{-2}$ \\
$t \bar{t} H$   &  5.3 $\cdot 10^{-5}$ &  2.8 $\cdot 10^{-3}$ &  2.7 $\cdot 10^{-2}$ &  5.0 $\cdot 10^{-2}$ \\
\hline
\hline
$H$             &   1.5 $\cdot 10^{1}$ &   3.8 $\cdot 10^{1}$ &   7.6 $\cdot 10^{1}$ &   9.6 $\cdot 10^{1}$ \\
$H H$           &  5.0 $\cdot 10^{-3}$ &  7.3 $\cdot 10^{-2}$ &  4.3 $\cdot 10^{-1}$ &  7.5 $\cdot 10^{-1}$ \\
$H H H$         &  3.6 $\cdot 10^{-7}$ &  3.1 $\cdot 10^{-5}$ &  8.4 $\cdot 10^{-4}$ &  2.3 $\cdot 10^{-3}$ \\
\hline
$H W W$         &  3.5 $\cdot 10^{-3}$ &  1.4 $\cdot 10^{-1}$ &   1.7 $\cdot 10^{0}$ &   3.8 $\cdot 10^{0}$ \\
$H Z Z$         &  2.5 $\cdot 10^{-5}$ &  4.9 $\cdot 10^{-4}$ &  3.6 $\cdot 10^{-3}$ &  5.9 $\cdot 10^{-3}$ \\
\hline
\hline
$W W$           &   2.2 $\cdot 10^{1}$ &   1.4 $\cdot 10^{2}$ &   5.2 $\cdot 10^{2}$ &   8.1 $\cdot 10^{2}$ \\
$Z Z$           &  1.2 $\cdot 10^{-1}$ &  4.0 $\cdot 10^{-1}$ &  7.4 $\cdot 10^{-1}$ &  8.0 $\cdot 10^{-1}$ \\
\bottomrule
\bottomrule
\end{tabular}
\caption{
SM cross sections [fb] for sample $ZZ/Z\gamma/\gamma\gamma$  fusion processes (with interference) in $\mpmm$ collisions at representative collider energies [TeV].
}
\label{tab:ZZSM}
\end{center}
\end{table}

We continue our survey at a potential multi-TeV $\mpmm$ facility by now exploring processes mediated through the neutral gauge bosons $Z$ and $\gamma$.
For a subset of final states considered in section~\ref{sec:sm_ww} for $W^+W^-$ fusion that can instead proceed through $ZZ$, $Z\gamma$, and $\gamma\gamma$ fusion,
we report in table~\ref{tab:ZZSM} cross sections [fb] for representative collider energies.
As described in section~\ref{sec:sm_technicalities} we do not remove diagrams by hand and include $\gamma/Z$ interference.
To regulate phase space singularities,  a $p_T$ cut of 30 GeV is applied on outgoing charged leptons.

As foreseen from the scaling of the $ZZ$ luminosity in section~\ref{sec:ppvsmuon}, the cross sections for $ZZ/Z\gamma/\gamma\gamma$ fusion are smaller than for $WW$ by roughly an order of magnitude.
The exceptions to this are (i) $W^+W^-$ production, which is highly comparable to the $W^+W^-\to W^+W^-$ rate, and (ii) $ZZ$ production, which is about two orders of magnitude smaller than $W^+W^-\to ZZ$.
Despite being lower, these rates are not small enough to be neglected.
Indeed, $HH$ production already  reaches $\sigma\sim5$ ab at $\sqrt{s}=1\TeV$ and grows to be as large as $\sigma\sim430~(750)\ab$ at $\sqrt{s}=14~(30)\TeV$.
Moreover, the presence of final-state charged leptons from $Z/\gamma$ splittings, for example, could be exploited to obtain a full reconstruction of the event.
For some particular channels it may also be useful to have charged lepton pairs to better identify a new resonance signal or increase sensitivity to an anomalous coupling. 
A simple but important example that is applicable to both the SM and BSM is  the production of invisible final states,  for example the SM process $WW/ZZ \to H \to   4\nu$.
Whereas the $WW$ production mode would lead to a \textit{totally} invisible final state, the $ZZ$ mode gives a means to tag the process. Numerous  BSM examples can also be constructed.

\subsection{$W  Z$ and $W\gamma$ scattering}\label{sec:sm_wz}

\begin{table}[t!]
\begin{center}
\renewcommand*{\arraystretch}{0.95}
\begin{tabular}{l||cccccccc}
\toprule
\toprule
\multirow{1}{*}{$\sigma$ [fb]} &        \multicolumn{1}{c}{$\sqrt s=$ 1 TeV} &    \multicolumn{1}{c}{$\sqrt s=$ 3 TeV} &    \multicolumn{1}{c}{$\sqrt s=$ 14 TeV} &      \multicolumn{1}{c}{$\sqrt s=$ 30 TeV}\\
\midrule 
\midrule 
$W$&$9.9\cdot 10^2$& $2.4\cdot 10^3$& $4.6\cdot 10^3$& $5.7\cdot 10^3$\\
\hline
$WZ$&$5.8\cdot 10^0$& $5.0\cdot 10^1$& $2.3\cdot 10^2$& $3.7\cdot 10^2$\\
$WH$&$8.4\cdot 10^{-1}$& $7.2\cdot 10^0$& $3.3\cdot 10^1$& $5.5\cdot 10^1$\\
\hline
$WWW$&$1.4\cdot 10^{-1}$& $4.2\cdot 10^0$& $4.4\cdot 10^1$& $1.0\cdot 10^2$\\
$WZZ$&$1.8\cdot 10^{-2}$& $8.0\cdot 10^{-1}$& $1.0\cdot 10^1$& $2.3\cdot 10^1$\\
$WZH$&$1.7\cdot 10^{-3}$& $8.0\cdot 10^{-2}$& $1.1\cdot 10^0$& $2.5\cdot 10^0$\\
$WHH$&$9.5\cdot 10^{-5}$& $6.2\cdot 10^{-3}$& $9.7\cdot 10^{-2}$& $2.3\cdot 10^{-1}$\\
\hline\hline
$t\bar b$&$4.4\cdot 10^{-1}$& $2.9\cdot 10^0$& $9.5\cdot 10^0$& $1.3\cdot 10^1$\\
\hline
$t\bar b Z$&$1.3\cdot 10^{-3}$& $4.4\cdot 10^{-2}$& $4.1\cdot 10^{-1}$& $8.0\cdot 10^{-1}$\\
$t\bar b H$&$1.5\cdot 10^{-4}$& $6.6\cdot 10^{-3}$& $6.6\cdot 10^{-2}$& $1.3\cdot 10^{-1}$\\
$t\bar tW$&$1.0\cdot 10^{-3}$& $7.6\cdot 10^{-2}$& $9.0\cdot 10^{-1}$& $1.9\cdot 10^0$\\
\bottomrule
\bottomrule
\end{tabular}
\caption{Same as table~\ref{tab:ZZSM} but for $WZ/W\gamma$ fusion.}
\label{tab:chargedSM}
\end{center}
\end{table}

Turning away from final states with zero net electric charge, we now explore processes mediated by $W\gamma$ and $WZ$ fusion.
For several representative processes, we summarize in table~\ref{tab:chargedSM}  their cross sections at our benchmark muon collider energies.
We apply a $p_T$ cut of 30 GeV on outgoing charged leptons to regulate phase space singularities.
Once again, following simple scaling arguments of the EWA luminosities in section~\ref{sec:ppvsmuon},
we expect and observe that cross  sections here are somewhere between those of $WW$ and $ZZ$ fusion.

With the present VBF configuration, we find that the rates for $VVV$, $VVH$, and $VHH$ production (where $V=W/Z$) all exceed the $\sigma\sim1$ ab threshold at $\sqrt{s}=3\TeV$.
At $\sqrt{s}=1\TeV$, the $VHH$ processes are strongly phase space-suppressed.
At $\sqrt{s}=14\TeV$, we find that the $VVH$ and $VHH$ rates reach roughly the $\sigma\sim1~(0.1)$ fb level and more than double at $\sqrt{s}=30\TeV$.
Moreover, as the final states here are charged, the potential arises for qualitatively different signatures that cannot be produced via $s$-channel annihilations.
For example: 
\confirm{
processes such as single $W$ production (with $\sigma\sim\mathcal{O}(1-5)$ pb),
single top quark (with $\sigma\sim\mathcal{O}(0.5-10)$ fb),
as well as 
$WWW$ (with $\sigma\sim\mathcal{O}(0.1-100)$ fb)
}
all have appreciable cross sections for $\sqrt{s}=1-30\TeV$.
If one assumes $\mathcal{O}(1-100)\invab$ datasets, then in these cases, 
interesting, ultra rare and ultra exclusive decay channels can be studied.

\subsection{$W^+  W^+$ fusion}\label{sec:sm_samesign}

Finally, we conclude our EW VBF survey by briefly  exploring the case of same-sign muon collisions.
This setup allows the production of doubly charged final states and therefore, as we discuss in section~\ref{sec:bsm},  is a natural setup where one can study lepton number-violating processes~\cite{Cai:2017mow}. 
For concreteness, we consider $\mu^+\mu^+$ collisions and in table~\ref{tab:doublychargedSM} present  the cross sections for representative $VV$ and $VVH$ processes
at our benchmark collider energies.

We report that the production rates for $VV$ and $VVH$ are highly comparable to those for $W^+W^-$ fusion in table~\ref{tab:neutralSM}.
We anticipate this from CP invariance. This dictates that the $W^+W^-$ luminosity in $\mpmm$ collisions is the same at lowest order
as the $W^+W^+$ luminosity in $\mu^+\mu^+$ collisions.
Differences between the two sets of rates originate from differences between the $W^+W^-\to X$ and analogous $W^+W^+\to X'$ matrix elements.
In $W^+W^+$ scattering, only $t$-channel exchanges of gauge and scalar bosons are allowed as there does not exist a doubly charged state in the EW sector.
In $W^+W^-$ scattering, these $t$-channel diagrams interfere (constructively and destructively) with allowed $s$-channel diagrams.

\begin{table}[t!]
\begin{center}
\renewcommand*{\arraystretch}{0.95}
\begin{tabular}{l||cccccccc}
\toprule
\toprule
\multirow{1}{*}{$\sigma$ [fb]} &        \multicolumn{1}{c}{$\sqrt s=$ 1 TeV} &    \multicolumn{1}{c}{$\sqrt s=$ 3 TeV} &    \multicolumn{1}{c}{$\sqrt s=$ 14 TeV} &      \multicolumn{1}{c}{$\sqrt s=$ 30 TeV}\\
\midrule
\midrule
$W^+ W^+$&2.2$\cdot 10^{1}$&1.4$\cdot 10^{2}$&5.6$\cdot 10^{2}$&9.0$\cdot 10^{2}$\\
$W^+ W^+ Z$&1.2$\cdot 10^{-1}$&4.2$\cdot 10^{0}$&4.9$\cdot 10^{1}$&1.1$\cdot 10^{2}$\\
$W^+ W^+ H$&9.3$\cdot 10^{-3}$&3.1$\cdot 10^{-1}$&3.7$\cdot 10^{0}$&8.5$\cdot 10^{0}$\\
\bottomrule
\bottomrule
\end{tabular}
\caption{SM cross sections [fb] for sample $W^+W^+$  fusion processes in $\mu^+\mu^+$ collisions at representative collider energies [TeV].}\label{tab:doublychargedSM}
\end{center}
\end{table}

\section{Precision electroweak measurements}\label{sec:eft}
In this section we explore the potential of a muon collider to probe new physics indirectly.
As it is not realistic to be exhaustive, 
after summarizing the effective field theory formalism in which we work (section~\ref{sec:eft_intro}),
we select a few representative examples related to the Higgs boson (section~\ref{sec:eft_higgs}) and the top quark (section~\ref{sec:eft_top}). 

\subsection{SMEFT formalism}\label{sec:eft_intro}
Undertaking precision measurements of SM observables is  of utmost importance if nature features heavy resonances at mass scales that are just beyond  the kinematic reach of 
laboratory experiments.
Be it perturbative or non-perturbative, the dynamics of such new states could leave detectable imprints through their interactions among the SM particles.
This is especially the case for the heaviest SM particles if the new physics under consideration is related to the flavor sector or the spontaneous breaking  of EW symmetry.

Generically, two broad classes of observables, defined in different regions of phase space, can be investigated. 
The first are bulk, or inclusive,  observables for which large statistics are available and even small deviations from the null (SM) hypothesis are detectable.
The second are  tail, or exclusive, observables, where the effects of new physics can be significantly enhanced by energy, say with selection cuts, and compensate for lower statistics.

A simple yet powerful approach to interpret indirect searches for new, heavy particles in low-energy observables is the
SMEFT framework \cite{Grzadkowski:2010es,Aebischer:2017ugx,Brivio:2017btx}. 
The formalism describes a large class of models featuring states that live above the EW scale and provides a consistent, quantum field theoretic description of deformations of SM interactions.
This  is done while employing a minimal set of assumptions on the underlying, ultraviolet theory.  
In SMEFT, new physics is parametrized through higher dimensional, i.e., irrelevant, operators that augment the unbroken SM Lagrangian,
yet preserve the fundamental gauge symmetries of the SM
by only admitting operators that are both built from SM fields and invariant under  $\mathcal{G}_{\rm SM}=SU(3)_c\times SU(2)_L \times U(1)_Y$ gauge transformations.
Accidental symmetries of the SM, such as lepton and baryon number conservation, are automatically satisfied under certain stipulations~\cite{Kobach:2016ami,Helset:2019eyc}.
Additional global symmetries can also be imposed on the Lagrangian. 
In this work we assume the flavor 
symmetry,\footnote{The labels $l, e, d, u, q$ refer, respectively, to the left-handed lepton doublets, the right-handed leptons, the right-handed down-type quarks, the right-handed up-type quarks, and the left-handed quark doublets.}
$\mathcal{S}=U(3)_l \times U(3)_e \times U(3)_d \times U(2)_u \times U(2)_q$.
This  helps reduce the number of independent degrees of freedom while simultaneously   singling out the top quark as a window onto new physics.

\begin{table}
{\centering
\begin{tabular}{|ll|ll|}
    \hline
     $\Op{W}$&
     $\varepsilon_{\sss IJK}\,W^{\sss I}_{\mu\nu}\,
                             {W^{{\sss J},}}^{\nu\rho}\,
                             {W^{{\sss K},}}^{\mu}_{\rho}$&
     $\Op{t\phi}$&
     $\left(\pdp-\tfrac{v^2}{2}\right)
     \bar{Q}\,t\,\tilde{\phi} + \text{h.c.}$
     \tabularnewline
     $\Op{\phi W}$&
     $\left(\pdp-\tfrac{v^2}{2}\right)W^{\mu\nu}_{\sss I}\,
                                    W_{\mu\nu}^{\sss I}$&
     $\Op{tW}$&
     $i\big(\bar{Q}\sigma^{\mu\nu}\,\tau_{\sss I}\,t\big)\,
     \tilde{\phi}\,W^I_{\mu\nu}
     + \text{h.c.}$
     \tabularnewline
     $\Op{\phi B}$&
     $\left(\pdp-\tfrac{v^2}{2}\right)B^{\mu\nu}\,
                                    B_{\mu\nu}$&
     $\Op{tB}$&
     $i\big(\bar{Q}\sigma^{\mu\nu}\,t\big)
     \,\tilde{\phi}\,B_{\mu\nu}
     + \text{h.c.}$
     \tabularnewline
     \cline{3-4}
     $\Op{\phi WB}$&
     $(\phi^\dagger \tau_{\sss I}\phi)\,B^{\mu\nu}W_{\mu\nu}^{\sss I}\,$&
     $\Op{\phi Q}^{\sss(3)}$&
     $i\big(\phi^\dagger\lra{D}_\mu\,\tau_{\sss I}\phi\big)
     \big(\bar{Q}\,\gamma^\mu\,\tau^{\sss I}Q\big)$
     \tabularnewline

     $\Op{\phi D}$&
     $(\phi^\dagger D^\mu\phi)^\dagger(\phi^\dagger D_\mu\phi)$&
     $\Op{\phi Q}^{\sss(1)}$&
     $i\big(\phi^\dagger\lra{D}_\mu\,\phi\big)
     \big(\bar{Q}\,\gamma^\mu\,Q\big)$
     \tabularnewline
     $\Op{\phi d}$&
     $(\varphi^\dagger\varphi)\square(\varphi^\dagger\varphi)$ &
     $\Op{\phi t}$&
     $i\big(\phi^\dagger\lra{D}_\mu\,\phi\big)
     \big(\bar{t}\,\gamma^\mu\,t\big)$
      \tabularnewline

     $\Op{\phi}$&
     $\left(\pdp-\tfrac{v^2}{2}\right)^3$&     
      &
     
     \tabularnewline
      \hline
  \end{tabular}

\caption{\label{tab:OP_DEF}
SMEFT operators at dimension-six relevant for the Higgs boson and the top quark in EW observables, in the so-called Warsaw basis~\cite{Grzadkowski:2010es},
and where a $U(3)^3\times U(2)^2$ flavor symmetry is assumed.  
$Q,\,t$, and $b$ denote the third generation components of $q,\,u$, and $d$.
}
}
 \end{table}

Under these assumptions, then after neglecting the Weinberg operator at dimension five  
and truncating the EFT expansion at dimension six, the SMEFT Lagrangian is
\begin{equation}
\mathcal{L}_{ SMEFT} = \mathcal{L}_{ SM} + \frac{1}{\Lambda^2} \sum C_i \mathcal{O}_i \, .
\end{equation}
Here, $C_i$ are the dimensionless Wilson coefficients of the dimension-six operators $\mathcal{O}_i$.
In the absence of additional symmetries, such as the flavor symmetry $\mathcal{S}$ defined above,
the number of independent  $\mathcal{O}_i$ stands at  59 if one considers only one generation of fermions and 
 2499 with three generations.
In practice, one usually studies only a subset of operators in order to establish the sensitivity of a measurement.
Since we are mainly interested in the top quark and Higgs sectors, we consequentially retain only operators that explicitly involve top or Higgs fields and affect EW observables.
 The full list of operators that we consider is given in table~\ref{tab:OP_DEF}, where the following conventions are adopted:
 \begin{align}
   \phi^\dag {\overleftrightarrow D}_\mu \phi=\phi^\dag D_\mu\phi-(D_\mu\phi)^\dag\phi,
   &\qquad
   \phi^\dag \tau_{\sss K} {\overleftrightarrow D}^\mu \phi=
   \phi^\dag \tau_{\sss K}D^\mu\phi-(D^\mu\phi)^\dag \tau_{\sss K}\phi, 
   \\
   W^{\sss K}_{\mu\nu} = \partial_\mu W^{\sss K}_\nu 
   - \partial_\nu W^{\sss K}_\mu 
   + g \epsilon_{\sss IJ}{}^{\sss K} \ W^{\sss I}_\mu W^{\sss J}_\nu,
   &\qquad
   B_{\mu\nu} = \partial_\mu B_\nu - \partial_\nu B_\mu, \\
   D_\mu\phi = \left(\partial_\mu -  i \frac{g}{2} \tau_{\sss K} W_\mu^{\sss K} - i\frac12 g^\prime B_\mu\right)\phi.
      &\quad
 \end{align}
 Here, $\tau_I$ denotes the Pauli $\sigma$ matrices, and $\epsilon_{IJK}$ is  antisymmetric and normalized to unity.

\begin{table}[t]
\begin{center}
\resizebox{\textwidth}{!}{
\hspace*{-0.5cm}
\begin{tabular}{|c|c|c|c|c|c|}
\hline
\multirow{2}{*}{Operators} & \multicolumn{2}{c|}{ Limit on $C_i$ $\TeV^{-2}$} & \multirow{2}{*}{Operators} & \multicolumn{2}{c|}{ Limit on $C_i$ $\TeV^{-2}$}\\
\cline{2-3}\cline{5-6}
{}                  & Individual    & Marginalized &  {}                 & Individual    & Marginalized  \\
\hline
 $\Op{\phi D}$            & [-0.021,0.0055]~\cite{Ellis:2018gqa}  & [-0.45,0.50]~\cite{Ellis:2018gqa}     &   
 $\Op{t \phi}$            & [-5.3,1.6]~\cite{Hartland:2019bjb}    & [-60,10]~\cite{Hartland:2019bjb}\\
\hline
 $\Op{\phi d}$         & [-0.78,1.44]~\cite{Ellis:2018gqa}     & [-1.24,16.2]~\cite{Ellis:2018gqa}     &   
 $\Op{tB}$                & [-7.09,4.68]~\cite{Buckley:2015lku}   & $-$\\
\hline
 $\Op{\phi B}$            & [-0.0033,0.0031]~\cite{Ellis:2018gqa} & [-0.13,0.21]~\cite{Ellis:2018gqa}     &   
 $\Op{tW}$                & [-0.4,0.2]~\cite{Hartland:2019bjb}   & [-1.8,0.9]~\cite{Hartland:2019bjb}\\
\hline
 $\Op{\phi W}$            & [-0.0093,0.011]~\cite{Ellis:2018gqa}  & [-0.50,0.40]~\cite{Ellis:2018gqa}     &   
 $\Op{\phi Q}^{\sss (1)}$ & [-3.10,3.10]~\cite{Buckley:2015lku}   & $-$\\
\hline
 $\Op{\phi WB}$           & [-0.0051,0.0020]~\cite{Ellis:2018gqa} & [-0.17,0.33]~\cite{Ellis:2018gqa}     &   
 $\Op{\phi Q}^{\sss (3)}$ & [-0.9,0.6]~\cite{Hartland:2019bjb}   & [-5.5,5.8]~\cite{Hartland:2019bjb}\\
\hline
 $\Op{W}$                 & [-0.18,0.18]~\cite{Butter:2016cvz}    & $-$                                   &
 $\Op{\phi t}$            & [-6.4,7.3]~\cite{Hartland:2019bjb}   & [-13,18]~\cite{Hartland:2019bjb}\\
\hline
 {$\Op{\phi}$}                       & $-$                                    & $-$                                    &  
 {}           & {}    & {}\\
\hline
\end{tabular}
} 
\end{center}
\caption{
Limits on the Wilson coefficients $C_i$ [TeV$^{-2}$] for the SMEFT operators listed in table~\ref{tab:OP_DEF}.}
\label{tab:OP_CONSTR}
\end{table}

In the following we perform a simple sensitivity study focusing on the Higgs self-couplings and the top quark's EW couplings.
In table~\ref{tab:OP_CONSTR} we summarize  the current constraints on Wilson coefficients corresponding to the operators in table~\ref{tab:OP_DEF}.

\subsection{Higgs self-couplings at muon colliders}\label{sec:eft_higgs}
A precise determination of the Higgs boson's properties is one of the foremost priorities of the high-energy physics community~\cite{Strategy:2019vxc,EuropeanStrategyGroup:2020pow}. 
At the moment, measurements of the Higgs's couplings to the heaviest fermions and gauge bosons are in full agreement with the SM predictions.
However, there exist several couplings that have yet to be measured, and in some cases  bounds are only weakly constraining.
Among these are the Yukawa couplings to the first and second generation of fermions as well as the shape of the SM's scalar potential.
Subsequently, a determination of the Higgs's trilinear and the quartic  self-couplings, which are now fully predicted in the SM, 
would certainly help elucidate the EW symmetry breaking mechanism~\cite{Chung:2012vg} and its role in the thermal history of the universe.  
 
  Despite this motivation, measurements of the Higgs's self-interactions appears to be too challenging for the LHC, unless substantial deviations from the SM  
 exist~\cite{Sirunyan:2017guj, CMS:2018rig, CMS:2018ccd, ATLAS:2018otd, Aaboud:2018zhh, Aad:2019uzh, Aad:2019yxi, Sirunyan:2018iwt, CMS:2018dvu, ATLAS:2019pbo, ATL-PHYS-PUB-2014-019, ATL-PHYS-PUB-2017-001, Kim:2018cxf}.
 As such, conclusively measuring the Higgs's properties is among the most compelling motivations for constructing 
 a lepton collider at a c.m.~energy of a few hundred GeV.\footnote{It is remarkable that a 100 m radius circular muon collider could reach this energy~\cite{Delahaye:2019omf}.} 
 The case for higher energies is also well-founded.
 For example:
Higgs sensitivity studies for CLIC up to  $\sqrt{s}=3$ TeV~\cite{Roloff:2018dqu, Vasquez:2019muw, Roloff:2019crr, Liu:2018peg, Maltoni:2018ttu, deBlas:2019rxi} 
support the expectation that increasing collider energy  provides additional leverage for precision measurements through VBF channels.
Indeed, as already shown in Fig.~\ref{fig:SMh}, 
VBF processes emerge as the dominant vehicles for $H, HH,$ and even $HHH$ production at high-energy lepton colliders and surpass $s$-channel processes below $\sqrt{s}=3$ TeV. 
Likewise, at $\sqrt{s}=10$ TeV and  with a benchmark luminosity of $\mathcal{L}=10 \invab$, one anticipates $8\cdot 10^6$ Higgs bosons in the SM \cite{Delahaye:2019omf}.
As backgrounds are expected to be under good control, multi-TeV muon colliders essentially function as de facto Higgs factories.

Certainly, the limitations to extracting the Higgs's self-couplings at the LHC and $e^+e^-$ colliders motivate other opportunities, particularly those offered by muon colliders.
However, past muon collider studies on the Higgs have been limited in scope,
focusing largely on properties determination within the SM~\cite{Barger:1995hr,Han:2012rb,Greco:2016izi}
and
its minimal extensions~\cite{Barger:1995hr,Chakrabarty:2014pja,Buttazzo:2018qqp}.
Only recently have more robust, model-independent investigations been conducted~\cite{Ruhdorfer:2019utl,Chiesa:2020awd}. 
Expanding on this work, we  perform in this section a first exploratory study on determining the SM's full scalar potential in a model-independent fashion using SMEFT.

Within the SMEFT framework, three operators directly modify the Higgs potential: 
\begin{equation}
\mathcal{O}_{\phi}, \quad \mathcal{O}_{\phi d}, \quad\text{and}\quad \mathcal{O}_{\phi D}.
\label{eq:eft_higgs_dim6ops}
\end{equation} 
The first contributes to the Higgs potential's cubic and quartic terms and shifts the field's ($\varphi$'s) vev $v$.
The latter two modify the Higgs boson's kinetic term and a field redefinition is necessary to recover the canonical normalization.
All of these operators give a contribution to VBF production of $H, HH,$ and $HHH$  through the following Lagrangian terms:
\begin{align}
&\mathcal{O}_{\varphi}= \left(\varphi^\dagger \varphi - \frac{v^2}{2}\right)^3 \supset v^3 H^3 + \frac{3}{2}v^2 H^4, \label{eq:op_expansion_varphi}\\
&\mathcal{O}_{\varphi d}= \left(\varphi^\dagger\varphi\right)\Box \left(\varphi^\dagger\varphi\right) \supset 2v \left(H\Box H^2 + H^2\Box H\right) + H^2 \Box H^2, \\
& \mathcal{O}_{\varphi D}= \left(\varphi^\dagger D _\mu \varphi\right)^\dagger\left(\varphi^\dagger D^\mu\varphi\right) \supset \frac{v}{2}H \partial_\mu H \partial^\mu H +\frac{H^2}{4} \partial_\mu H \partial^\mu H.
\label{eq:op_expansion}
\end{align}

For conciseness, we investigate only the impact of $\mathcal{O}_{\phi}$ and $\mathcal{O}_{\phi d}$ on prospective  Higgs's self-coupling measurements.
We neglect $\mathcal{O}_{\phi D}$ since it also modifies couplings to gauge bosons and hence is already well-constrained by precision EW measurements.
(See table~\ref{tab:OP_CONSTR} for details.)
In the following, we consider a high-energy $\mpmm$ collider at a c.m.~energy of  $\sqrt{s}=3, 14$, and 30 TeV, 
with respective benchmark luminosities  $\mathcal{L}=6, 20$, and 100 $\invab$.

For the processes under consideration, we first discuss the impact of a single operator on inclusive cross sections while fixing all other higher dimensional Wilson coefficients to zero. 
Within the SMEFT, the total cross section $(\sigma)$ of a process can be expressed by
\begin{equation}
\sigma = \sigma_{ SM} + \sum_i c_i \sigma_{ Int}^i + \sum_{i,j} c_{i,j} \sigma_{ Sq}^{i,j} \, .
\label{eq:eft_xsec_smeft}
\end{equation}
Here the $\sigma_{ Int}^i$ are the leading corrections in the $\Lambda$ power counting to the SM cross sections $( \sigma_{ SM})$
and are given by the interference between SM amplitudes and SMEFT amplitudes at $\mathcal{O}(\Lambda^{-2})$.
The $\sigma_{ Sq}^{i,j}$ corrections are the square contributions at  $\mathcal{O}(\Lambda^{-4})$, and come purely from SMEFT amplitudes at $\mathcal{O}(\Lambda^{-2})$.
The indices $i,j$ run through the set of operators that directly affect the process.
We work under the assumption that the Wilson coefficients $C_i$ for operators in equation~\ref{eq:eft_higgs_dim6ops} are real.
This indicates that the coefficients in $\sigma$ correspond to $c_i = C_i$ and $c_{i,j} = C_i C_j$.
As a na\"ive measure of the sensitivity to the dimension-six operators $\mathcal{O}_i$
and considering only one operator at the time, we define the ratio
\begin{equation}
R(c_i)\equiv \frac{\sigma}{\sigma_{ SM}}= 1+ c_i\frac{\sigma_{ Int}^i}{\sigma_{ SM}} + c_{i,i}^2 \frac{\sigma_{ Sq}^{i,i}}{\sigma_{ SM}} = 1 + r_i + r_{i,i}.
\label{Eq:sens_ratio}
\end{equation}

\begin{figure}[t]
\centering\mbox{\subfigure[]{\includegraphics[width=.45\textwidth]{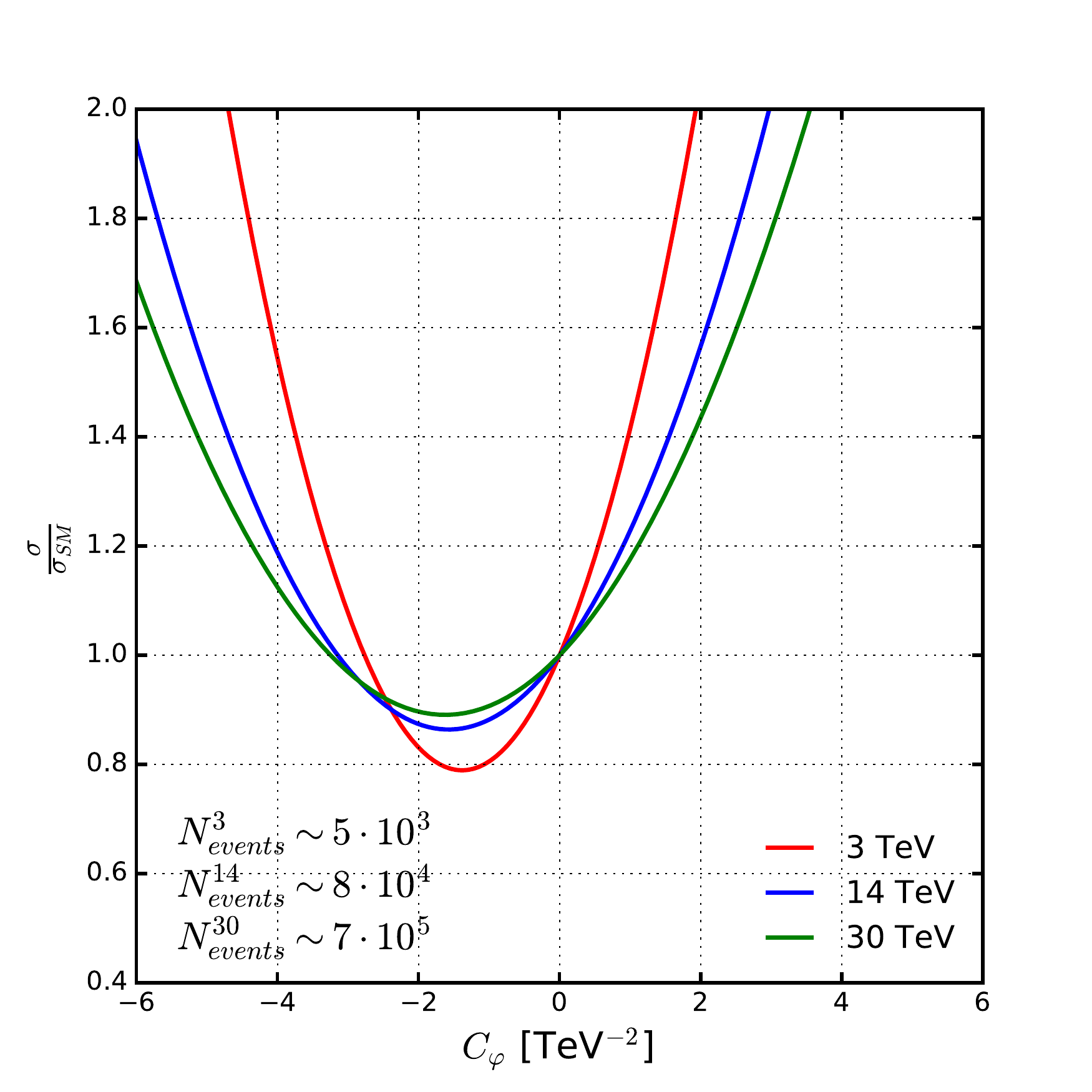}}}
\centering\mbox{\subfigure[]{\includegraphics[width=.45\textwidth]{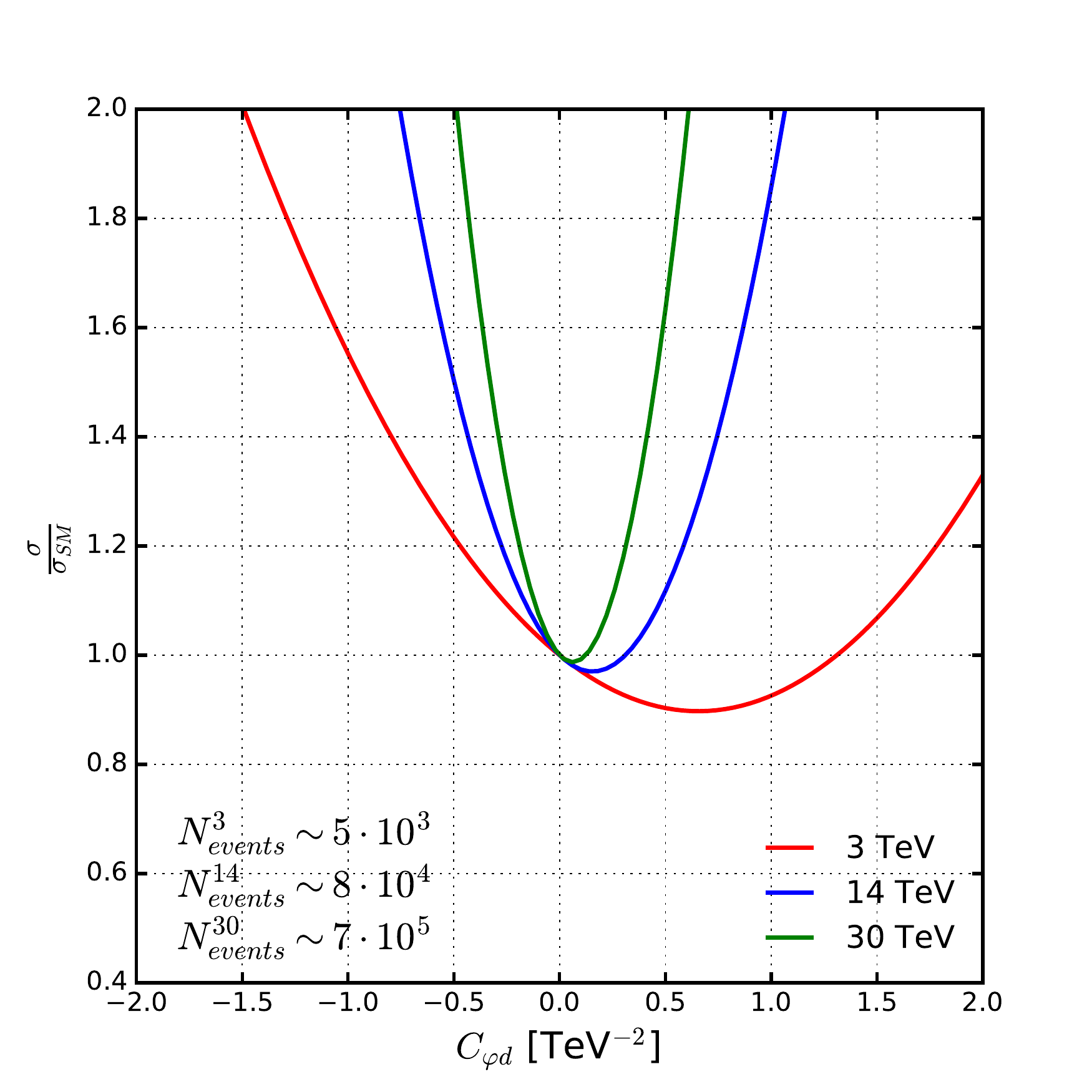}}}
\caption{Sensitivity to Higgs pair production from VBF 
as a function of the Wilson coefficients for (a) $C_{\varphi}$  and (b) $C_{\varphi d}$ (right panel) at $\sqrt{s}=3$ TeV (red), 14 TeV (blue) and 30 TeV (green).}
\label{fig:WW_HH}
\end{figure}

\begin{figure}[t]
\centering\mbox{\subfigure[]{\includegraphics[width=.45\textwidth]{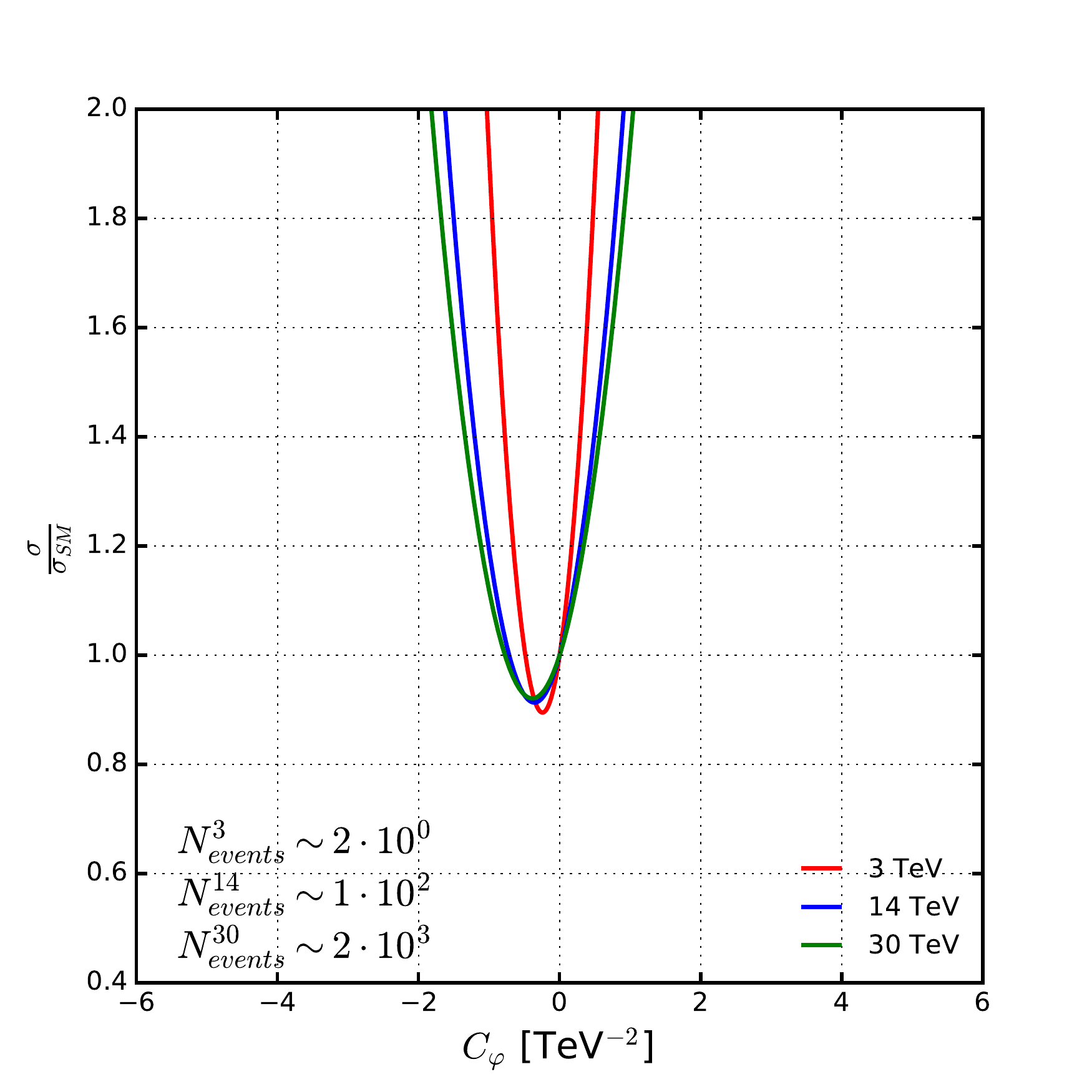}}}
\centering\mbox{\subfigure[]{\includegraphics[width=.45\textwidth]{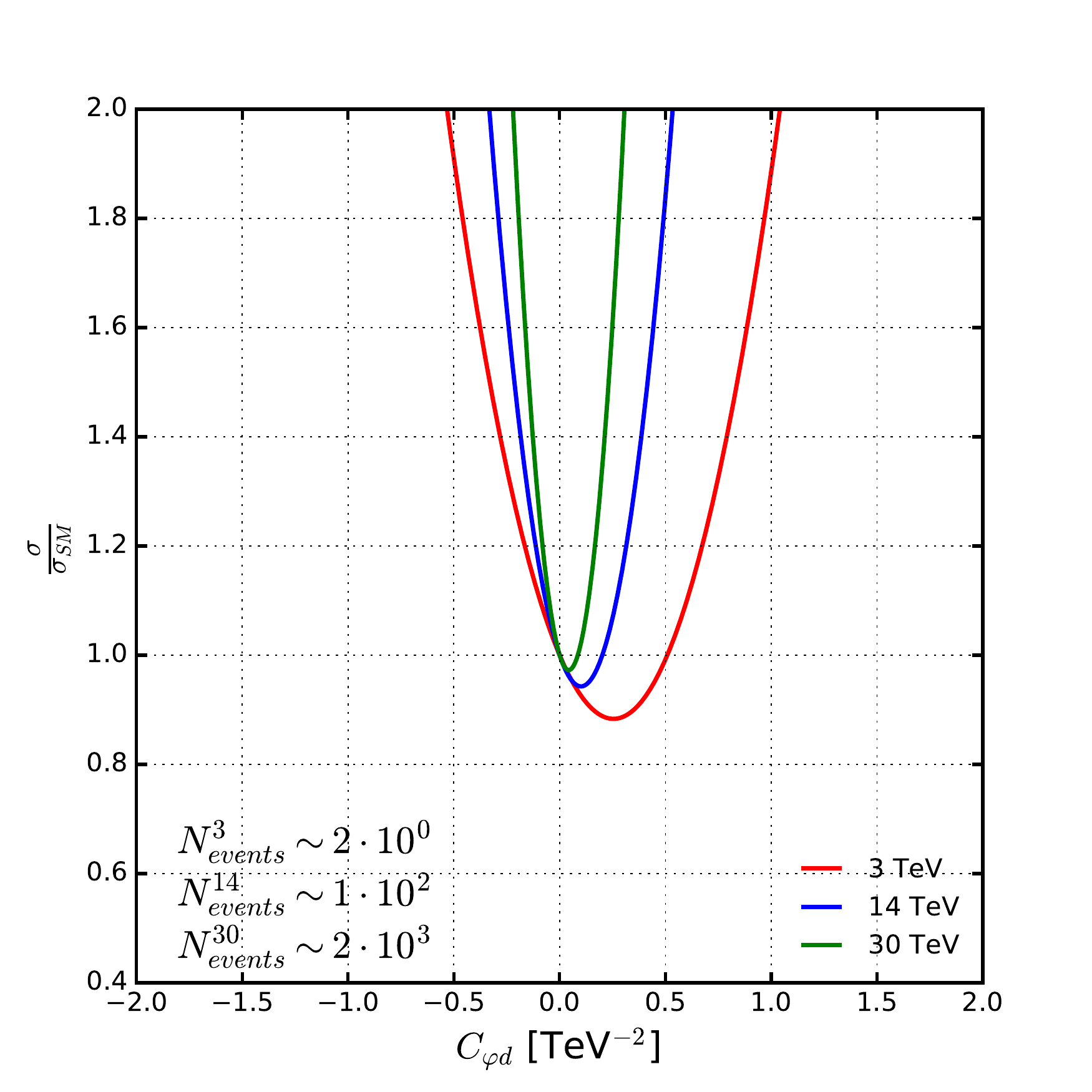}}}
\caption{Same as figure~\ref{fig:WW_HH} but for triple Higgs production from VBF.}
\label{fig:WW_HHH}
\end{figure}

In figures~\ref{fig:WW_HH} and \ref{fig:WW_HHH}, we respectively plot the sensitivity ratio, as defined in equation~\ref{Eq:sens_ratio},
for $HH$ and $HHH$ production from VBF in  $\mpmm$ collisions 
as a function of Wilson coefficients for operators (a) $\mathcal{O}_{\varphi}$ and (b) $\mathcal{O}_{\varphi d}$,
for representative collider energies $\sqrt{s}=3$ (red), $14$ (blue) and $30\TeV$ (green).
Immediately, one sees that the two operators affect the ratio $R(c_i)=\sigma/\sigma_{\rm SM}$, and hence prospects for measuring the Higgs's self couplings, in qualitatively different ways.
To explore this, we first note that $\mathcal{O}_{\varphi}$ in equation~\ref{eq:op_expansion_varphi} only shifts the Higgs's trilinear and quartic couplings. 
The operator does not generate an additional energy dependence in the squared matrix element,
apart from that which could be obtained by spoiling  SM unitarity cancellations. 
As a result, the highest sensitivity to $\mathcal{O}_{\varphi}$ is reached near threshold production.
Increasing $\sqrt{s}$ actually results in losing sensitivity to $HH$ production.
Similarly for $HHH$ production, no significant impact on the cross section ratio is observed with increasing the collider energy, 
only a gain in the total number of events stemming from an increasing production rate.
For the particular case of $HHH$ production at $\sqrt{s}=3\TeV$, the cross section is negligible and no measurement for this process can be undertaken. 
Independent of shifts to $R(c_i)$, it is important  to point out that the higher the event rate the more feasible it becomes to study differential distributions of the above processes.
Generically, an increased number of events allow us to more fully explore, and therefore exploit, regions of phase space that are more sensitive to BSM.

Contrary to $\mathcal{O}_{\varphi}$, the operator  $\mathcal{O}_{\varphi d}$ introduces a kinematical $p^2$ dependence in interaction vertices. 
As a consequence, the impact of $\mathcal{O}_{\varphi d}$ grows stronger and stronger as collider energy increases, potentially leading to a substantial gain in sensitivity. 
The imprint of this behavior is visible  in the fact that the $c_i$ interference term between the SM and new physics becomes negligible as  probing energy goes up. 
In this limit, the squared $c_{i,i}$ term dominates as na\"ively expected from power counting at higher energies. 
This follows from the purely new physics contributions  in SMEFT forcing $R(c_i)$ to grow at most as $(E/\Lambda)^4$,
 while the linear $c_i$ contributions force $R(c_i)$ to grow at most as $(E/\Lambda)^2$.
Leaving aside questions of the EFT's validity when $(E/\Lambda)^4$ corrections exceed those at $(E/\Lambda)^2$,
our point is that it is clear that sensitivity to $\mathcal{O}_{\varphi}$ and $\mathcal{O}_{\varphi d}$ are driven by complementary phase space regions. 

As a final comment, we would like to note that while the study of individual SMEFT operators can give important and useful information,
in a realistic BSM scenario, multiple operators would simultaneously contribute to a given observable.
In this more complicated scenario, correlations and numerical cancellations among operators appear, and phenomenological interpretations become more nuanced, more difficult.
If we nevertheless put ourselves in the scenario where a measured cross section $(\sigma)$ is consistent with the SM, 
then we can still define a simplified estimate of the experiment's  constraining power.
In particular, we define the  space of Wilson coefficients 
that predicts a cross section indistinguishable from SM expectation at the 95\% confidence level (CL) by the following:
\begin{equation}
\frac{S}{\sqrt{B}} = \frac{|\mathcal{L} \cdot (\sigma - \sigma_{SM})|}{\sqrt{\mathcal{L} \cdot \sigma_{SM}}} \le 2 \, .
\label{eq:sig_back}
\end{equation}
Here $\sigma$ is the same SMEFT cross section as defined in equation~\ref{eq:eft_xsec_smeft}.
The number of background events $B$ is the SM expectation $(\sigma_{SM})$ at a given luminosity $\mathcal{L}$,
and the number of signal events $S$ is determined from the net difference between SMEFT and SM expectations.

\begin{figure}[t]
\centering\mbox{\subfigure[]{\includegraphics[width=.45\textwidth]{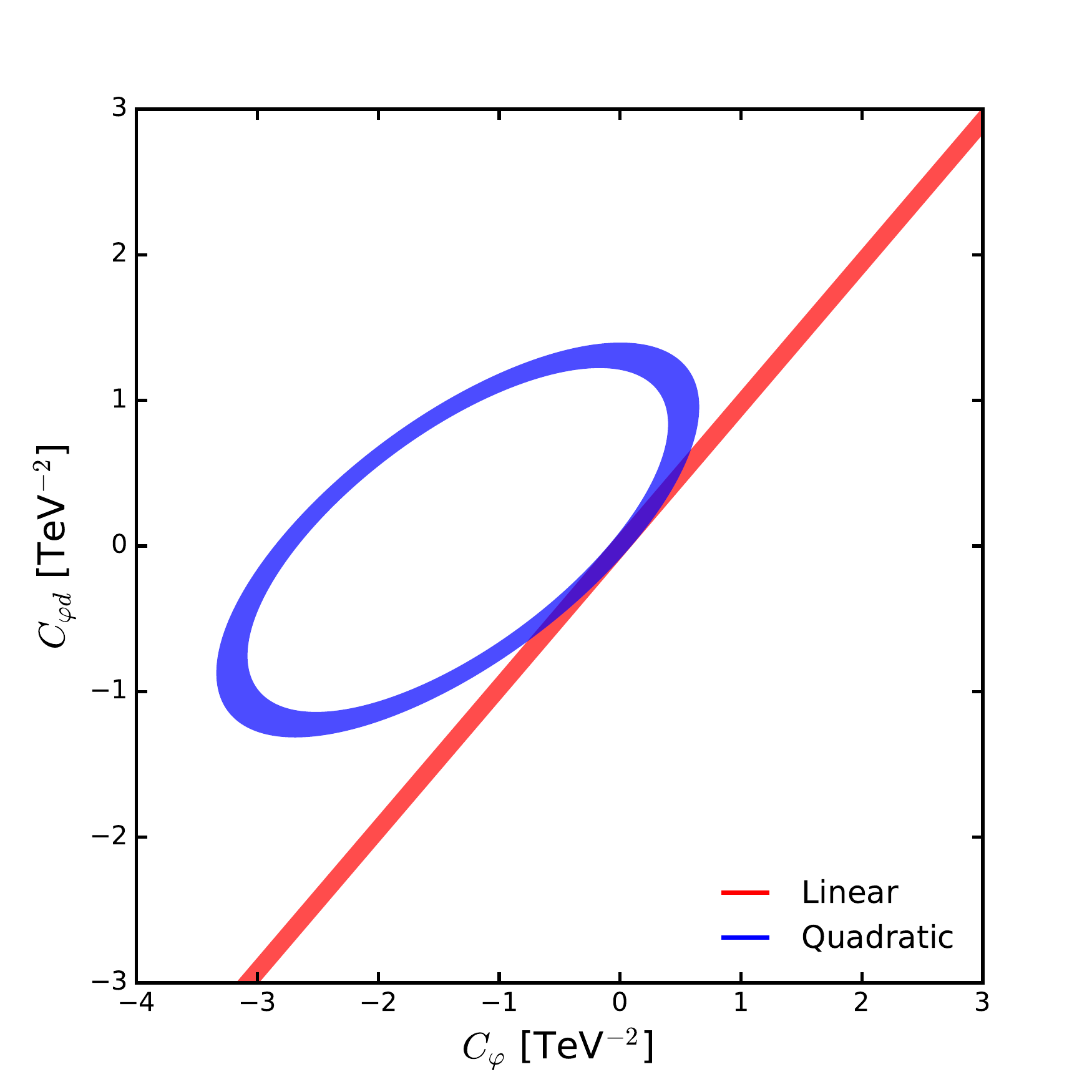}}}
\centering\mbox{\subfigure[]{\includegraphics[width=.45\textwidth]{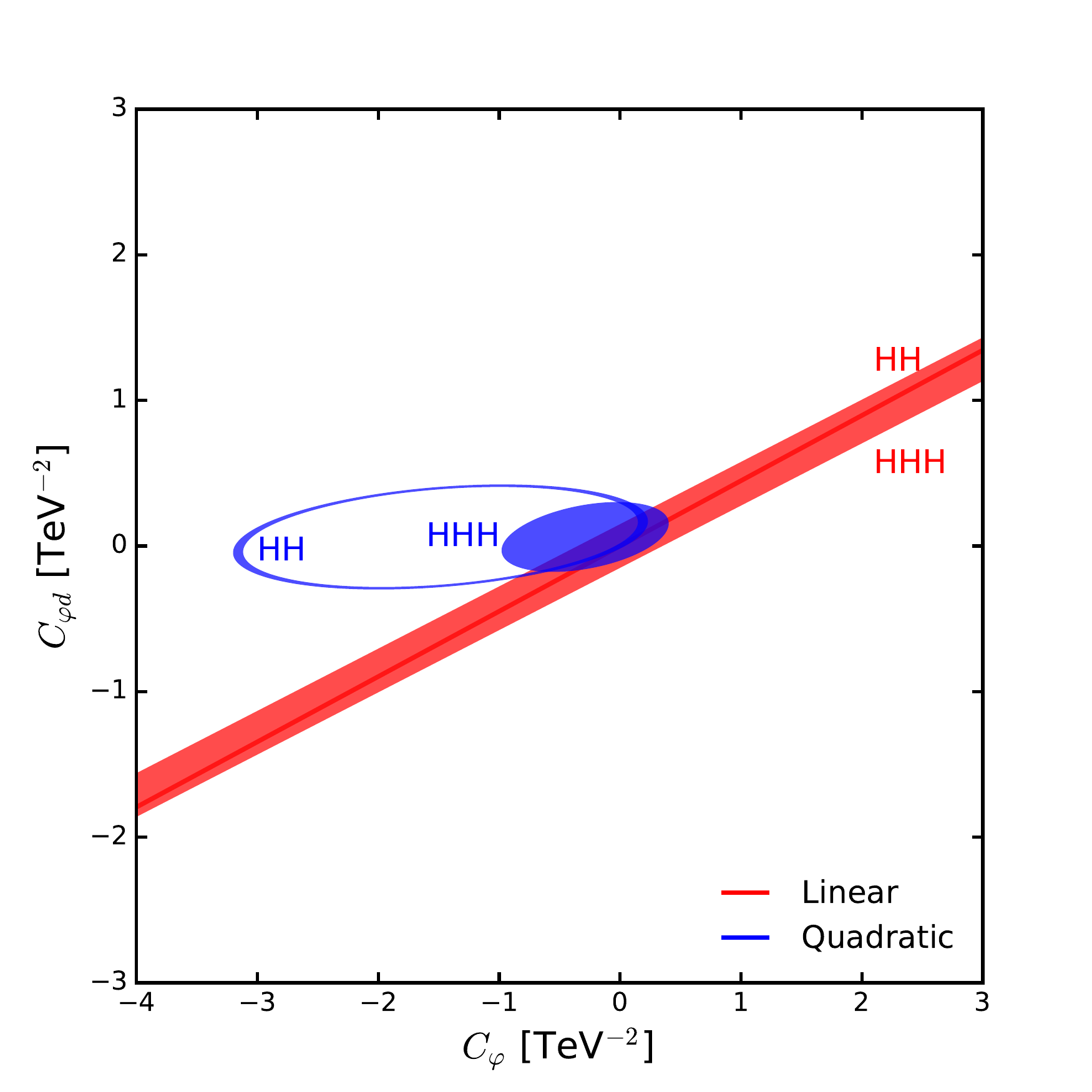}}}
\caption{
Allowed Wilson coefficient space under hypothesizes measurements of (a) $HH$ at $\sqrt{s}= 3$ TeV and (b) both $HH$ plus $HHH$ at 14 TeV,
for when only linear $c_i$ corrections  to  cross sections are retained (red band) and when quadratic $c_{i,j}$ contributions are also included (blue band). 
}
\label{fig:2d_lim}
\end{figure}

If we restrict ourselves to the two aforementioned operators, 
then equation~\ref{eq:sig_back} identifies an annulus or a disk in the 2D parameter space of Wilson coefficients. 
Hence, by combining observables one can gain constraining power by breaking such degeneracies.
To see this explicitly, we show in Fig.~\ref{fig:2d_lim} the 2D contour of allowed Wilson coefficients for $\mathcal{O}_{\varphi}$ and $\mathcal{O}_{\varphi d}$ 
at (a) $\sqrt{s}=3\TeV$ via $HH$ production  and (b) $14\TeV$ via both $HH$ and $HHH$ production.
(As mentioned above, the $HHH$ rate at 3 TeV is insignificant and hence omitted here.)
Solutions to equation \ref{eq:sig_back} are provided under the assumption that only linear $(c_i)$ corrections to $\sigma$ are retained (red band)
as well as when quadratic $(c_{i,j})$ corrections are included (blue band).
We also report the projected, marginalized limits on the two Wilson coefficients in table~\ref{tab:2d_lim}.
Clearly, the lower energy machine leaves a much larger volume of parameter space unconstrained.
In the 3 TeV case, the absence of a second measurement leads to a flat constraint  in the linear case,
which  suggests an  impossibility of conclusively constraining the parameter space. 
Moreover, this represents a strong case for measuring the triple Higgs production at lepton colliders in order to pin down the Higgs's self-couplings.
From this perspective, we argue that a $\sqrt{s}=14$ TeV lepton (muon or electron) collider would be ideal over lower energy scenarios.
Such a machine allows us to take advantage of both double and triple Higgs production, 
and at last measure  the SM's scalar potential.

It is important to point out that in order to have realistic assessments, one would need to perform a global study that includes multiple processes and operators together. 
In particular, while $\mathcal{O}_{\varphi}$ affects only $HH$ and
$HHH$ production, $\mathcal{O}_{\varphi d}$ also shifts the $HWW$ coupling. This means that the total rate of single Higgs production
is affected by $\mathcal{O}_{\varphi d}$ and therefore one can constrain the corresponding Wilson coefficient. Even if the sensitivity to the operator
is lower and does not grow with energy, the high statistics foreseen in multi-TeV lepton colliders is such that
this operator will be heavily constrained by the inclusive measurement of single Higgs production. Assuming the aforementioned luminosities,
we can estimate $95\%$ confidence level limits for $\mathcal{O}_{\varphi d}$ to be roughly equal to $[-0.01, 0.01] \TeV^{-2}$ at 3 TeV and  $[-0.004, 0.004] \TeV^{-2}$ at 14 TeV. Nonetheless, we found instructive to include this operator in the study, given the high sensitivity
(see Fig.~\ref{fig:WW_HH} and Fig.~\ref{fig:WW_HHH}) caused by derivative couplings that lead to unitarity violating effects. Despite this,
we notice that higher limits can be obtained in single Higgs production and therefore only $\mathcal{O}_{\varphi}$ (and potentially dimension 8 operators) will be relevant for $HH$ and $HHH$ production.

In order to offer a comparison with other hypothetical future collider proposals, we quote here the projections from combined results at FCC-ee$_{240}$, FCC-ee$_{365}$, FCC-eh and FCC-hh, as reported in Ref.~\cite{deBlas:2019rxi}.
The first two are $e^+e^-$ colliders with $\mathcal{L}= 5$, $1.5\invab$ at $\sqrt{s}=240$, $365\GeV$ respectively.
The third is an $e^\pm p$ collider with $\mathcal{L}= 2\invab$ at $\sqrt{s}=3.5\TeV$, 
while the last is a $pp$ collider with $\mathcal{L}= 30\invab$ at $\sqrt{s}=100\TeV$.
Under these scenarios, the projected individual bounds at 68\% CL for operators we consider are
\begin{equation}
C_{\varphi} \sim [-0.79, 0.79] \textrm{ TeV$^{-2}$} \,  \quad\text{and}\quad C_{\varphi d} \sim [-0.03, 0.03] \textrm{ TeV$^{-2}$} \, .
\end{equation}
At a $\sqrt{s}=14\TeV$ muon collider, we report that the anticipated sensitivity on the individual operators at 68\% CL from measuring
single Higgs production, as well as from double and triple Higgs production are
\begin{equation}
C_{\varphi} \sim [-0.02, 0.02] \textrm{ TeV$^{-2}$} \, \quad\text{and}\quad C_{\varphi d} \sim [-0.002, 0.002] \textrm{ TeV$^{-2}$} \, .
\end{equation}
The difference is roughly a factor of $40$ for $C_{\varphi}$ and a factor $15$ for $C_{\varphi d}$.
In the absence of $HH$ production, the results here are comparable to those reported elsewhere~\cite{Chiesa:2020awd}.
This na\"ive comparison again shows the potential of a high-energy lepton collider in studying EW physics, allowing us 
to reach a precision that is certainly competitive with what attainable at other proposed colliders.

\begin{table}[!t]
\begin{center}
\begin{tabular}{|c|c|c|}
\hline
 & 3 TeV  & 14 TeV \\
$C_{\varphi}$ & [-3.33, 0.65] & [-0.66, 0.23]\\
$C_{\varphi d}$ & [-1.31, 1.39] & [-0.17, 0.30] \\
\hline
\end{tabular}
\end{center}
\caption{Marginalized projected limits at 95\% confidence level on the Wilson coefficients in TeV$^{-2}$.}
\label{tab:2d_lim}
\end{table}

\subsection{Top electroweak couplings at muon colliders}\label{sec:eft_top}

Due to its ultra heavy mass and complicated decay topologies, 
the era of precision top quark physics has only recently begun in earnest at the LHC.
This is despite the particle's discovery decades ago
and rings particularly true for the quark's neutral EW interactions~\cite{Giammanco:2017xyn}.
For example:
The associated production channel $t\overline{t}Z$ was only first observed using the entirety of the LHC's Run I dataset~\cite{Aad:2015eua,Khachatryan:2015sha}.
Likewise, the single top channel $tZ$ was  observed only for the first time during the Run II program~\cite{Aaboud:2017ylb,Sirunyan:2018zgs}.
And importantly, only recently has the direct observation of $t\bar{t}H$  production process
 confirmed that the top quark's Yukawa coupling to the Higgs boson is $\mathcal{O}(1)$~\cite{CMS:2018rbc, Sirunyan:2018mvw, Sirunyan:2018shy, Aaboud:2017jvq, Aaboud:2017rss}. 
Since a precision program for measuring the top quark's EW couplings is still in its infancy, there exists a margin for $\mathcal{O}(10\%)$ deviations from SM expectations. 
This makes it of stark importance to understand how to best measure these couplings, as searching for deviations could lead to new physics.

On this pretext, Ref.~\cite{Maltoni:2019aot} studied a class of $2 \to 2$ scattering processes involving the top quark and the EW sector within the SMEFT framework.
There, the authors performed  a systematic analysis of unitarity-violating effects induced by higher dimensional operators. 
By considering $2 \to 2$ scattering amplitudes embedded in physical processes at present and future colliders, specific processes were identified that exhibited a distinct sensitivity to new physics.
Among these processes, VBF at future lepton colliders stands out.
The Wilson coefficients belonging to the operators in table~\ref{tab:OP_DEF} that impact VBF processes and involve the top quark are not strongly constrained.
Hence, an improved measurement of these channels is important for the indirect tests of a plethora of BSM models.

In the context of a multi-TeV muon collider and following the proposal of Ref.~\cite{Henning:2018kys}, 
in this subsection we consider and compare the constraining potential of $2 \to 3$ processes on anomalous couplings of the top quark.
Even though such processes feature more complex Feynman diagram topologies and additional phase space suppression,
their utility within the SMEFT framework stems from also featuring  
higher-point (higher-leg) contact interactions  with a stronger power-law energy dependence at tree-level. 
In addition, a larger number of diagrammatic topologies translates into more possibilities to insert dimension-six operators,
 which, roughly speaking,  may trigger larger deviations from the SM.  
 (Though arguably larger cancellations are also possible.)
 For rather understandable limitations, such as finite computing resources, such considerations were not widely investigated before.
 
As an example, we consider the operator  $\Op{t W}$ from table~\ref{tab:OP_DEF}. 
For the case of $W^+ W^- \to t \bar{t}$ scattering, 
this operator generates the four-point contact vertex 
\begin{align}
\Op{t W} ~=~  i\big(\bar{Q}\sigma^{\mu\nu}\,\tau_{\sss I}\,t\big)\, 
     \tilde{\phi}\,W^I_{\mu\nu}
     + \text{H.c.} ~
   \supset \bar{t} \sigma^{\mu\nu} t \, v \, W_{\mu} W_{\nu} \, + \text{H.c.}
\end{align}
Here, one has to pay a vev penalty of $(v/\Lambda)$,  where the $v$ originates   from the Higgs doublet $\varphi$, 
and thereby makes the term effectively a dimension-five contact term.
On the other hand, by extending the final state with a Higgs field one can saturate the operator:
\begin{equation}
\Op{t W} \supset \bar{t} \sigma^{\mu\nu} t \, H \, W_{\mu} W_{\nu} \, + \text{H.c.}
\end{equation}
Remarkably, instead of $(v/\Lambda)$, one is ``penalized''  by a factor of $(E/\Lambda)$, 
where the energy dependence originates from the three-body phase space volume.
This mechanism is rather generic and hence can be exploited for other operators and multiplicities 
in order to maximize the energy growth of amplitudes,
and therefore the sensitivity to new physics. 

\begin{figure}[!t]
\centerline{\subfigure[]{\includegraphics[width=.925\textwidth]{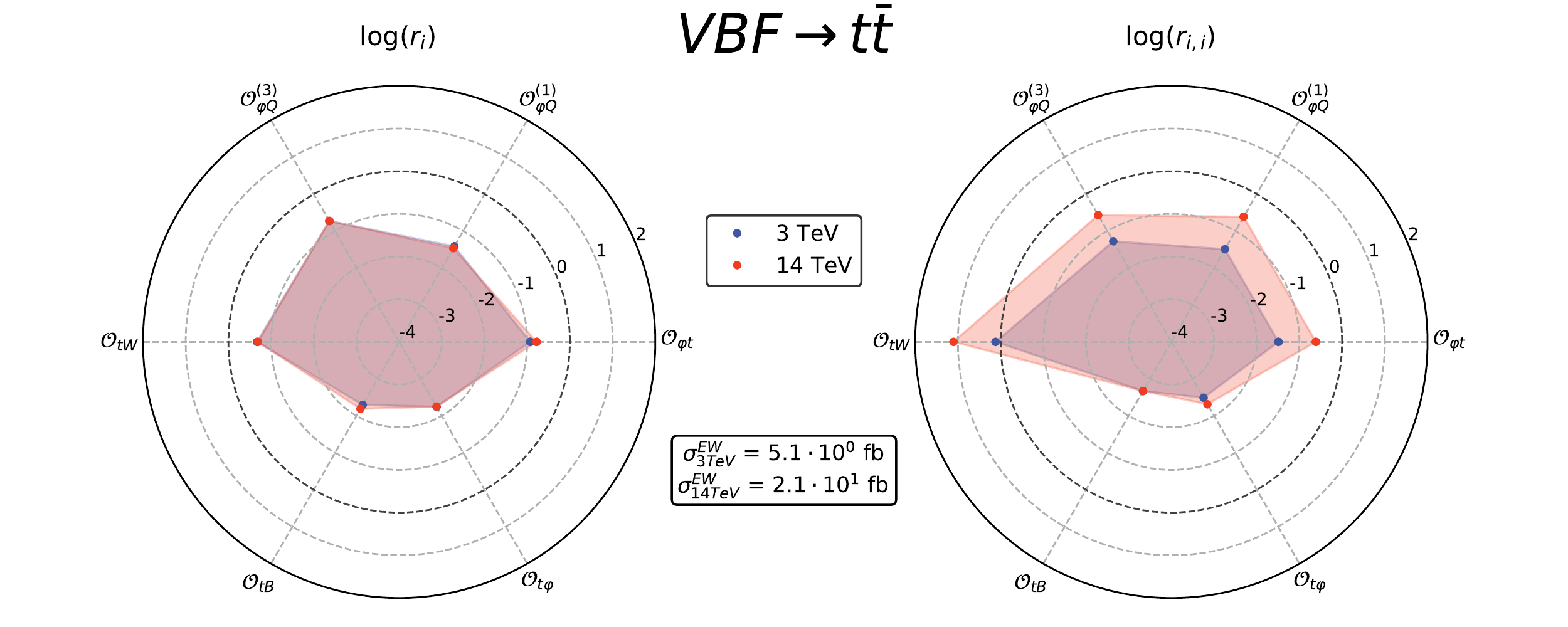}		\label{fig:radars_ttX}}}
\centerline{\subfigure[]{\includegraphics[width=.925\textwidth]{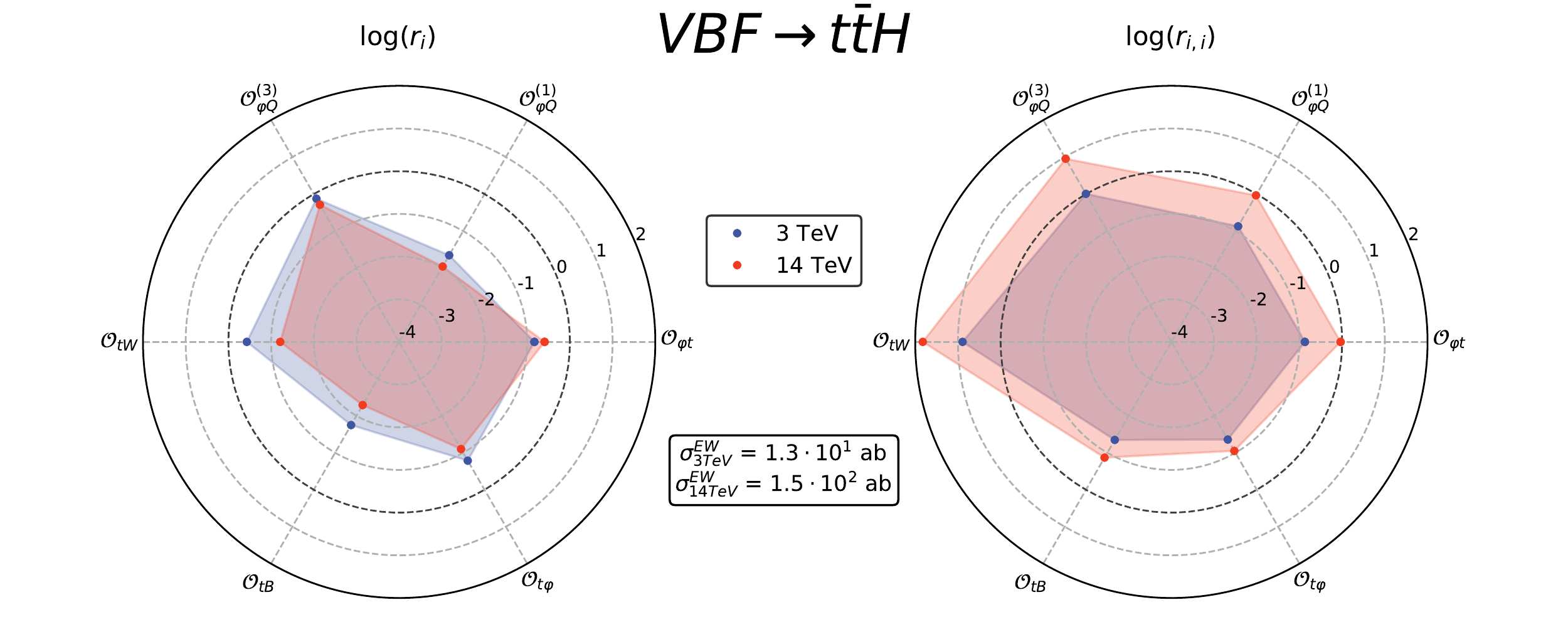}	\label{fig:radars_ttH}}}
\centerline{\subfigure[]{\includegraphics[width=.925\textwidth]{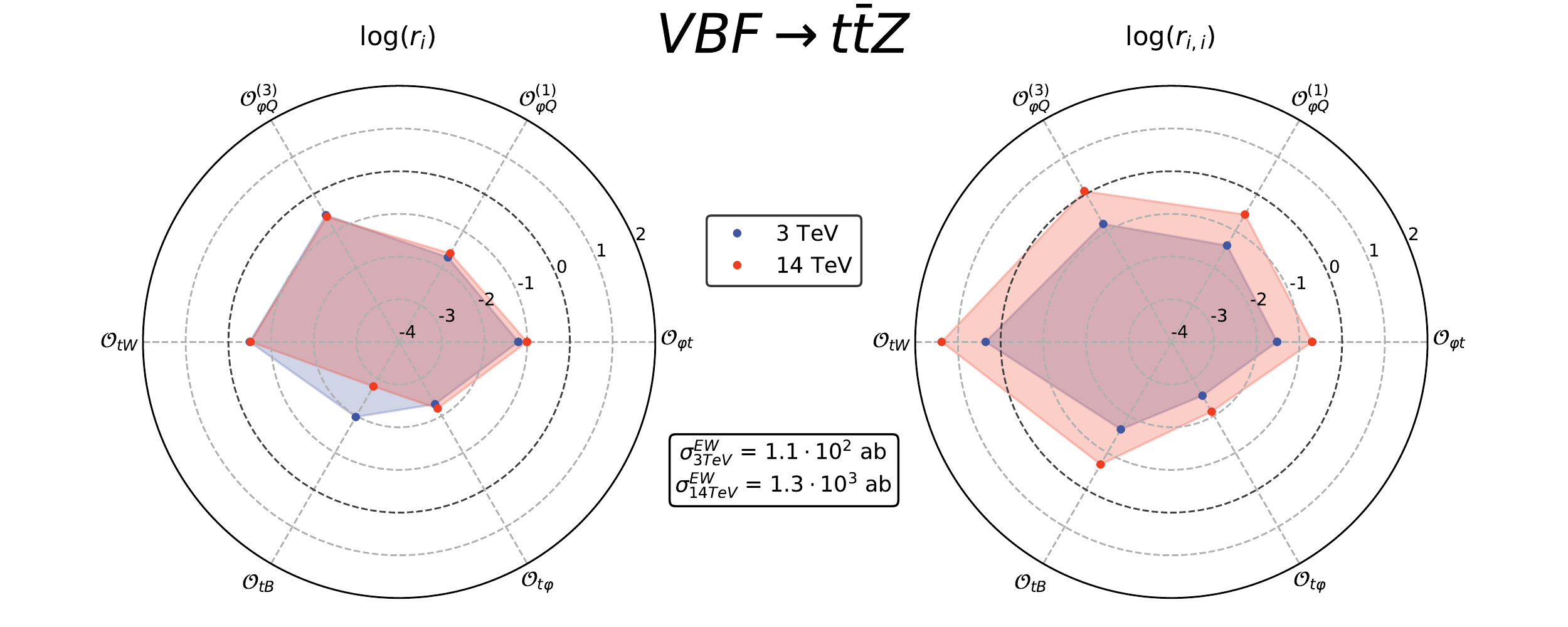}	\label{fig:radars_ttZ}}}
\caption{Impact of dimension-six operators (polar direction)  on 
(left) the interference term $\vert r_i\vert $ and (right) the quadratic term $r_{i,i}$  (radial direction in logarithmic scale) from the ratio $R$ in equation~\ref{Eq:sens_ratio} 
for the EW VBF $\to t \bar{t} (H/Z)$ processes  at a lepton collider of $\sqrt{s}=3\TeV$ (blue dots) and $14\TeV$ (red dots),
assuming a Wilson coefficient of $1 \TeV^{-2}$.}
\label{Fig:radars}
\end{figure}

For concreteness, we compare the $2\to2$ production of $t\bar{t}$ from  VBF
to the $2 \to 3$ associated production of $t\bar{t}H$ and $t\bar{t} Z$ from  VBF.
For each process we present in Fig.~\ref{Fig:radars} the 
ratio coefficients $\vert r_i \vert$ and $r_{i,i}$ of $R(c_i)$ as defined in equation~\ref{Eq:sens_ratio}, in the compact, radar plot format.
More specifically, for several SMEFT operators (presented in the polar direction) we plot
(left) the absolute value of the interference term $r_i$ at $\mathcal{O}(\Lambda^{-2})$
and (right) the quadratic term $r_{i,i}$ at $\mathcal{O}(\Lambda^{-4})$ in the radial direction (in logarithmic scale).
We representatively fix each Wilson coefficient to $C_\mathcal{O}=1$ TeV$^{-2}$ and consider collider energies $\sqrt{s}=3\TeV$ (blue dots) and $\sqrt{s}=14\TeV$ (red dots).
Contours at $r_i,~r_{i,i}=1$ are bolded for clarity. 
Also reported in the figure are the total cross sections [fb] predicted in the SM.

We observe in the $t\bar{t}$ case (Fig.~\ref{fig:radars_ttX}) that the sensitivity to the operators under consideration is somewhat marginal.
For both the linear (left) and quadratic (right) ratios deviations reach at most $\mathcal{O}(10\%)$.
The exception is $\mathcal{O}_{tW}$, which features an $r_{i,i}$ term that can reach $\mathcal{O}(1-10)$ at $\sqrt{s}=3-14\TeV$.
For all operators, linear contributions do not vary appreciably when passing from a c.m.~energy of 3 TeV to 14 TeV. 
On the other hand, the quadratic terms exhibit an overall growth, just not a dramatic one.
The smallness of $\vert r_i\vert$ contributions suggests that considering   higher multiplicity processes, such as $t\overline{t}H$ and $t\overline{t}Z$, could prove more sensitive to new physics, 
despite na\"ive phase space suppression.

Adding a Higgs boson (Fig.~\ref{fig:radars_ttH}) or a Z boson (Fig.~\ref{fig:radars_ttZ}) in the final state has a noticeable, 
quantitative impact on the overall behavior of ratio coefficients in the radar plots.
When looking at the linear interference terms, it is surprising to see that many of the operators' contributions decrease when going to higher energies.
On the other hand, a sensitivity gain is unambiguous for all the operators in the quadratic case, which reach as much as $\mathcal{O}(100)$.
The behavior of interference is often more subtle to predict.
Being non-positive definite, cancellations can and do readily take place depending on the specific phase space region that is considered. 
In particular, we infer that at higher energies these cancellations are enhanced, leading effectively to a lower sensitivity at the inclusive level.

Generically, each operator and process has a cancellation pattern of its own,
which is also reinforced by the linear independence of  SMEFT operators.
Hence, designing a single recipe for every operator to invert cancellations with the aim of fully exploiting the increased sensitivity to energy is complicated.
On the other hand, dedicated studies could lead to the discovery of a most sensitive (or a highly optimized)
 phase space region for a specific set of operators, enhancing the possibility to detect new physics.

While being more difficult to measure,
these $2\to3$ processes offer an overall improvement to sensitivity with respect to  $2\to2$ production of $t \bar{t}$.
This is both from the energy-growing perspective and from an absolute one. 
In essence, our very preliminary study here suggests that having a multi-TeV muon collider would benefit us for at least two reasons: 
(i) Due to phase space enhancements $(E/\Lambda)$, 
a higher energy collider would allow us to  take advantage of larger deviations from SM expectations, and hence higher sensitivity to SMEFT operators.
(ii) The growth in the inclusive VBF cross section would allow us to have enough statistics to precisely measure higher multiplicity final states
that would otherwise be infeasible even at $\sqrt{s}=3\TeV$.
 For example: we  compare the $\sim 100$ $t \bar{t} H$ events at 3 TeV  to the $\sim 3000$ at 14 TeV, 
 assuming the benchmark luminosities considered  ($\mathcal{L}=6\invab$ and $20\invab$, respectively). 
The program to precisely determine the top quark's EW interactions    would therefore benefit greatly from a potential future muon collider
by allowing us to take into account new processes that could help break degeneracies among SMEFT operators and learn about the dynamics of EW symmetry breaking.

\section{Searches for new physics}\label{sec:bsm}
Like hadron beams, muon beams emit significantly less synchrotron radiation than their electronic counterpart due to the muon's much larger mass.
As a result, $\mpmm$ colliders can reach {partonic} c.m.~energies that far exceed conventional $e^+e^-$ facilities, such as LEP II, and potentially even $pp$ colliders;
see section~\ref{sec:ppvsmuon} for further details.
Thus, in addition to the abundance of achievable SM measurements described in sections~\ref{sec:sm} and \ref{sec:eft}, 
a muon collider is able to  explore new territory in the direct search for new physics.

In this section, we present a survey of  BSM models and the potential sensitivity of a $\mpmm$ collider.
Explicitly, we consider the $s$-channel annihilation and VBF processes
\begin{equation}
\mpmm ~\to~ X \quad\text{and}\quad
\mpmm ~\to~ X \ell \ell'.
\end{equation}
Here, $\ell\in\{\mpm,\overset{(-)}{\nu_\mu}\}$ and $X$ is some BSM final state, which may include SM particles.
We focus on the complementarity of the two processes because while $s$-channel annihilation grants accesses to the highest available c.m.~energies, 
it comes at the cost of a cross section suppression that scales as $\sigma\sim1/s$ when far above production threshold.
On the other hand, in VBF,  the emission of transversely polarized, $t$-channel bosons gives rise to logarithmic factors that grow with the available collider energy.
Thus,  VBF  probes a continuum of mass scales while avoiding a strict $1/s$-suppression, but at the cost of EW coupling suppression.

To investigate this interplay, for each scenario, we consider the mass and collider ranges:
\begin{align}
m_X\in[0.4,4]\;\textrm{TeV}
\quad\text{and}\quad
\sqrt s\in [1,30]\;\textrm{TeV}.
\end{align}

We limit our study to $\sqrt{s}\leq30\TeV$ due to the emergence of EW Sudakov logarithms in the VBF channels that scale as $\sigma_{\rm VBF}\sim\alpha_W^k\log^k(s/M_V^2)$, for $V=W,Z$.
These logarithms can potentially spoil the perturbative reliability of cross sections at LO and necessitate resummation of EW Sudakov factors~\cite{Bauer:2016kkv,Chen:2016wkt,Bauer:2017isx,Manohar:2018kfx,Han:2020uid}.
While important for reliable predictions at higher $\sqrt{s}$, such resummation is beyond the present scope.
For the various BSM scenarios, we assume benchmark values for relevant couplings.
We  omit generator-level phase space cuts where possible but stipulate them when needed to regulate matrix elements.
In the following, we present the production rates of new processes.
As a standard candle reference, in most scenarios, we also plot SM $H$ production via $W^+W^-$ fusion (black, solid curve).

We start our survey in section~\ref{sec:bsm_scalar} with minimally extending the SM by a scalar that is a singlet under the SM's gauge symmetries.
We then move onto the production of scalars in the context of the Two Higgs Doublet Model in section~\ref{sec:bsm_2hdm}, and the Georgi-Machacek Model in section~\ref{sec:bsm_gm}.
In section~\ref{sec:bsm_mssm}, we investigate the production of sparticles in the context of the Minimal Supersymmetric Standard Model.
We also consider representative phenomenological models describing the production of 
leptoquarks in section~\ref{sec:bsm_LQ}, heavy neutrinos in section~\ref{sec:bsm_heavyN}, and
vector-like quarks in section~\ref{sec:bsm_vlq}.
We give an overview of this survey in section~\ref{sec:bsm_overview}.
A detailed comparison of  $s$-channel and VBF production mechanisms in BSM searches at multi-TeV muon colliders is deferred to section~\ref{sec:bsm_vbf}.

\subsection{Scalar singlet extension of the Standard Model}\label{sec:bsm_scalar}

The scalar sector of the SM consists of a single scalar $SU(2)_L$ doublet with a nonzero $U(1)_Y$ charge.
While this is the minimal scalar content that supports the generation of fermion and weak boson masses through EWSB,
the measured couplings of the $M_H\approx125\GeV$ Higgs boson uphold this picture~\cite{ATLAS:2018doi,Sirunyan:2018koj}. 
Theoretical motivation for extending this scalar sector, however, is well-established and the phenomenology of these scenarios have been studied extensively.
For reviews, see Refs.~\cite{Gunion:1989we,Branco:2011iw,Morrissey:2012db,Ivanov:2017dad,Khan:2015ipa,Kanemura:2016sos,Ilnicka:2018def} and references therein.

One of the simplest extensions that respects the SM symmetries 
is the addition of a single, real scalar $(\sigma)$ that is neutral under all SM charges but carries an odd $\mathbb{Z}_2$ parity.
Such scenarios have received recent attention~\cite{Buttazzo:2018qqp,Ruhdorfer:2019utl}
as simplified models through which one can explore the sensitivity of multi-TeV muon colliders to new  scalars.
In light of LHC data, the phenomenology of a singlet scenario is categorized by whether $\sigma$ acquires a nonzero vev:
In the so-called inert scenario, $\sigma$ does not acquire a vev, interacts at tree level only with the SM Higgs boson $(H)$, 
and impacts $H$'s coupling to fermions and bosons at loop level~\cite{Craig:2013xia}.
If instead $\sigma$ acquires a vev, then it mixes with the SM Higgs, 
which in turn modifies $H$'s coupling to SM particles at tree-level.

\begin{figure}[t!]
\centering\mbox{
\subfigure[]{\includegraphics[width=.42\textwidth]{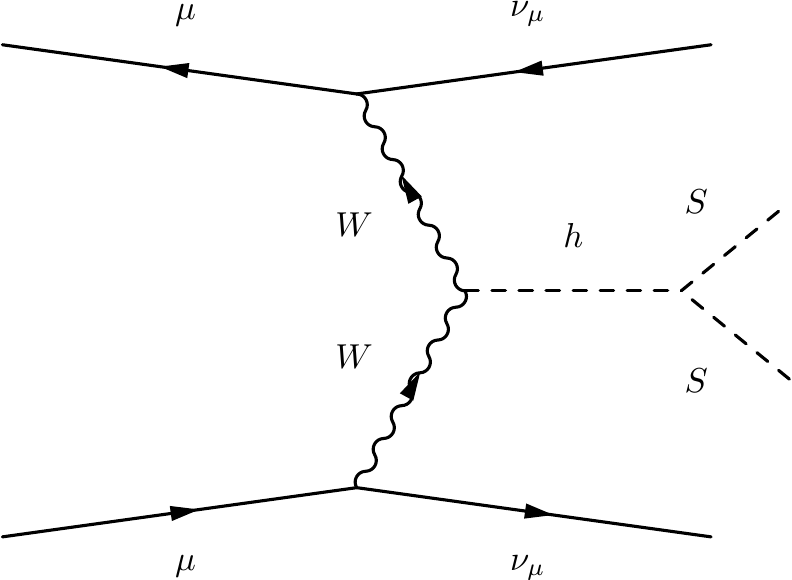}\label{fig:bsm_vbfsing_diag}}
\hspace{0.75cm}
\subfigure[]{\includegraphics[width=.52\textwidth]{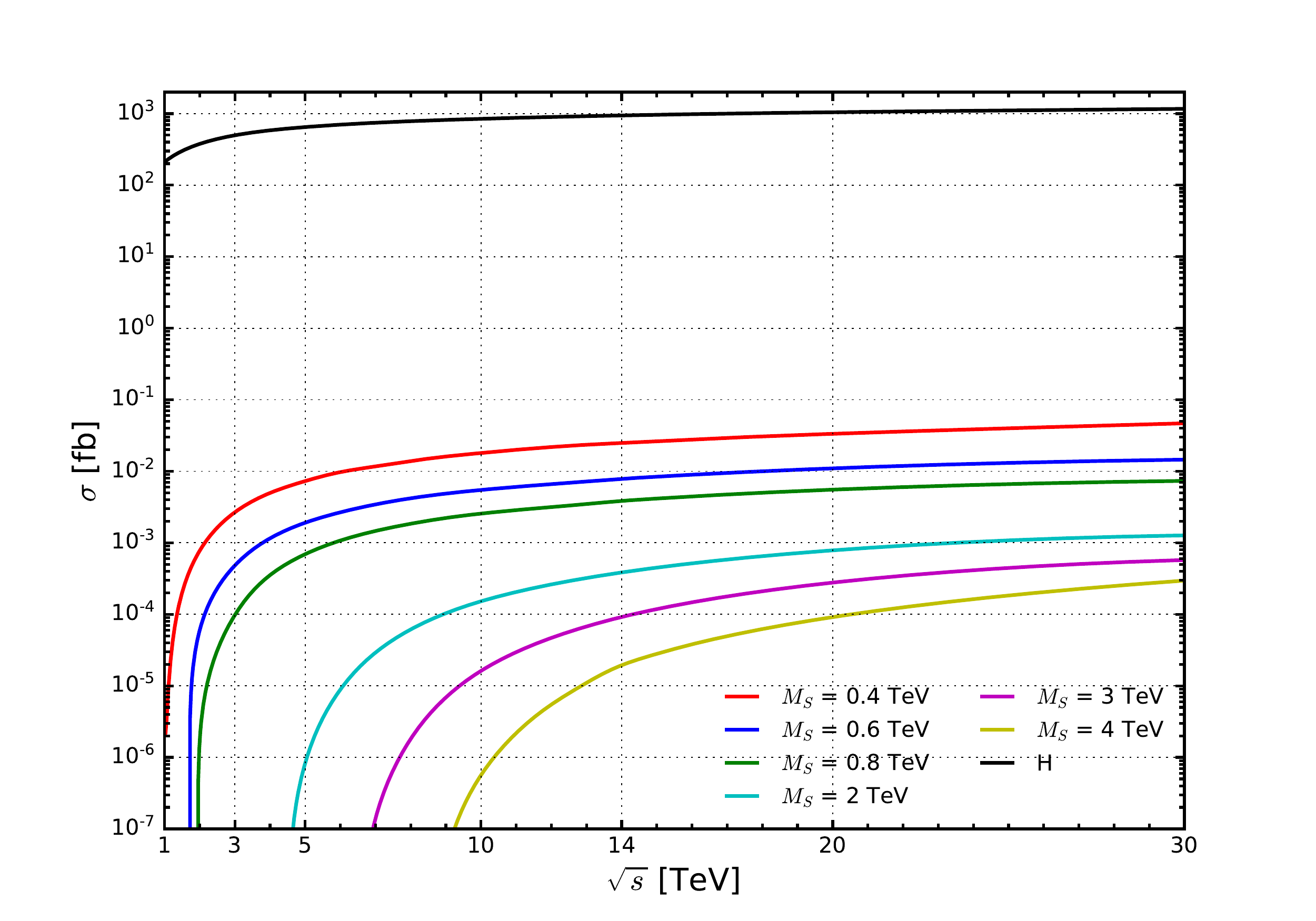}\label{fig:bsm_vbfsing_xsec}}
}
\caption{
(a) Diagrammatic representation of $SS$ pair production through $W^+W^-$ fusion in a scalar singlet extension of the SM.
(b) $SS$ pair production cross section [fb] via EW VBF  in $\mpmm$ collisions as a function of collider energy $\sqrt{s}$ [TeV] for representative coupling inputs.
Also shown for comparison is the $H$ production process via EW VBF (black curve) in the SM.
}\label{fig:bsm_vbfsing}
\end{figure}

We investigate the muon collider sensitivity to the SM with an extra scalar singlet by considering the case where the vev of $\sigma$ is nonzero, i.e., $\langle \sigma \rangle \approx v_\sigma +\sigma_0$.
The (unbroken) Lagrangian that describes such a scenario, including the $\mathbb{Z}_2$ symmetry, is given by
\begin{equation}
\mathcal{L}=\mathcal{L}_{\rm SM}+\frac{1}{2}\partial_\mu \sigma\, \partial_\mu \sigma-\frac{1}{2}m_\sigma^2\, \sigma^2-\frac{\lambda_\sigma}{4!}\sigma^4-\frac{\kappa_\sigma}{2} \sigma^2\, \varphi^\dagger\varphi\,,
\end{equation}
where $\mathcal{L}_{\rm SM}$ is the full SM Lagrangian.
After both the SM doublet $\varphi$ and $\sigma$ acquire their respective vevs, $v$ and $v_\sigma$, 
a mass-mixing term between $\sigma_0$ and the neutral part of the doublet $\phi_0$,
and  proportional to $\delta m^2\propto\kappa_\sigma v v_\sigma$,  is generated.
Rotating $\phi_0$ and $\sigma_0$  from the gauge basis and into the mass basis by an angle $\theta$,
we obtain the mass eigenstates $H$ and $S$ with mass eigenvalues $M_H$ and $M_S$.
The coupling of the lightest neutral scalar, which we assume is $H$, to SM fermions and gauge bosons is suppressed  relative to the SM by a factor of $\cos\theta$. 
Owing to strong constraints on anomalous Higgs couplings~\cite{ATLAS:2018doi,Sirunyan:2018koj}, 
one scalar is aligned closely with the SM Higgs, which we assign to $H$, implying $\cos\theta\simeq1$.
The bare parameters $m_\sigma, \lambda_\sigma, \kappa_\sigma$, can subsequently be exchanged for the physical parameters $M_S, v_\sigma, \theta$,
which therefore permits us to express the trilinear scalar interactions as:
\begin{eqnarray}
\lambda_{hhh}&=&-\frac{3M_H^2 }{v\,v_\sigma}(v_\sigma \cos^3\theta+v \sin^3\theta)\,,\\
\lambda_{sss}&=&\frac{3M_S^2}{v\,v_\sigma}(v \cos^3\theta - v_\sigma \sin^3\theta)\,,\\
\lambda_{hss}&=&-\frac{(M_H^2+2M_S^2)}{2v\,v_\sigma} \sin2\theta(v\cos\theta+v_\sigma\sin\theta)\,,\\
\lambda_{hhs}&=&\frac{(2M_H^2+M_S^2)}{2v\,v_\sigma} \sin2\theta(v_\sigma\cos\theta-v\sin\theta)\,.
\end{eqnarray} 
The non-inert singlet scenario\footnote{Similarly, the inert singlet scenario is available using the \texttt{SM\_Plus\_Scalars\_UFO} UFO libraries~\cite{vanderBij:2006ne}.} 
is implemented in the \texttt{Minimal Dilaton Model} UFO libraries by Ref.~\cite{Abe:2012eu},
and hence can be simulated using general purpose event generators.

$S$ production in $\mpmm$ collisions  can proceed through several mechanisms, including $W^+W^-$ fusion, as shown in figure~\ref{fig:bsm_vbfsing_diag},
which is mediated by an $s$-channel $H$ boson.
As shown above, for a given $v_\sigma$ and $\theta$, the  $\lambda_{hss}$ coupling is related to the $H-S$ mass difference.
Assuming the fixed, baseline mass splitting of  Ref.~\cite{Abe:2012eu},
we show in figure~\ref{fig:bsm_vbfsing_xsec} the $SS$ pair production cross section [fb] via EW VBF  as a function of collider energy $\sqrt{s}$ [TeV].

In the numerical analysis we have assumed $\theta=v/v_\sigma$ and $v_\sigma=20 M_S$.
For $M_S = 0.4- 0.8 \TeV$, we see that the VBF process rate spans roughly $\sigma\sim 10^{-3}-10^{-2}$ fb for $\sqrt{s}=5-30\TeV$. For $M_S = 2-4 \TeV$, 
we observe that the corresponding rates reach the order of $10^{-4} -10^{-3}$ fb at $\sqrt{s}=30\TeV$.
By comparing these numbers with the SM productions of $H$ via VBF over the whole range of collider energies, we find that the latter are several order of magnitude larger, spanning $\sigma\sim 100-1000$ fb.

\subsection{Two Higgs Doublet Model}\label{sec:bsm_2hdm}

\begin{figure}[t!]
\centering\mbox{
\subfigure[]{\includegraphics[width=.42\textwidth]{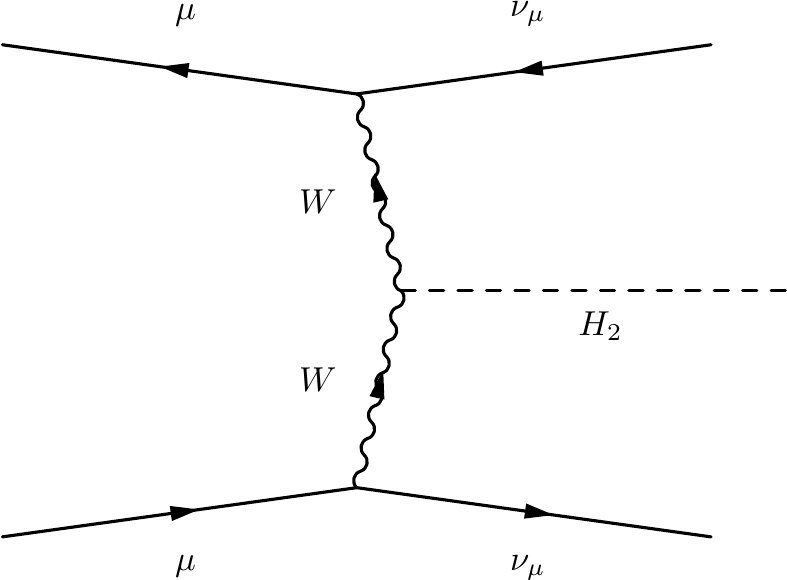}
\label{fig:bsm_vbf2hdm_diag}
}
\hspace{0.75cm}
\subfigure[]{\includegraphics[width=.52\textwidth]{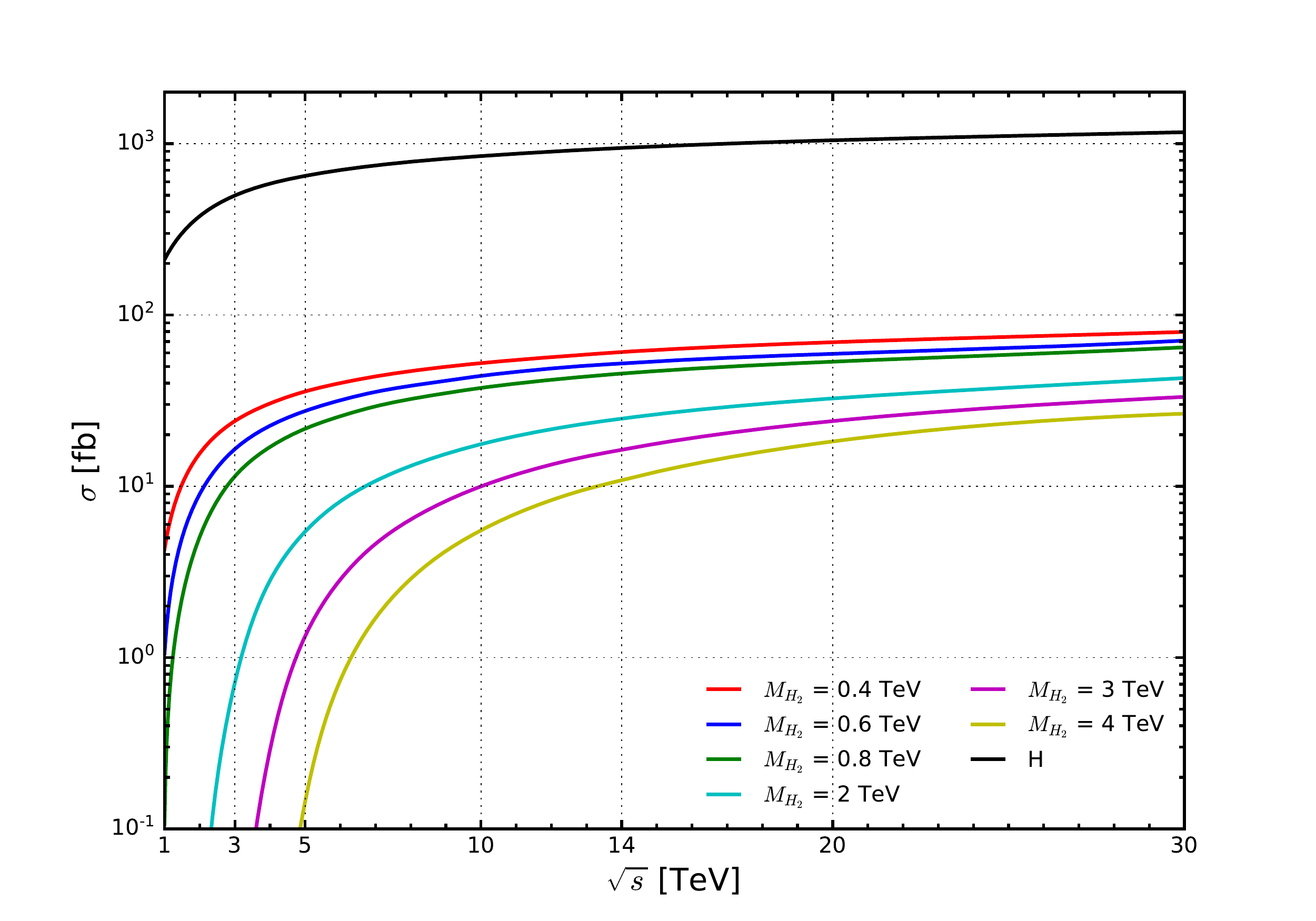}
\label{fig:bsm_vbf2hdm_xsec}
}
}
\caption{
(a) Diagrammatic representation of  $H_2$ production through $W^+W^-$ fusion in the CP conserving 2HDM.
(b) $H_2$ production cross section [fb] via VBF  in $\mpmm$ collisions as a function of collider energy $\sqrt{s}$ [TeV] for representative $H_2$ mass $(M_{H_2})$.
}\label{fig:bsm_vbf2hdm}
\end{figure}

If a new neutral scalar does indeed exist, rather than being a SM singlet as posed in section~\ref{sec:bsm_scalar}, it may actually be a component of a second scalar $SU(2)_L$ doublet.
Such scenarios, known in the literature as Two Higgs Doublet  Models (2HDMs), have been extensively reviewed~\cite{Gunion:1989we,Branco:2011iw,Crivellin:2013wna,Craig:2013hca}, 
particularly for their necessity to realize Supersymmetry in nature.
We consider the benchmark, CP-conserving 2HDM, the scalar potential of which is
\begin{align}
V&=\mu_1 \varphi_1^\dagger \varphi_1 + \mu_2 \varphi_2^\dagger \varphi_2 + \left(\mu_3 \varphi_1^\dagger \varphi_2 + {\rm H.c.}\right) + \lambda_1 \left(\varphi_1^\dagger \varphi_1\right)^2 + \lambda_2 \left(\varphi_2^\dagger \varphi_2\right)^2 \nonumber\\
&+\lambda_3 \left(\varphi_1^\dagger \varphi_1\right) \left(\varphi_2^\dagger \varphi_2\right) + \lambda_4 \left(\varphi_1^\dagger \varphi_2\right) \left(\varphi_2^\dagger \varphi_1\right) + \left(\lambda_5 \left(\varphi_1^\dagger \varphi_2\right)^2 + {\rm H.c.}\right)\nonumber\\
&+ \varphi_1^\dagger \varphi_1\left(\lambda_6 \left(\varphi_1^\dagger \varphi_2\right) + {\rm H.c.}\right) + \varphi_2^\dagger \varphi_2\left(\lambda_7 \left(\varphi_1^\dagger \varphi_2\right) + {\rm H.c.}\right).
\end{align}
Here, the couplings $\lambda_i$ are real and  the scalar $SU(2)_L$ doublets $\varphi_1$ and $\varphi_2$ are given by
\begin{equation}
\varphi_1\equiv \left(\begin{array}{c}-ih_1^+\\\frac{h^0_1+i a_1+v_1}{\sqrt 2}\end{array}\right) \qquad\text{and}\qquad \varphi_2\equiv \left(\begin{array}{c}h_2^+\\\frac{h^0_2+i a_2+v_2}{\sqrt 2}\end{array}\right).
\end{equation}
After $\varphi_1$ and/or $\varphi_2$ acquire vacuum expectation values, EW is broken and fields with identical quantum numbers mix.
More specifically, the charged scalars and neutral, CP-odd scalars mix into the EW Goldstone bosons $G^\pm,~G^0$ and the physical states $H^\pm,~A^0$.
Likewise, the neutral, CP-even scalars mix by an angle $\theta$ into the physical states $H_1$ and $H_2$.
Here, $H_1$ is identified as the observed, SM-like Higgs  with $M_{H_1}\approx 125\GeV$ and $H_2$ is heavier with $M_{H_2} > M_{H_1}$.
In terms of mass eigenstates, $h_1^0$ and $h_2^0$ are given explicitly by
\begin{eqnarray}
\begin{pmatrix} h^0_1 \\ h^0_2 \end{pmatrix} = \begin{pmatrix}
  \cos\theta & \sin\theta \\ 
-\sin\theta & \cos\theta \end{pmatrix} \begin{pmatrix} H_1 \\ H_2 \end{pmatrix}.
\end{eqnarray}

Among the simplest processes we can analyze at a muon collider is resonant production of $H_2$ from $W^+W^-$ fusion, 
which we show diagrammatically in figure~\ref{fig:bsm_vbf2hdm_diag}.
To estimate the sensitivity  to this process, we consider the 2HDM
in its CP-conserving scenario, as implemented in the \texttt{2HDM} model file \cite{Degrande:2014vpa}.
We show in figure~\ref{fig:bsm_vbf2hdm_xsec} the $H_2$ production cross section [fb] via EW VBF as a function of collider energy $\sqrt{s}$ [TeV] for representative $H_2$ mass $(M_{H_2})$.
\confirm{For $M_{H_2}=400-800\GeV$, we find that cross sections span approximately $\sigma\approx0.1-100$ fb for $\sqrt{s}=1-30\TeV$.
For $M_{H_2}=2-4\TeV$, we find that rates can reach several tens of fb at $\sqrt{s}=30\TeV$.
Over the entire range of collider energies, we see that the SM production of $H$ is  over an order of magnitude larger, reaching $\sigma\sim100-1000$ fb. 
 }
 
\subsection{Georgi-Machacek Model}\label{sec:bsm_gm}

Another possibility at a future muon facility is the VBF production of electrically charged scalars.
These, of course, do not exist in the SM nor  in the simplest, na\"ive extensions of the SM scalar sector.
In models such as the Georgi-Machacek (GM) model \cite{Georgi:1985nv} 
and the Type II Seesaw model for neutrino masses~\cite{Konetschny:1977bn,Schechter:1980gr,Cheng:1980qt,Lazarides:1980nt,Mohapatra:1980yp},
VBF production of singly charged $(H^\pm)$ and doubly charged $(H^{\pm\pm})$ scalars is possible due to the existence of scalar triplet representations of $SU(2)_L$
with nonzero hypercharge. (Higher $SU(2)_L\otimes U(1)_Y$ representations also permit scalars with even larger electric charges.)

For present purposes, we focus on the feasibility of seeing exotically charged scalars from the GM 
model\footnote{While it is also possible to model the Type II Seesaw with the \texttt{TypeIISeesaw} UFO libraries~\cite{Fuks:2019clu}, we do not anticipate a  qualitative difference in sensitivity from the GM case.}.
Broadly speaking, the model extends the SM with a real and a complex triplet with hypercharge $Y=0$ and $1$, respectively. 
If the vevs of the triplets' neutral components are aligned, then  tree-level, custodial symmetry is respected 
and  strong constraints on the $\rho$ parameter are alleviated~\cite{Gunion:1989we,Chen:2005jx,Han:2005nk,Chen:2008jg,Perez:2008ha,Kanemura:2012rs,Das:2016bir,Ismail:2020zoz}.
More specifically, the GM scalar sector 
consists of the usual SM complex doublet $(\phi^+,\phi^0)$ with $Y = 1/2$, 
a real $SU(2)_L$ triplet $(\xi^+,\xi^0,\xi^-)$ with $Y = 0$, 
and  a complex $SU(2)_L$ triplet $(\chi^{++},\chi^+,\chi^0)$ with $Y=1$.  
Writing the doublet and triplets in the form of a bi-doublet $(\Phi)$ and bi-triplet $(X)$, we have
\begin{eqnarray}
	\Phi &=& \left( \begin{array}{cc}
	\phi^{0*} &\phi^+  \\
	-\phi^{+*} & \phi^0  \end{array} \right) 
	\quad\text{and}\quad
	X =
	\left(
	\begin{array}{ccc}
	\chi^{0*} & \xi^+ & \chi^{++} \\
	 -\chi^{+*} & \xi^{0} & \chi^+ \\
	 \chi^{++*} & -\xi^{+*} & \chi^0  
	\end{array}
	\right).
	\label{eq:PX}
\end{eqnarray}

\begin{figure}[t!]
\centering\mbox{
\subfigure[]{\includegraphics[width=.42\textwidth]{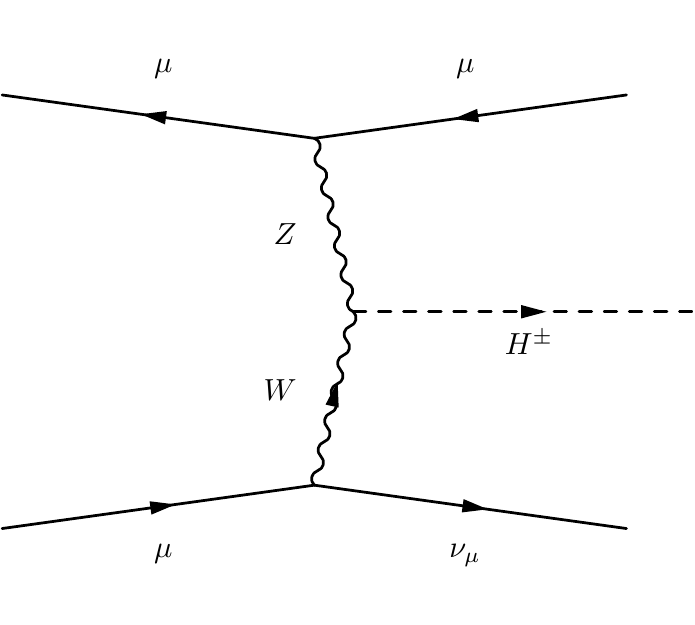}\label{vbfgm_hpx_diag}}\hspace{0.75cm}
\subfigure[]{\includegraphics[width=.52\textwidth]{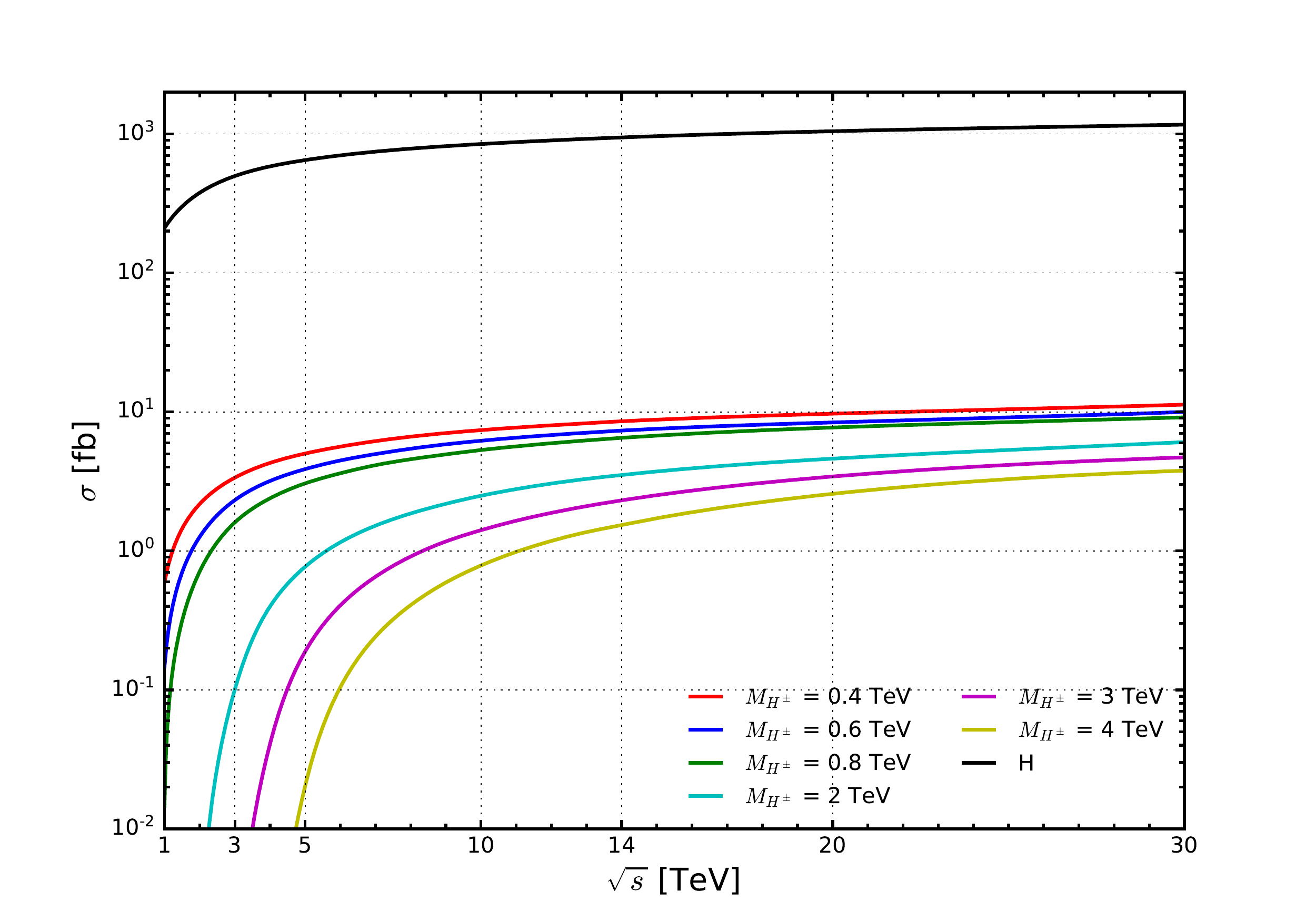}\label{vbfgm_hpx_xsec}}
}
\mbox{
\subfigure[]{\includegraphics[width=.42\textwidth]{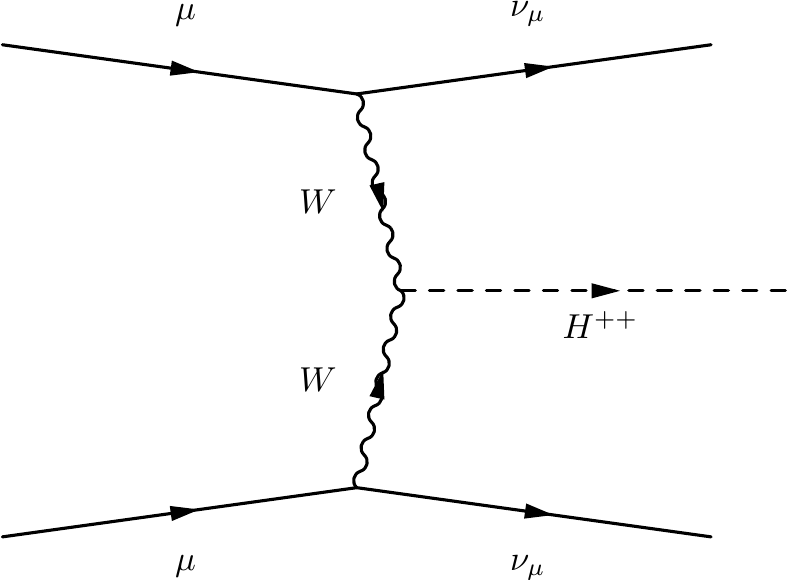}\label{vbfgm_hpp_diag}}\hspace{0.75cm}
\subfigure[]{\includegraphics[width=.52\textwidth]{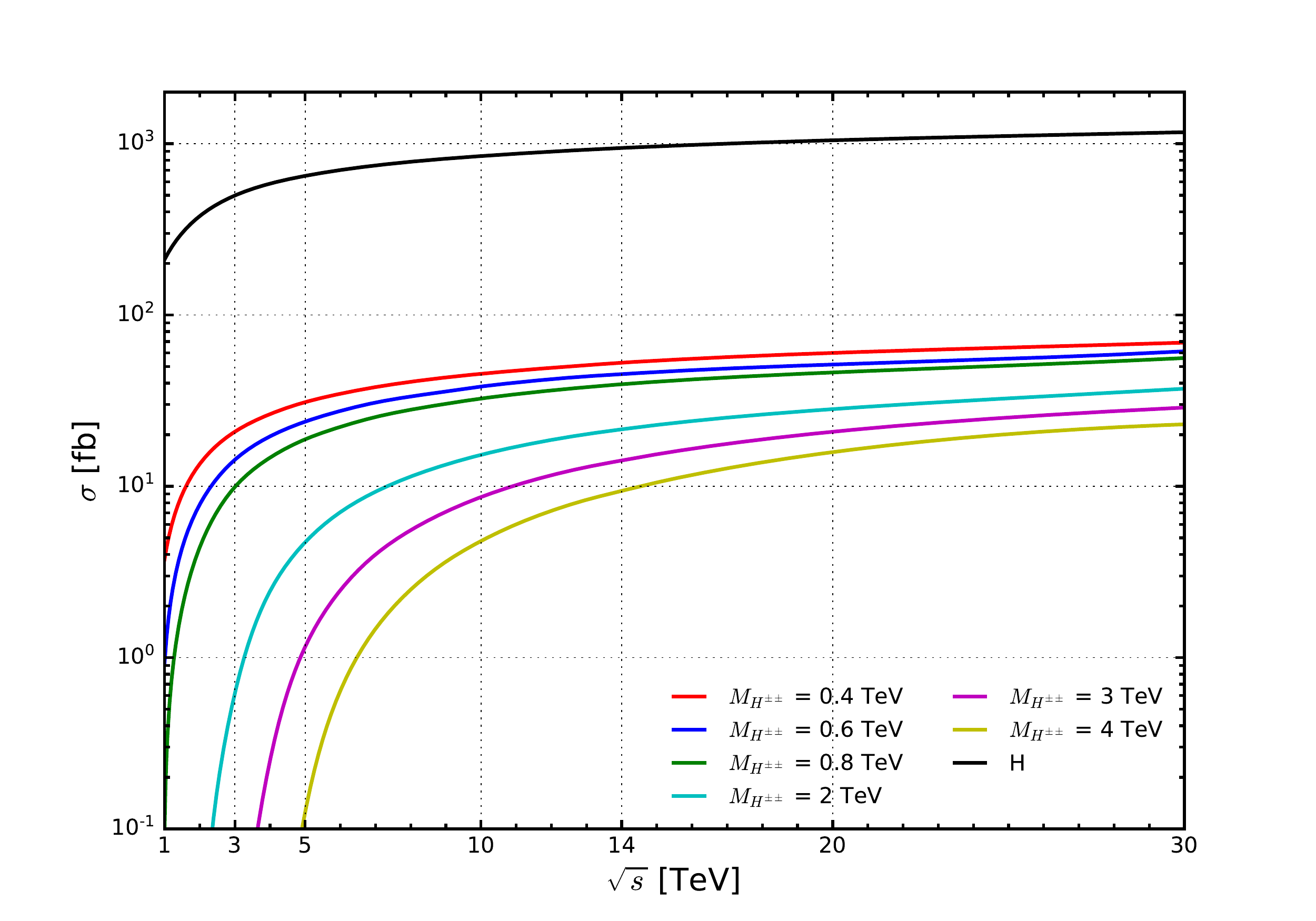}\label{vbfgm_hpp_xsec}}
}
\caption{
(a) Diagrammatic representation of  $H^\pm$ through EW VBF in the GM model, in $\mpmm$ collisions.
(b) The $H^\pm$ production rate [fb] via EW VBF  in $\mpmm$ collisions as a function of collider energy $\sqrt{s}$ [TeV] for representative $M_{H^\pm}$.
(c,d) Same as (a,b) but for   $H^{++}$ in $\mu^+\mu^+$ collisions.
}\label{vbfgm}
\end{figure}

For our numerical results, we consider the decoupling limit of the GM model as implemented in the \texttt{GM\_UFO} UFO~\cite{Hartling:2014zca,Degrande:2015xnm}.
The  (unbroken) scalar potential is given by
\begin{align}
	V(\Phi,X) &=  \frac{\mu_2^2}{2}  \text{Tr}(\Phi^\dagger \Phi) 
	+  \frac{\mu_3^2}{2}  \text{Tr}(X^\dagger X)  
	+ \lambda_1 [\text{Tr}(\Phi^\dagger \Phi)]^2  
	+ \lambda_2 \text{Tr}(\Phi^\dagger \Phi) \text{Tr}(X^\dagger X)   \nonumber \\
          &  + \lambda_3 \text{Tr}(X^\dagger X X^\dagger X)  
          + \lambda_4 [\text{Tr}(X^\dagger X)]^2 
           - \frac{\lambda_5}{4} \text{Tr}( \Phi^\dagger \tau_I \Phi \tau_J) \text{Tr}( X^\dagger t_I X t_J) 
           \nonumber \\
           &  - \frac{M_1}{4} \text{Tr}(\Phi^\dagger \tau_I \Phi \tau_J)(U X U^\dagger)_{IJ}  
           -  M_2 \text{Tr}(X^\dagger t_I X t_J)(U X U^\dagger)_{IJ}.
\end{align} 
Here $\tau_I$ are the Pauli $\sigma$ matrices. 
The matrices $t_I$ and $U$ are \cite{Hartling:2014zca}
\begin{equation}
	t_1= \frac{1}{\sqrt{2}} \left( \begin{array}{ccc}
	 0 & 1  & 0  \\
	  1 & 0  & 1  \\
	  0 & 1  & 0 \end{array} \right), \quad  
	  t_2= \frac{1}{\sqrt{2}} \left( \begin{array}{ccc}
	 0 & -i  & 0  \\
	  i & 0  & -i  \\
	  0 & i  & 0 \end{array} \right), \quad 
	t_3= \left( \begin{array}{ccc}
	 1 & 0  & 0  \\
	  0 & 0  & 0  \\
	  0 & 0 & -1 \end{array} \right),
\end{equation}
\begin{equation}
	 U = \left( \begin{array}{ccc}
	- \frac{1}{\sqrt{2}} & 0 &  \frac{1}{\sqrt{2}} \\
	 - \frac{i}{\sqrt{2}} & 0  &   - \frac{i}{\sqrt{2}} \\
	   0 & 1 & 0 \end{array} \right).
	 \label{eq:U}
\end{equation}
After aligning all states into their mass eigenstates, we are left with $H^\pm$ and $H^{\pm\pm}$, 
in addition to a number of neutral scalar and pseudoscalar states that we do not consider.
In order to keep a consistent measure of collider sensitivity, we restrict ourselves to EW VBF production of $H^\pm$ and $H^{\pm\pm}$.
In figure~\ref{vbfgm_hpx_diag}, we show a diagrammatic representation of the singly charged scalar $H^\pm$ produced resonantly through EW boson fusion,
and present in figure~\ref{vbfgm_hpx_xsec} the production cross section [fb] as a function of collider energy $\sqrt{s}$ [TeV] for representative masses $(M_{H^\pm})$.
For relatively light $M_{H^\pm}<1\TeV$, we find that resonant production rates are as low as $\sigma\sim0.01-1$ fb at $\sqrt{s}=2\TeV$ and can reach as high as $\sigma\sim5-10$ fb at $\sqrt{s}=30\TeV$.
For the relatively heavy $M_{H^\pm}=2-4\TeV$, rates can reach up to several fb at the largest $\sqrt{s}$ we consider.
In figures~\ref{vbfgm_hpp_diag} and \ref{vbfgm_hpp_xsec}, we show the same for $H^{++}$ in $\mu^+\mu^+$ collisions. 
For the same mass and collider scales, we find that the resonant production rates of  $H^{++}$ are a factor of a few larger than for  $H^{\pm}$.
We attribute this to the fact that the  $W\ell\nu$ coupling in the SM is larger than the $Z\ell\ell$ coupling.

\subsection{Minimal Supersymmetric Standard Model}\label{sec:bsm_mssm}

In the SM, the Higgs boson possesses no symmetry that protects or stabilizes its mass against quantum corrections that naturally drive the mass away from the EW scale and toward the scale of new physics.
As such, supersymmetric extensions of the SM (SUSY) are well-motivated theoretical scenarios.
Under SUSY, the so-called hierarchy problem is softened or removed by
hypothesizing that SM particles, along with missing members of a multiplet, 
belong to nontrivial representations of the  Poincar\'e group~\cite{Nilles:1983ge,Haber:1984rc,Martin:1997ns,Baer:2006rs}.
This leads to the existence of a new degree of freedom for each  SM one that is mass-degenerate but with opposite spin-statistics,
and that order-by-order contribute oppositely to quantum corrections of the Higgs's mass.
The lack of experimental evidence for  
superpartners~\cite{Baer:2006rs,Tanabashi:2018oca,Aad:2019pfy,Sirunyan:2019ctn,Aad:2019ftg,CMS:2019tlp,Aad:2019qnd,Aad:2019vvi,Sirunyan:2019glc,Sirunyan:2020ztc},
however, suggests that if SUSY is realized at a certain scale it is broken at the EW scale.

\begin{figure}[h!]
\centering\mbox{
\subfigure[]{\includegraphics[width=.42\textwidth]{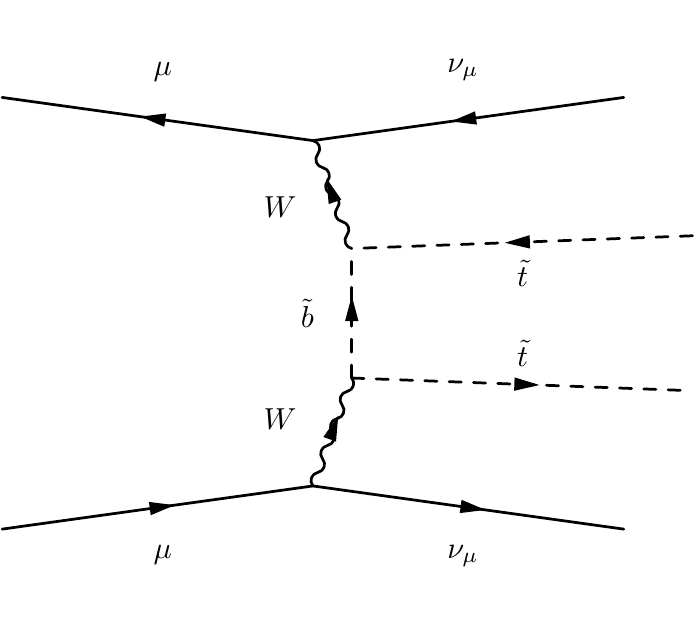}\label{vbfmssm_stop_diag}}
\hspace{0.75cm}
\subfigure[]{\includegraphics[width=.52\textwidth]{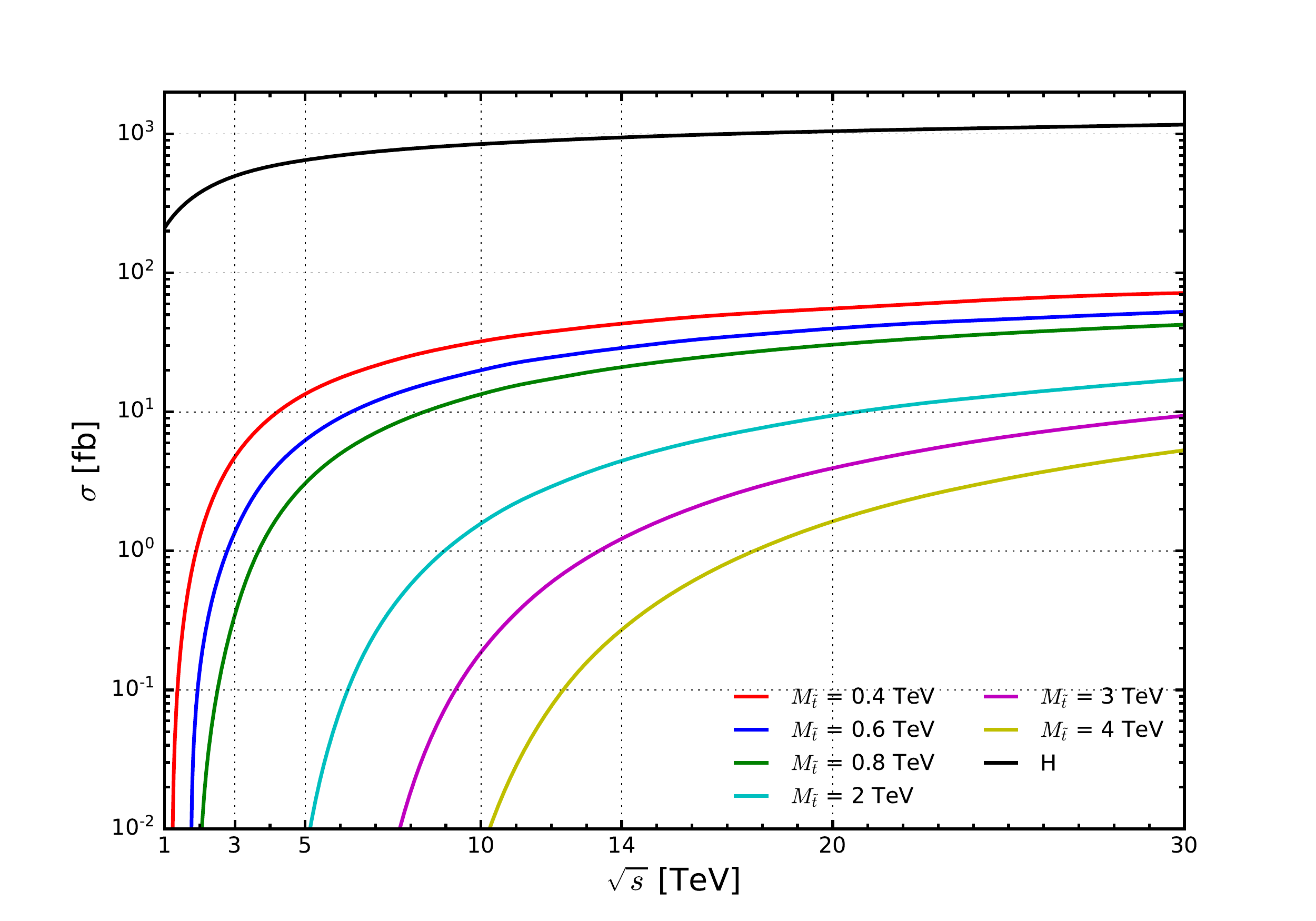}\label{vbfmssm_stop_xsec}}
}
\mbox{
\subfigure[]{\includegraphics[width=.42\textwidth]{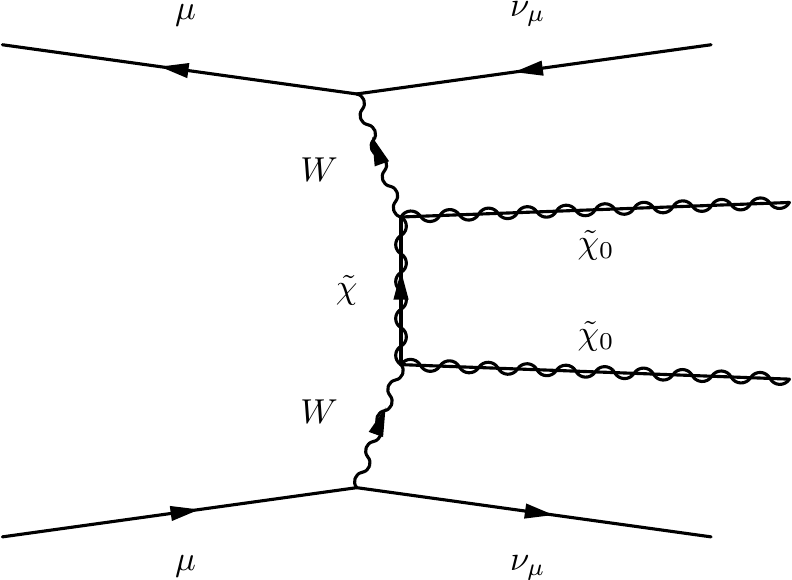}\label{vbfmssm_nuino_diag}}
\hspace{0.75cm}
\subfigure[]{\includegraphics[width=.52\textwidth]{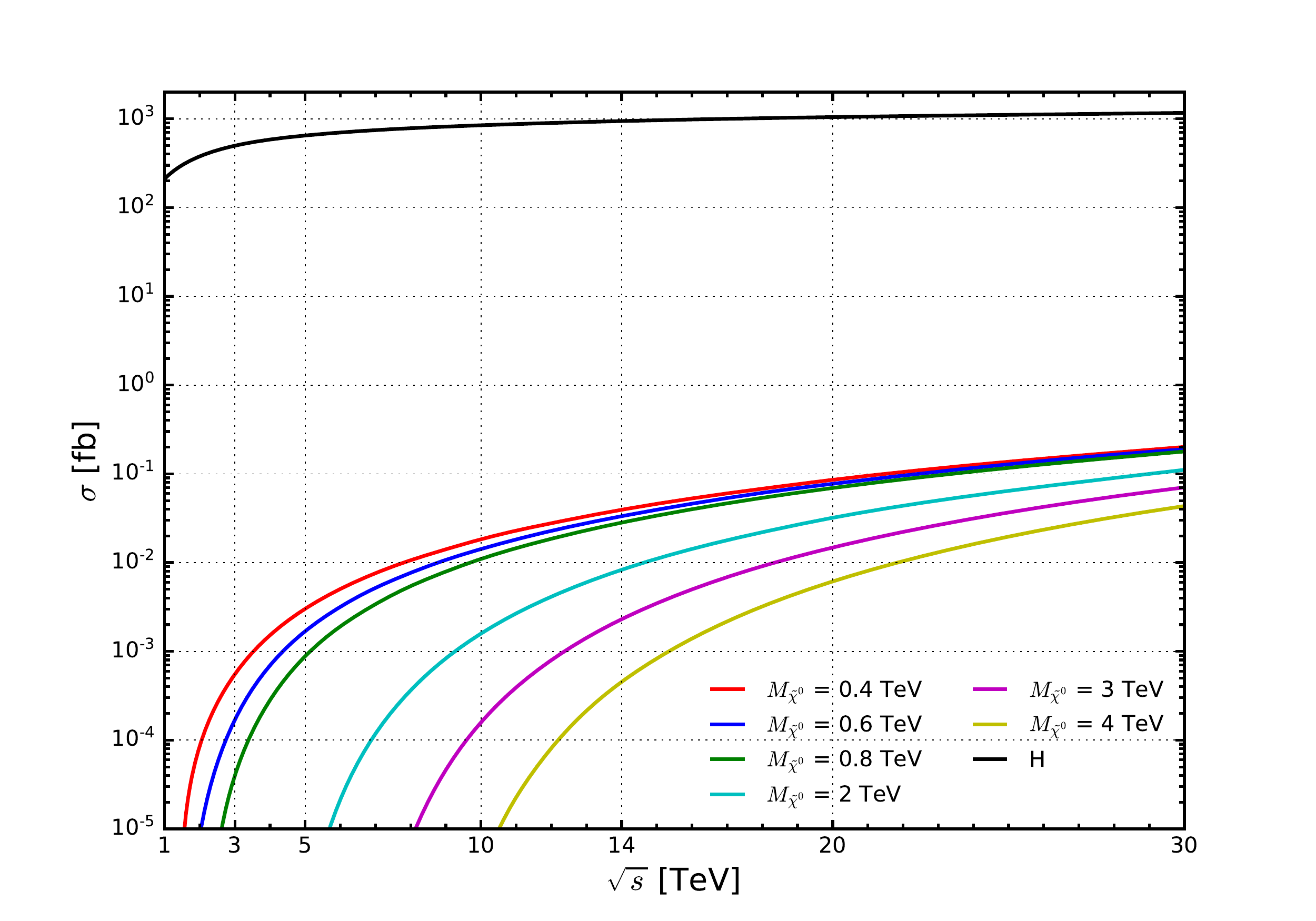}\label{vbfmssm_nuino_xsec}}
}
\mbox{
\subfigure[]{\includegraphics[width=.42\textwidth]{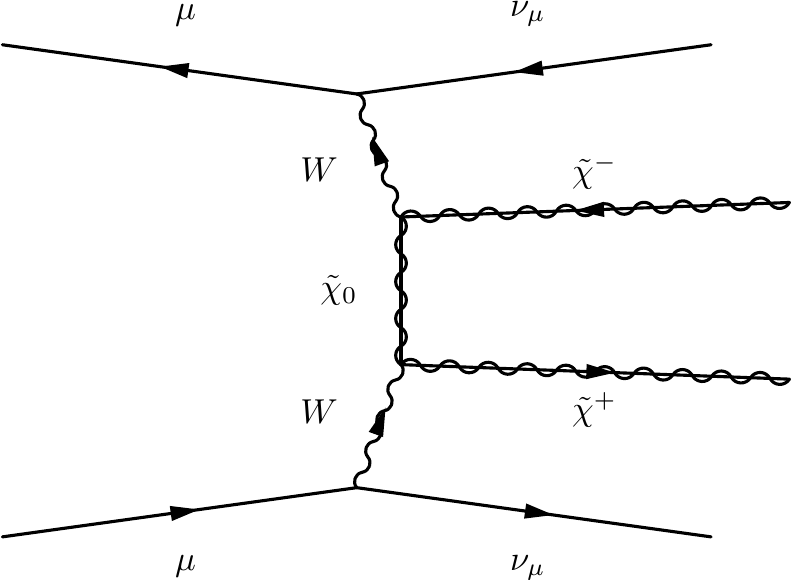}\label{vbfmssm_chino_diag}}
\hspace{0.75cm}
\subfigure[]{\includegraphics[width=.52\textwidth]{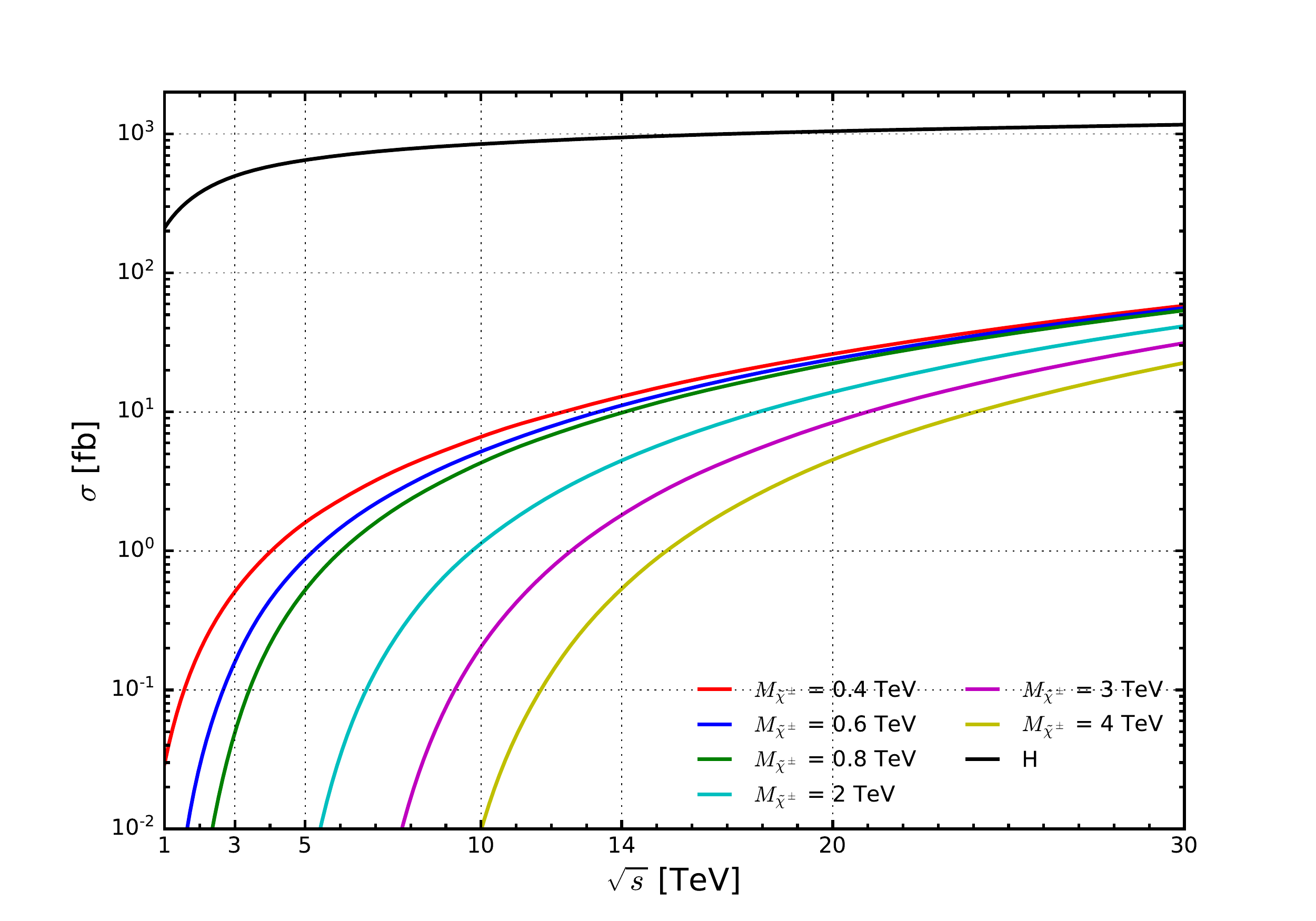}\label{vbfmssm_chino_xsec}}
}
\caption{
Same as figure~\ref{fig:bsm_vbf2hdm} but in the MSSM for 
(a,b) stop pair production,
(c,d) neutralino pair production, and
(e,f) chargino pair production.
}
\label{vbfmssm}
\end{figure}

While many variations of SUSY exist and are actively investigated,
the Minimal Supersymmetric Standard Model (MSSM) is the simplest supersymmetric model supported by phenomenology~\cite{Nilles:1983ge,Haber:1984rc,Martin:1997ns,Baer:2006rs}.
In it, the holomorphicity of the superpotential and anomaly cancellation require that two Higgs superfields be present
(implying also that the MSSM is a supersymmetric extension of the 2HDM).
The superpotential of the MSSM is given by
\begin{equation}
\mathcal{W}_{MSSM}=y_u \, \bar{u} Q H_u - y_d \, \bar{d} Q H_d - y_e \bar{e} L H_d + \mu H_u H_d,
\end{equation} 
where $H_u,\,H_d,\,Q,\,L,\,\bar{u},\,\bar{d},\,\bar{e}$, are the chiral superfields to which the Higgs and fermions belong.
Apart from these terms are the vector superfields containing gauge bosons and gauginos as well as the K\"ahler potential, which describes particles' kinetic terms.
In studies and tests of the MSSM, one often also considers $R$-parity, 
defined for each particle as
\begin{equation}
\mathcal{P}_R = (-1)^{3(B-L)+2s},
\end{equation}   
where $B$, $L$, and $s$ are the baryon number, lepton number, and spin of the particle.
By construction, all SM particles (and 2HDM scalars) have $\mathcal{P}_R=+1$, whereas their superpartners have $\mathcal{P}_R=-1$.
A consequence of $R$ parity is that the lightest supersymmetric particle is stable and, if it is electrically neutral, it is a good dark matter (DM) candidate.

Generically, scalar superpartners of quarks and leptons (squark and sleptons) with the same electric charge and color quantum numbers mix.
In the MSSM, this results in two $6\times 6$ mixing matrices for the squarks (one each for the up and down sectors) 
and a $3\times 3$ mixing matrix for charged sleptons. 
(Neutrinos are natively massless in the MSSM as they are in the SM.)
The neutral and charged superpartners of SM scalar and vector bosons also mix.
The mass eigenstates, denoted by $\tilde\chi^0_k$ and $\tilde\chi^\pm_k$,
are given as linear combinations of the fields $\{\tilde B,\tilde W^0,\tilde H^0_d,\tilde H^0_u\}$ and $\{\tilde W^+,\tilde H_u^+,\tilde W^-,\tilde H_d^-\}$, respectively.
Despite extensive searches for these states~\cite{Nilles:1983ge,Haber:1984rc,Martin:1997ns,Baer:2006rs,Tanabashi:2018oca},
including direct searches at the LHC~\cite{Aad:2019pfy,Sirunyan:2019ctn,Aad:2019ftg,CMS:2019tlp,Aad:2019qnd,Aad:2019vvi,Sirunyan:2019glc,Sirunyan:2020ztc},
evidence for the MSSM at the weak scale has yet be established.
If  the MSSM, or any variation of SUSY, is realized at the EW- or TeV-scale, 
then a multi-TeV muon collider could be an optimal machine to discover missing superparticles or study the spectrum properties.

To investigate the sensitivity of muon colliders to the MSSM,  we  consider the benchmark, simplified scenario where 
generation-1 and -2 sfermions decouple while generation-3 squarks mix in pairs, $(\tilde t_R,\,\tilde t_L)$ and $(\tilde b_R,\,\tilde b_L)$.
We use the \texttt{MSSM}~UFO libraries as developed by Ref.~\cite{Duhr:2011se},
and vary masses while keeping mass-splittings and couplings fixed.

In figure~\ref{vbfmssm} we show diagrammatically and numerically pair production of (a,b) top squarks, (c,d) neutralinos, and (c,d) charginos through VBF in $\mu^+\mu^-$ collisions.
Starting with figure~\ref{vbfmssm_stop_xsec}, we have the $\tilde{t}\overline{\tilde{t}}$ production cross section [fb] as a function of $\sqrt{s}$ [TeV], for representative stop masses.
For lighter stops with $m_{\tilde{t}}\lesssim1\TeV$, we see that cross sections span \confirm{$\sigma\sim0.01-1$ fb at $\sqrt{s}\sim2\TeV$ and reach $\sigma\sim50-75$ fb at $\sqrt{s}\sim30\TeV$}.
For heavier stops with $m_{\tilde{t}}=2-4\TeV$,  production rates reach $\sigma\sim5-20$ fb at $\sqrt{s}\sim30\TeV$.

In figure~\ref{vbfmssm_nuino_xsec}, we show the same information but for $\tilde{\chi}^0\tilde{\chi}^0$.
Overall, the picture is bleaker.
For lighter neutralinos with \confirm{$m_{\tilde{\chi}^0}\lesssim1\TeV$, pair production rates through weak boson fusion remain below $\sigma\sim0.01$ fb for collider energies below $\sqrt{s}\sim7-10\TeV$.} 
They reach just below \confirm{$\sigma\sim0.2$ fb at $\sqrt{s}\sim30\TeV$}.
For heavier neutralinos with $m_{\tilde{\chi}^0}=2-4\TeV$, we see that cross sections remain below \confirm{$\sigma\sim0.1$ fb for  $\sqrt{s}\lesssim30\TeV$}.

In figure~\ref{vbfmssm_chino_xsec}, we again show the same information but for $\tilde{\chi}^+{\tilde{\chi}^-}$.
We find that the outlook is somewhere between the previous cases.
For lighter charginos with \confirm{$m_{\tilde{\chi}^\pm}\lesssim1\TeV$}, pair production rates quickly reach about \confirm{$\sigma\sim0.01$ fb for  $\sqrt{s}\sim2-4\TeV$} 
and  about \confirm{$\sigma\sim75$ fb at $\sqrt{s}\sim30\TeV$}.
For heavier charginos with \confirm{$m_{\tilde{\chi}^\pm}\lesssim2-4\TeV$}, rates reach $\sigma\sim0.01-1$ fb when \confirm{$\sqrt{s}\sim7-12\TeV$},
and span roughly \confirm{$\sigma\sim20-40$ fb for the highest $\sqrt{s}$} considered.


\subsection{Vector leptoquarks}\label{sec:bsm_LQ}

\begin{figure}[t!]
\centering\mbox{
\subfigure[]{\includegraphics[width=.42\textwidth]{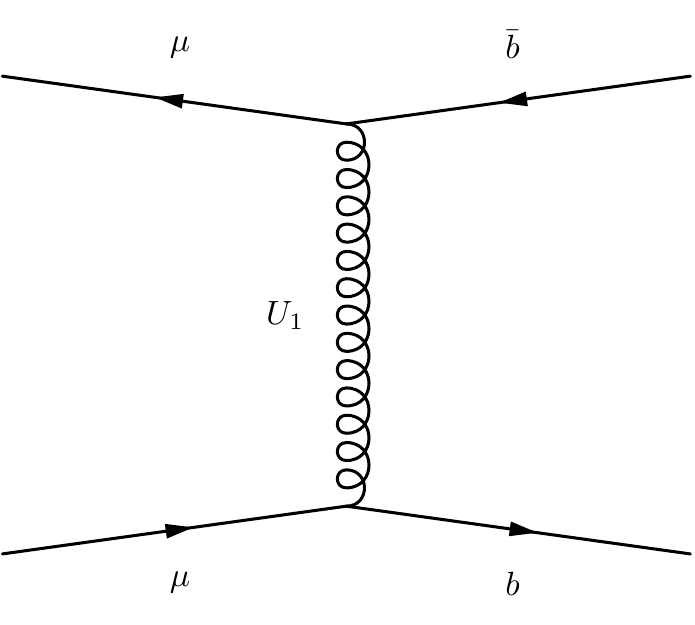}\label{vbfveclq_diag}}
\hspace{0.75cm}
\subfigure[]{\includegraphics[width=.52\textwidth]{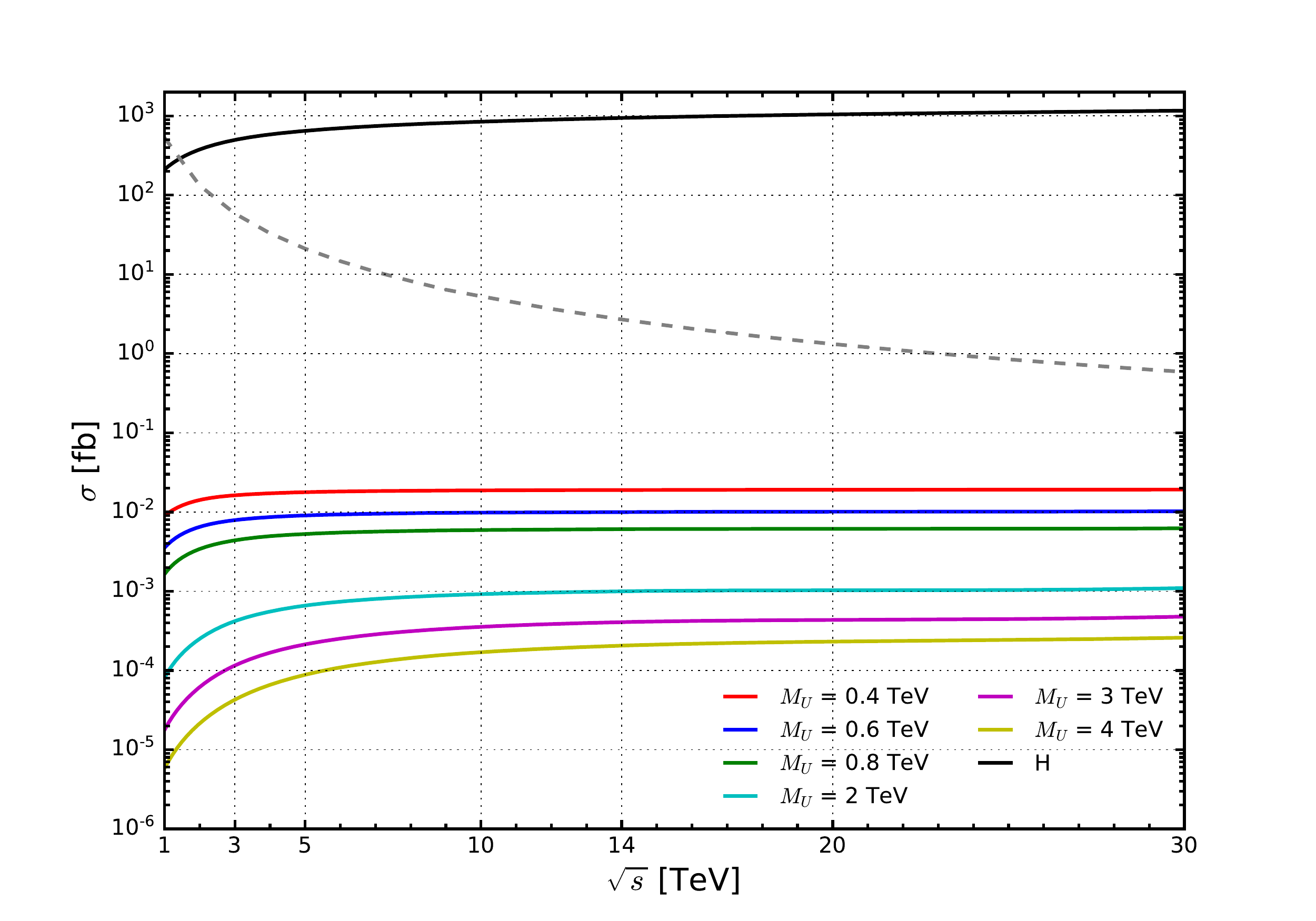}\label{vbfveclq_xsec}}
}
\caption{
(a) Diagrammatic representation of $b\overline{b}$ production in $\mu^+\mu^-$ collisions via the $t$-channel exchange of the vector leptoquark $U_1^\mu$.
(b) The associated cross section [fb] as a function of collider energy $\sqrt{s}$ [TeV] for representative $M_{\rm U}$.
Also shown is SM $\mu^+\mu^- \to b\overline{b}$ production (dashed curve).
}
\label{vbfveclq}
\end{figure}

The existence of leptoquarks, i.e., scalar and vector bosons with nonzero baryon and lepton numbers, 
that also possess SM gauge charges have long been predicted
due to their necessity in certain grand unified 
theories (GUTs)~\cite{Pati:1974yy,Georgi:1974sy,Fritzsch:1974nn,Dimopoulos:1979es,Senjanovic:1982ex,Schrempp:1984nj,Hewett:1988xc,Frampton:1989fu}.
Though not conclusively  established~\cite{Dorsner:2016wpm,Tanabashi:2018oca},
the possibility  of leptoquarks is a viable solution to longstanding anomalies observed across several flavor experiments~\cite{Lees:2013uzd,Aaij:2014ora,Aaij:2015yra,Hirose:2016wfn,Aaij:2017deq,Aaij:2017vbb}.
These anomalies suggest a violation of lepton flavor universality  beyond what is allowed by  neutrino oscillations.
Hence, discovering and measuring properties of leptoquarks constitute an intriguing prospect at current and future experiments.
For  reviews on the topic, see Refs.~\cite{Dorsner:2016wpm,Cerri:2018ypt} and references therein.

While the spectrum of leptoquark models is vast, especially interesting options are those featuring vector leptoquarks due to their direct role in GUTs 
and recent demonstrations of their ultraviolet completions~\cite{Barbieri:2015yvd,Buttazzo:2016kid,Barbieri:2016las,DiLuzio:2017vat}.
For our purposes, we consider the concrete example~\cite{Baker:2019sli} where the vector leptoquark $U_1^\mu$ arises from the enlarged gauge group
\begin{equation}
\mathcal{G}_{\rm NP} = { SU}(4)\times { SU}(2)_L\times { U}(1)_{T_R^3},
\end{equation}
which itself is a subgroup of the Pati-Salam group $\mathcal{G}_{{\rm PS}} = SU(4)\times SU(2)_L\times SU(2)_R$~\cite{Pati:1974yy}. 
In this case, $U_1^\mu$ is in the $(\mathbf 3, \mathbf 1, 2/3)$ representation of the SM gauge group.
At low energies, the relevant Lagrangian (before EWSB) can be described phenomenologically  by~\cite{Baker:2019sli}:
\begin{align}
\mathcal{L}_{U_1}&=-\frac{1}{2}\,U_{1\,\mu\nu}^\dagger\, U_1^{\mu\nu}+M_U^2\,U_{1\,\mu}^\dagger\, U_1^{\mu}-ig_s\,(1-\kappa_U)\,U_{1\,\mu}^\dagger\,T^a\,U_{1\,\nu}\,G^{a\,\mu\nu}
\label{eq:bsm_LagLeptoquark}
\\
&\quad-ig_Y\,\frac{2}{3}\,(1-\tilde\kappa_U)\,U_{1\,\mu}^\dagger\,U_{1\,\nu}\,B^{\mu\nu}+\frac{g_U}{\sqrt{2}}\,[U_1^\mu\,(\beta_L^{ij}\,\bar q^i_L\gamma_\mu\ell_L^j+\beta_R^{ij}\,\bar d^i_R\gamma_\mu e_R^j)+{\rm H.c.}].\nonumber
\end{align}
Here, $G^{\mu\nu}=T^a G^{a\mu\nu}$ and $B^{\mu\nu}$ are the QCD and hypercharge field strengths, with associated gauge couplings $g_s$ and $g_Y$.
$U^{\mu\nu}_1$ and $M_U$ are the field strength and mass of $U_1$.
$\kappa_U$ and $\kappa_{\tilde{U}}$ are anomalous couplings that vanish in gauged leptoquark models.
$q_L,~\ell_L,~d_R,~e_R$ are the SM chiral fermion fields in the flavor basis,
and $g_U$ is a flavor-universal $q-\ell-U$ coupling strength while $\beta^{ij}$ absorbs possible flavor dependencies.

In view of the aforementioned flavor anomalies, we assume that leptoquarks, if they indeed exist,
couple mainly to generation-3 fermions with the possible extension to muons.
Hence, to explore the sensitivity of multi-TeV muon colliders, we consider the process
\begin{equation}\label{eq:veclq}
\mu^+ \mu^- \to b\,\bar b
\end{equation} 
mediated by a $t$-channel exchange of the vector leptoquark $U_1^\mu$,
as shown in figure~\ref{vbfveclq_diag}.
We work in the framework of equation~\ref{eq:bsm_LagLeptoquark} as implemented into the \texttt{LeptoQuark} FeynRules UFO model~\cite{Baker:2019sli}.
For our purposes, the relevant parameters are $g_U$ and $\beta_{L/R}^{ij}$ and we assume the default values of the model file.
We report our results in figure~\ref{vbfveclq_xsec}, where we show the $\mu^+ \mu^- \to b\,\bar b$ 
cross section [fb] as a function of collider energy $\sqrt{s}$ [TeV] for representative $M_{U}$.
Also shown is the SM $\mu^+\mu^- \to b\overline{b}$ production rate (grey, dash curve).
For both light and heavy $U_1^\mu$ masses, we observe only a mild dependence on collider energy.
More specifically, for $M_{U} = 0.4 - 0.8\TeV$, we find cross sections are roughly \confirm{$\sigma\sim\mathcal{O}(0.01)$ fb for $\sqrt{s}\sim2-30\TeV$}.
For heavier masses in the range of $M_{U} = 2 - 4\TeV$, we see that cross sections span \confirm{$\mathcal{O}(10^{-4})-\mathcal{O}(10^{-3})$ fb for 
collider energies of $\sqrt{s}\sim5-30\TeV$}.

\subsection{Heavy Dirac and Majorana neutrinos}\label{sec:bsm_heavyN}

In the SM, neutrinos are massless fermions. 
Neutrino oscillation data~\cite{Ahmad:2002jz,Ashie:2005ik}, however, 
unambiguously demonstrate that they in fact possess exceptionally tiny masses, with $m_{\nu_k} < \mathcal{O}(1)$ eV~\cite{Aker:2019uuj}.
If neutrinos are Majorana fermions, then their mass are also related to the breaking of lepton number conservation~\cite{Schechter:1981bd,Hirsch:2006yk,Duerr:2011zd,Moffat:2017feq}, 
an accidental symmetry in the SM.
In order to reconcile these observations with the SM paradigm, neutrino mass models, collectively known as Seesaw models, 
hypothesize the existence of new particles that necessarily~\cite{Ma:1998dn} couple to SM leptons and the Higgs.
If kinematically accessible, such states may be discovered at collider experiments through spectacular processes that violate lepton flavor and lepton number conservation;
for comprehensive reviews, see Refs.~\cite{Cai:2017jrq,Cai:2017mow}.

\begin{figure}[t!]
\centering\mbox{
\subfigure[]{\includegraphics[width=.42\textwidth]{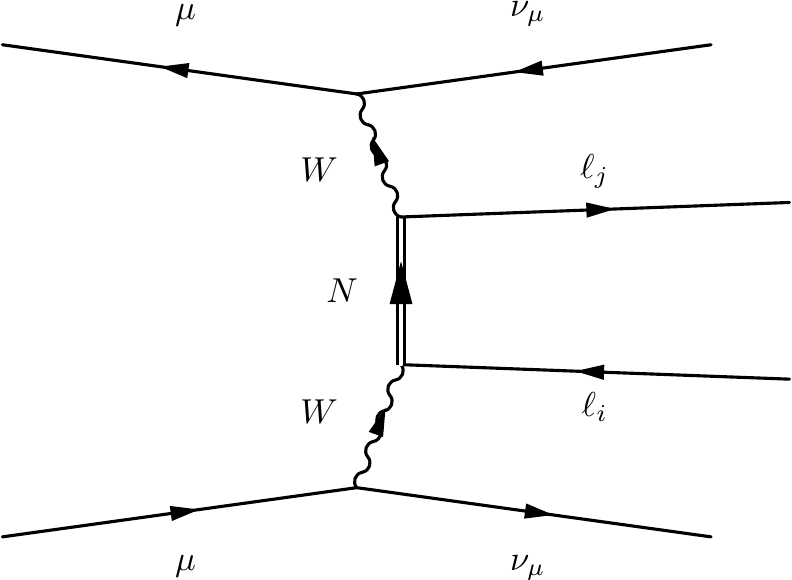}\label{vbfhn_diag}}
\hspace{0.75cm}
\subfigure[]{\includegraphics[width=.52\textwidth]{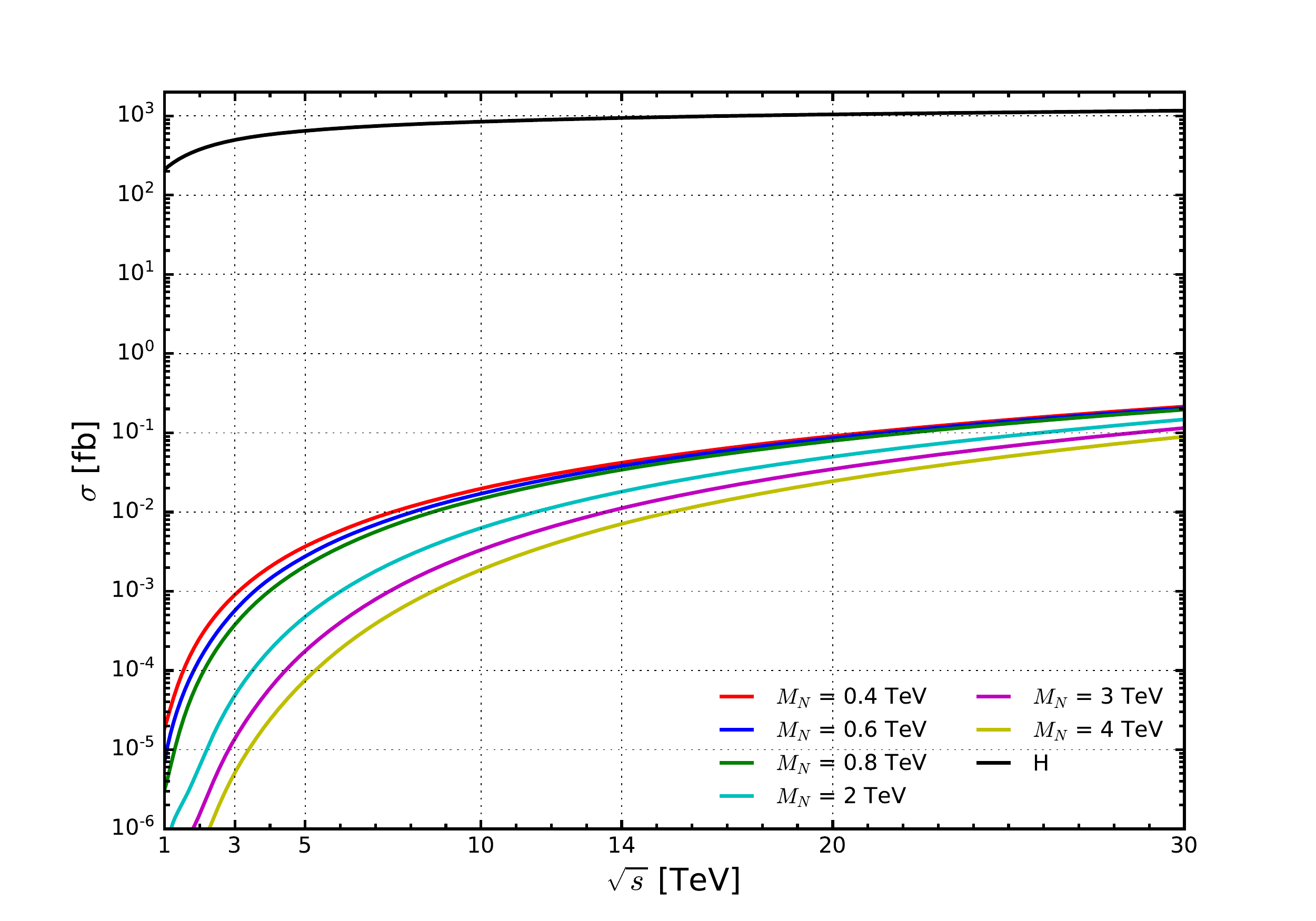}\label{vbfhn_xsec}}
}
\caption{
(a) Diagrammatic representation of $\ell^+_i\ell^-_j$ production via $t$-channel exchange of a heavy neutrino $N$.
(b) The cross section [fb] as a function of collider energy $\sqrt{s}$ [TeV] for mass $M_{N}$.
}
\label{vbfhn}
\end{figure}

A commonality of many Seesaw models is the existence of heavy neutrino mass eigenstates $N_{m'}$ that can be either (pseudo-)Dirac or Majorana.
These states couple to the SM sector through mixing with SM neutrinos and/or new gauge couplings.
For our purposes, we neglect subtleties related to decoupling of lepton number-violating processes in simplified models with only heavy neutrinos~\cite{Pilaftsis:1991ug,Kersten:2007vk,Moffat:2017feq},
and consider the well-studied~\cite{delAguila:2008cj,Atre:2009rg,Pascoli:2018heg} Phenomenological Type I Seesaw benchmark model,
as implemented in the \texttt{HeavyN} UFO libraries of Ref.~\cite{Alva:2014gxa,Degrande:2016aje}.
In this model, neutrino flavor eigenstates $\nu_\ell$ can be expressed generically~\cite{Atre:2009rg} in terms of light and heavy mass eigenstates by the decomposition
\begin{equation}
\nu_\ell = \sum_{m=1}^3 U_{\ell m} \nu_m ~+~ \sum_{m'=4}^6 V_{\ell m'} N_{m'} \approx \sum_{m=1}^3 U_{\ell m} \nu_m ~+~  V_{\ell m'=4} N.
\label{eq:nuDecomposition}
\end{equation}
In the last expression we assumed that active-sterile mixing $V_{\ell N}$ is dominated by the lightest, heavy mass eigenstate $(m'=4)$, 
which we relabel as $N\equiv N_{m'=4}$.
The relevant interaction Lagrangian coupling $N$ to the SM Weak bosons after EWSB  is 
\begin{eqnarray}
\mathcal{L}_N^{\rm Int.} 
\approx	& -& \cfrac{g}{\sqrt{2}} 			\sum_{\ell=e}^\tau \overline{N} V^*_{\ell 4} W_\mu^+ \gamma^\mu  P_L  \ell^- 
  	 -\cfrac{g}{2\cos\theta_W} 	\sum_{\ell=e}^\tau \overline{N} V^*_{\ell 4} Z_\mu \gamma^\mu  P_L  \nu_\ell  \nonumber\\
 	& -& \cfrac{g}{2M_W} h 		\sum_{\ell=e}^\tau \overline{N} V^*_{\ell 4} M_{N}  P_L  \nu_\ell 
 	 + {\rm H.c.}
	 \label{eq:vbfhn_lag}
\end{eqnarray}
Here, $g\approx0.65$ is the $SU(2)$ coupling constant, $\theta_W$ is the weak mixing angle, 
and $P_{L/R}=(1\mp\gamma^5)/2$ are the usual  chiral projection operators for four-component fermions.

While there exists a number of processes in which heavy neutrinos can participate, 
we focus on the production of oppositely charged lepton pairs through $W^+W^-$ scattering:
\begin{equation}
W^+ W^- \to \ell^+_i \ell^-_j,
\end{equation}
as show in figure~\ref{vbfhn_diag}.
This signature  complements conventional channel, including the $s$-channel $N\ell$ and $N\nu$ processes and $W^\pm\gamma\to N\ell^\pm$ fusion,
due to its particular sensitivity to active-sterile mixing, which scales as $\sigma_{\mu\mu}\sim\vert V_{\ell_i N}V_{\ell_j N}^* \vert^2$,
and not requiring that $N$ be on-shell.
Furthermore, observing this process for $\ell_i\neq \ell_j$ would give a clear indication of charged lepton flavor violation and provide guidance on the structure of neutrino  mixing.

 In figure~\ref{vbfhn_xsec}, we show the cross section [fb] for the flavor-conserving process,
\begin{equation}
\mu^+\mu^- \to \nu_\mu\overline{\nu}_\mu \mu^+ \mu^-,
\end{equation} 
mediated by a heavy $t$-channel neutrino, for  representative mass $M_{N}$, and as a function of collider energy $\sqrt{s}$ [TeV].
For concreteness, we take {$\vert V_{\mu N}\vert=0.1$.}
  As in the leptoquark case in section~\ref{sec:bsm_LQ}, we observe only a slight rate dependence over a large range of neutrino masses.
  For $m_N = 0.4-4\TeV$, we find that cross sections reach the \confirm{$\sigma\sim10^{-4}$ fb threshold at about $\sqrt{s}=1-5\TeV$.}
  For much larger collider energies, we observe that \confirm{scattering rates can reach up to $\sigma\sim0.1-0.2$ fb for collider energies as large as $\sqrt{s}=30\TeV$.}

\subsection{Vector-like quarks}\label{sec:bsm_vlq}

A more curious aspect of the SM is the existence of three copies, or generations, of matter.
While at least three generations are necessary for CP violation in the quark sector~\cite{Kobayashi:1973fv},
no first-principle argument establishes this to be the case.
Moreover, as additional chiral generations are  constrained by flavor and Higgs data~\cite{Eberhardt:2010bm,Djouadi:2012ae,Eberhardt:2012gv},
 if more copies do exist, such matter particles likely belong to different EW representations or possess new quantum numbers.
One such example: vector-like fermions, which are characterized by their left- and right-handed chiral components possessing identical gauge transformations 
but may nonetheless carry the same gauge charges as SM particles after EWSB.

As discussed in section~\ref{sec:bsm_heavyN}, vector-like electrons and neutrinos are key ingredients of neutrino mass models. 
In addition, vector-like quarks (VLQ) offer viable, non-supersymmetric solutions to the Higgs mass hierarchy and  dynamical EWSB~\cite{ArkaniHamed:2002qy,Schmaltz:2005ky,Martin:2009bg}.
The phenomenology of such models is rich, well-documented~\cite{Han:2003wu,Hubisz:2004ft,Perelstein:2005ka,Aguilar-Saavedra:2013qpa,Buchkremer:2013bha},
and has led to LHC searches for vector-like top and  bottom quarks
in a variety of final states~\cite{Aaboud:2018xpj,Aaboud:2018pii,Aaboud:2018wxv,Sirunyan:2018qau,Sirunyan:2019sza,Sirunyan:2019xeh}.

\begin{figure}[t!]
\centering\mbox{
\subfigure[]{\includegraphics[width=.42\textwidth]{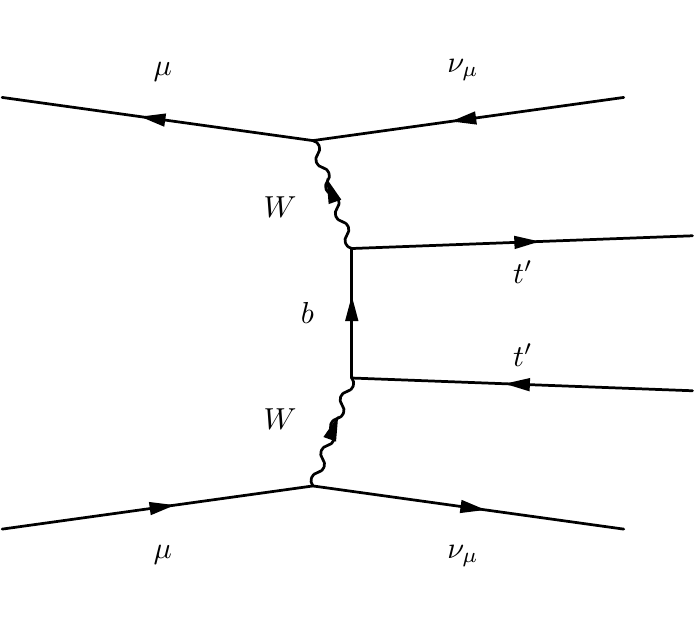}\label{vbfvlq_diag}}
\hspace{0.75cm}
\subfigure[]{\includegraphics[width=.52\textwidth]{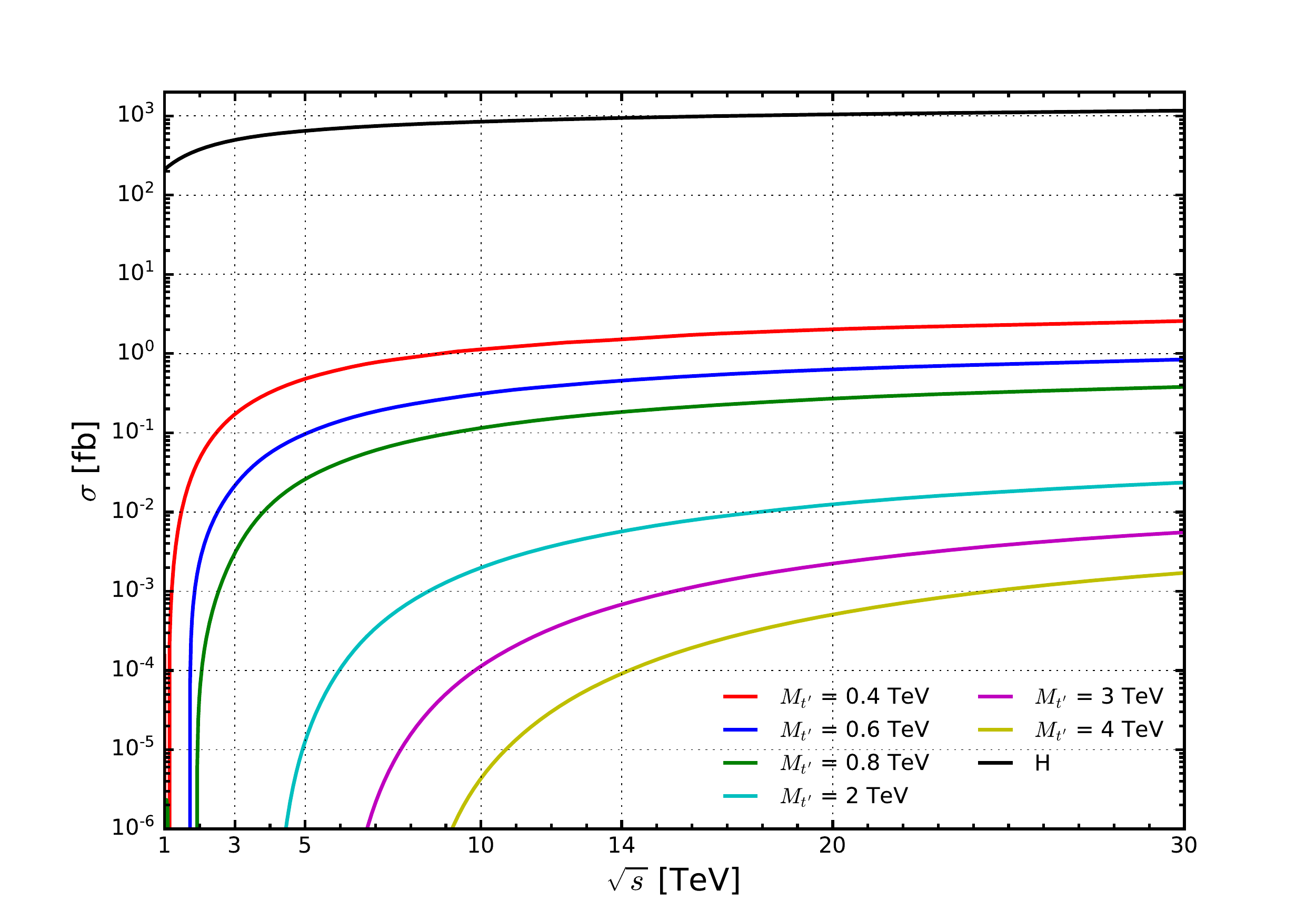}\label{vbfvlq_xsec}}
}
\caption{
Same as figure~\ref{fig:bsm_vbfsing} but for the VLQ pair $t'\overline{t'}$, as described by equation~\ref{eq:topL_proc}.
}
\label{vbfvlq}
\end{figure}

For fermionic top partners $t'$, i.e., a VLQ with the same quantum numbers as the top quark after EWSB,
the effective Lagrangian describing $t'$ can be parametrized by ``decomposing'' the top quark  further into two mass eigenstate:
\begin{equation}
t(M_t\approx173\GeV) \to t  \approx t(M_t\approx173\GeV) + \kappa t'(M_{t'}) + \mathcal{O}(\kappa^2).
\end{equation}
Here, $\kappa$ is a small, model-dependent mixing parameter and the abuse of notation is obvious.
While the Lorentz structure of the gluon and photon interactions with $t'$ are dictated by gauge invariance, those of the EW bosons are less restricted.
Generically, the EW couplings of a single $t'$ with  $u$- and $d$-flavored, SM quarks can be written as~\cite{Buchkremer:2013bha}:
\begin{align}
\mathcal{L}_{t'-single} &=  \kappa_W V_{L/R}^{4i} \frac{g}{\sqrt{2}}\; [\bar{t'}_{L/R} W_\mu^+ \gamma^\mu d^i_{L/R} ]  + \kappa_Z V_{L/R}^{4i} \frac{g}{2 c_W} \; [\bar{t'}_{L/R} Z_\mu \gamma^\mu u^i_{L/R} ]  \nonumber \\
&- \kappa_H V_{L/R}^{4i} \frac{M_{t'}}{v}\; [\bar{t'}_{R/L} H u^i_{L/R} ] + {\rm H.c.}
\label{eq:topL}
\end{align}
Here, $M_{t'}$ is the mass of the VLQ, 
$V_{L/R}^{4i}$ is model-dependent and accounts for any potential flavor mixing,
the index $i$ runs over the three SM generations,
and the parameters $\kappa_V$ ($V = W$, $Z$, $H$) encode anomalous couplings to the EW bosons.

To investigate the sensitivity of multi-TeV muon colliders to VLQs, we consider $t'\overline{t'}$ pair production from $W^+W^-$ fusion,
as shown in figure~\ref{vbfvlq_diag} and given by
\begin{equation}
\mu^- \mu^+ \to \nu_\mu \overline{\nu}_\mu t' \overline{t'}.
\label{eq:topL_proc}
\end{equation}
Using equation~\ref{eq:topL} as implemented in the \texttt{VLQ} UFO libraries by Ref.~\cite{Buchkremer:2013bha},
we show in figure~\ref{vbfvlq_xsec} the $W^+W^- \to t' \overline{t'}$ cross section [fb] in 
$\mpmm$ collisions as a function of collider energy $\sqrt{s}$ [TeV], for representative $M_{t'}$.
We assume the \confirm{default couplings} of Ref.~\cite{Buchkremer:2013bha}.

Overall, we observe a large variation of production rates as a function of mass and collider energy.
For lighter $t'$, with $M_{t'} = 0.4-0.8\TeV$, we find that cross sections remain below the \confirm{$\sigma\sim10^{-4}$ fb level for $\sqrt{s}=2-3\TeV$},
 but quickly grow with increasing $\sqrt{s}$.
For the same mass range, rates reach roughly \confirm{$\sigma\sim0.5-5$ fb by $\sqrt{s}=30\TeV$.}
For heavier $t'$ with $M_{t'}=2-4\TeV$, we see that the rate growth is  milder,
with the  \confirm{$\sigma\sim10^{-4}$ fb threshold  achieved at $\sqrt{s}\sim7-15\TeV$.}
By  \confirm{$\sqrt{s}=30\TeV$, rates reach up to $\sigma\sim10^{-3}-10^{-2}$ fb.}

\subsection{Overview of vector boson fusion sensitivity}\label{sec:bsm_overview}

\begin{figure}[t!]
\centering\mbox{{\includegraphics[width=.70\textwidth]{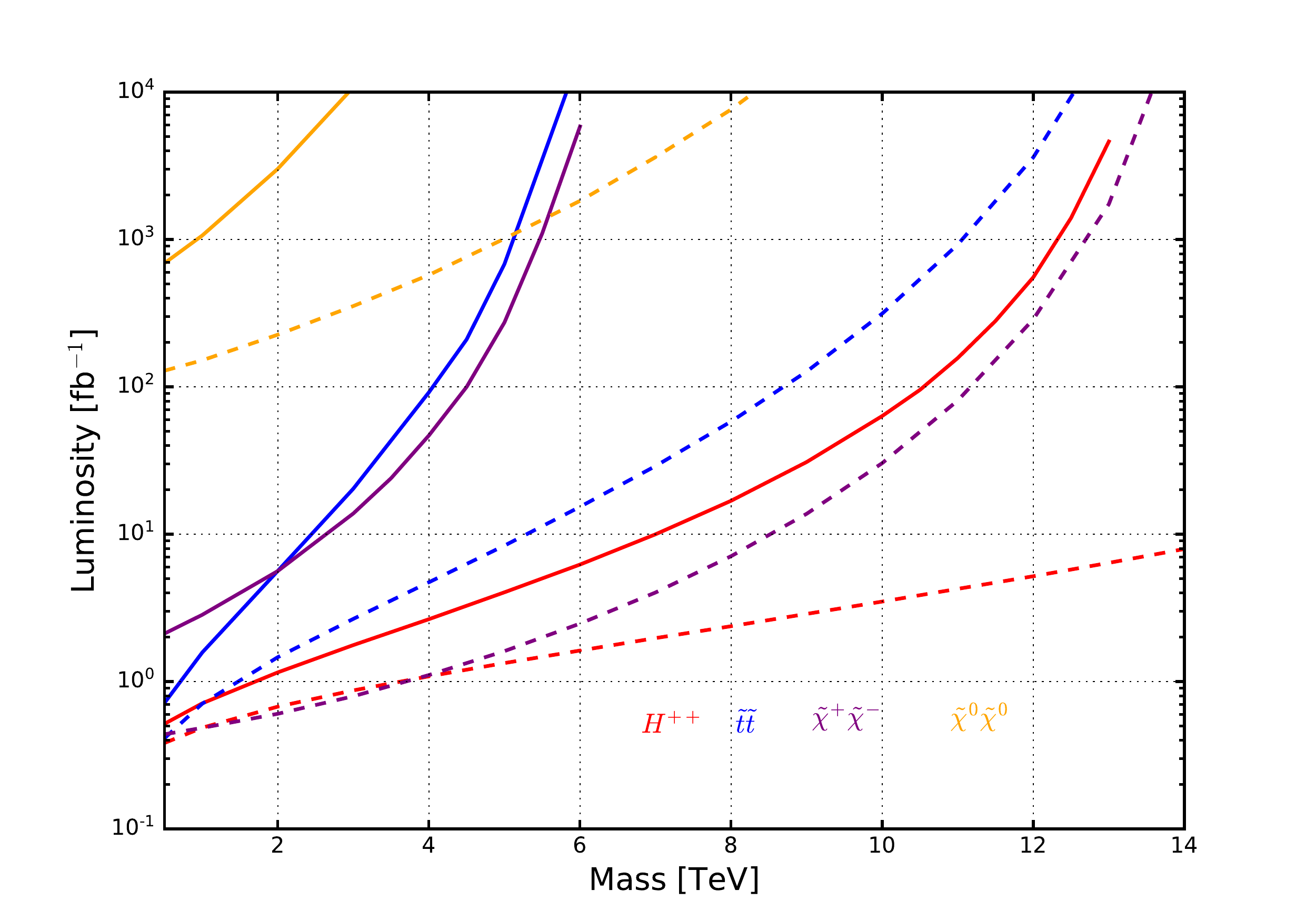}}}
\caption{
Required luminosity [fb] for a  $5\,\sigma$ discovery of 
$H^{++}$ (red) in the GM model;
$\tilde{t}\overline{\tilde{t}}$ (blue), 
$\tilde{\chi}^+\tilde{\chi}^-$ (purple), and
$\tilde{\chi}^0\tilde{\chi}^0$ (yellow)  from in the MSSM,
using VBF in $\sqrt{s}=14\TeV$ (solid) and $30\TeV$ (dashed)  muon collisions.
}
\label{fig:sensi}
\end{figure}

In this section we investigated the sensitivity of EW VBS to a variety of BSM scenarios at multi-TeV muon colliders.
In order to give an overview picture of this reach, we present in figure~\ref{fig:sensi} the
requisite integrated luminosity $\mathcal{L}$ [fb$^{-1}$] for a $5\sigma$ discovery as a function of new particle mass in $\sqrt{s}=14\TeV$ (solid) and $30\TeV$ (dashed) muon collisions.
We consider specifically 
the doubly charged Higgs $H^{++}$ (red) from the GM model (see section~\ref{sec:bsm_gm});
$\tilde{t}\overline{\tilde{t}}$ (blue), 
$\tilde{\chi}^+\tilde{\chi}^-$ (purple), and
$\tilde{\chi}^0\tilde{\chi}^0$ (yellow) pairs from the MSSM (see section~\ref{sec:bsm_mssm}).
As dedicated signal and background analyses are beyond the scope of this document, we crudely assume a zero background hypothesis and full signal acceptance.
We therefore also use
as a simple measure of statistical significance $(\mathcal{S})$ the formula,  $\mathcal{S}=\sqrt{\mathcal{L}\times\sigma}$.

As a general feature, we see that less integrated luminosity is needed to achieve the same discovery at
higher collider energies (dashed lines) than is needed at lower collider energies (solid lines).
For example: 
For $\tilde{\chi}^0\tilde{\chi}^0$ pair production with $M=2\TeV$, about 
 \confirm{$\mathcal{L}\approx 3000~(200)\invfb$ at $\sqrt{s}=14~(30)\TeV$ are needed to reach $5\sigma$.}
Similarly, for $\tilde{\chi}^\pm$ pair production with $M=5$ TeV, one can pass the $5\sigma$
threshold with roughly \confirm{$\mathcal{L}\approx 250~(1.5)\invfb$.}
For $H^{++}$ of mass $M=10\TeV$, one would need \confirm{about $\mathcal{L}=60~(3.5)\invfb$ at  $\sqrt{s}=14~(30)\TeV$.}

While highly intuitive for $pp$ colliders, this behavior is somewhat a novelty for lepton colliders because typical, $s$-channel annihilation processes 
exhibit cross sections that \textit{decrease} with \textit{increasing} collider energy.
Hence, for $s$-channel annihilations, one typically  needs more data at higher collider energies to achieve the same discovery potential.
We attribute this improved outlook to the increasing likelihood for forward, initial-state EW boson radiation at higher collider energies.
That is to say, the opening and increasing importance of EW vector boson fusion channels.
In terms of the parton luminosity language of section~\ref{sec:ppvsmuon},
a higher collider energy translates to a larger EW boson parton luminosity.
For a fixed ``partonic'' scattering rate, this  leads to an increased, beam-level cross section, and therefore higher sensitivity.
In this sense, multi-TeV lepton colliders start resembling proton colliders, and effectively become high-luminosity, weak boson colliders.
While remaining  in the context of the above BSM scenarios, we now explore this perspective further.

\section{New physics processes  at muon colliders: annihilation vs fusion}\label{sec:bsm_vbf}

As we have shown here and throughout previous sections, VBF production cross sections $(\sigma^{\rm VBF})$ grow with increasing $\sqrt{s}$,
a phenomenon that follows from the propensity for forward emission of transverse gauge bosons at  increasing collider energies.
While the precise dependence of $\sigma^{\rm VBF}$ on collider energies of course 
depends  on the precise BSM signature, for example on the particles involved, their underlying dynamics, and their kinematics,
it nevertheless contrasts with $s$-channel, annihilation processes.
These processes feature cross sections $(\sigma^{s-ch.})$ that instead decrease with collider energy as $\sigma^{s-ch.}\sim1/s$,
when well above kinematic thresholds.
Hence, just as in the SM, we find a commonality in all VBF process here: 
assuming fixed model inputs, then for sufficiently high collider energies, VBF cross sections exceed those of analogous, $s$-channel production modes.

\begin{figure}[t!] 
\centering
\subfigure[]{\includegraphics[width=.49\textwidth]{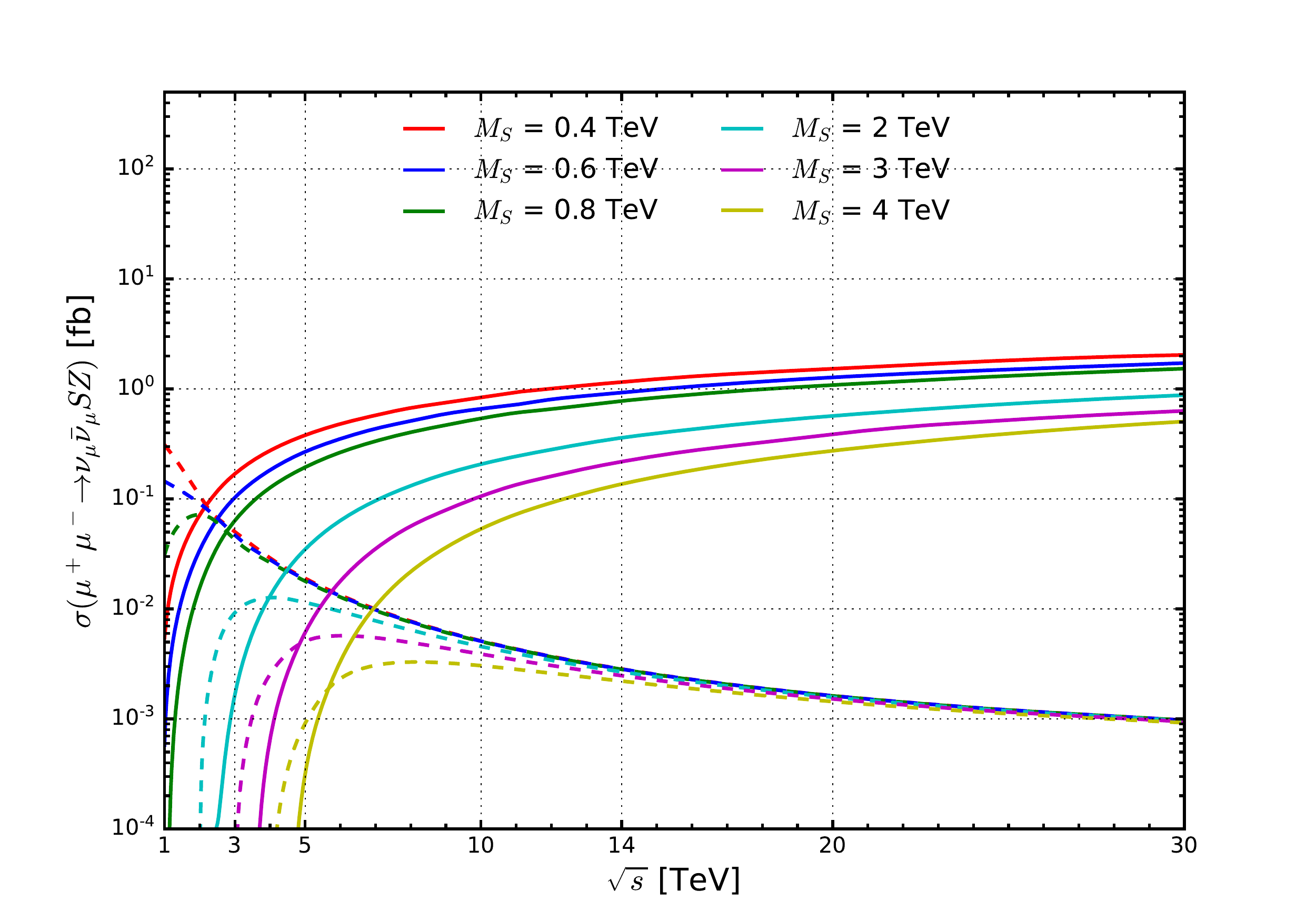}\label{fig:vbfschh2s_sing}}
\subfigure[]{\includegraphics[width=.49\textwidth]{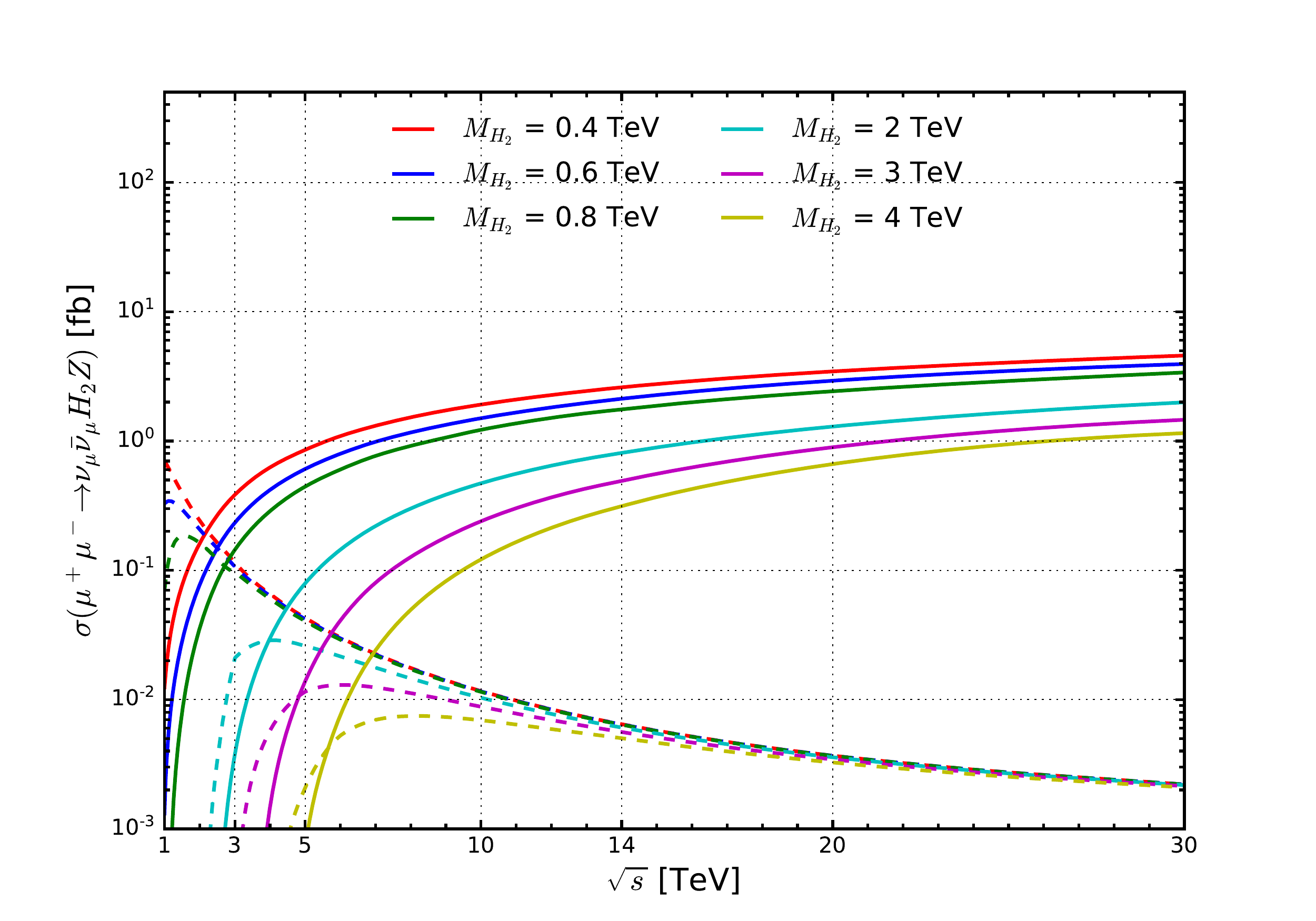}\label{fig:vbfschh2s_2hdm}}
\subfigure[]{\includegraphics[width=.49\textwidth]{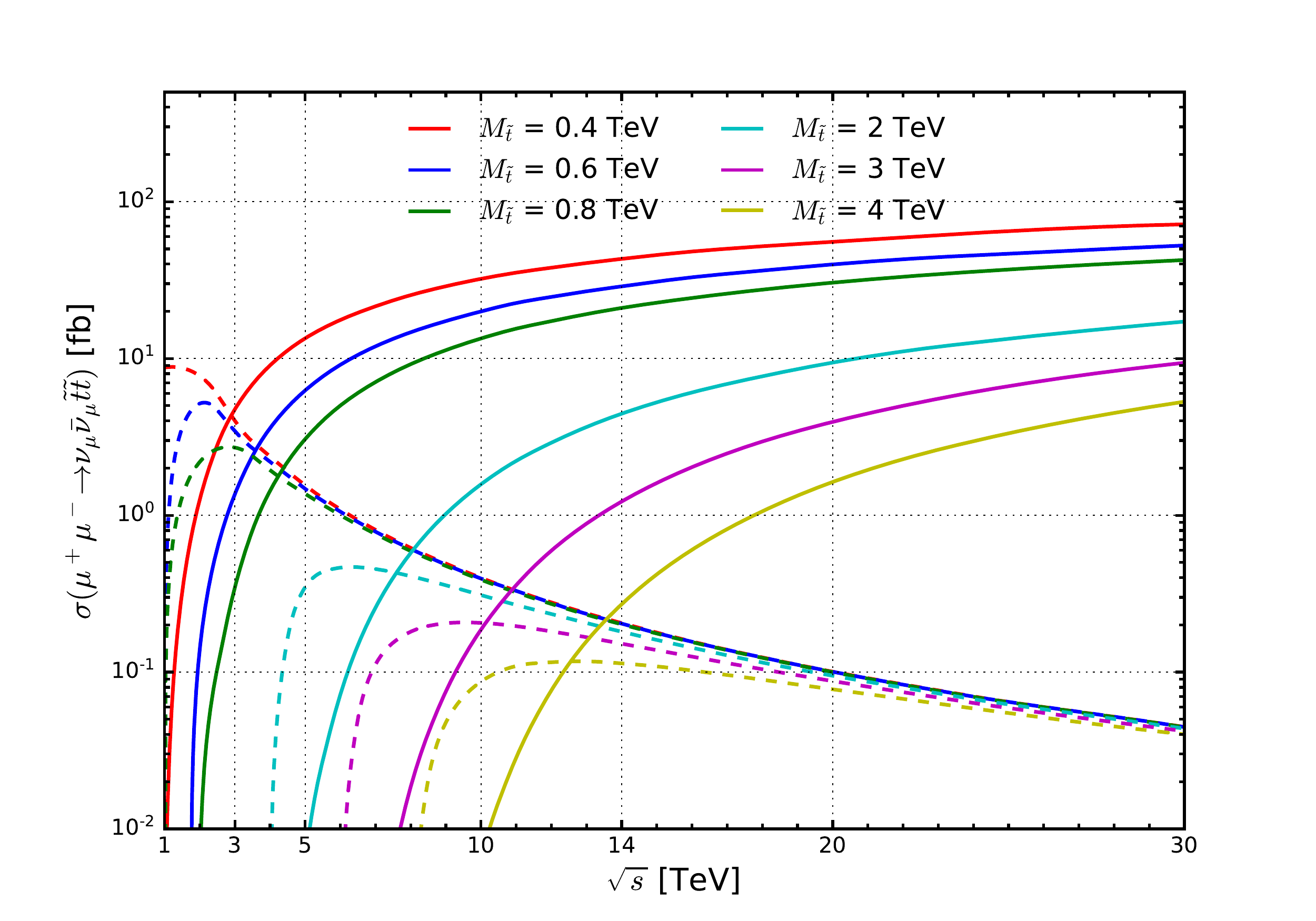}\label{fig:vbfschsttp_stop}}
\subfigure[]{\includegraphics[width=.49\textwidth]{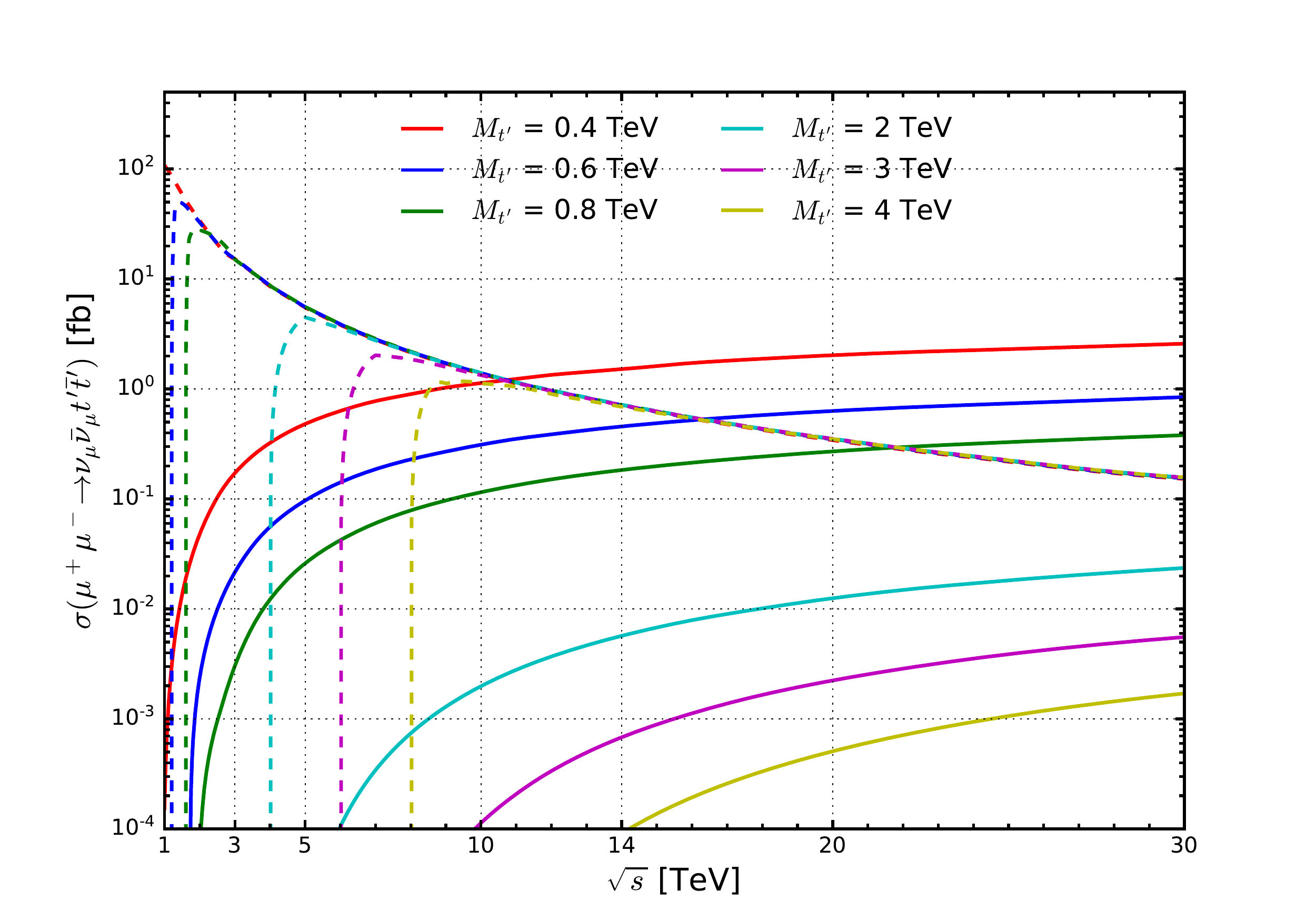}\label{fig:vbfschsttp_tptp}}
\subfigure[]{\includegraphics[width=.49\textwidth]{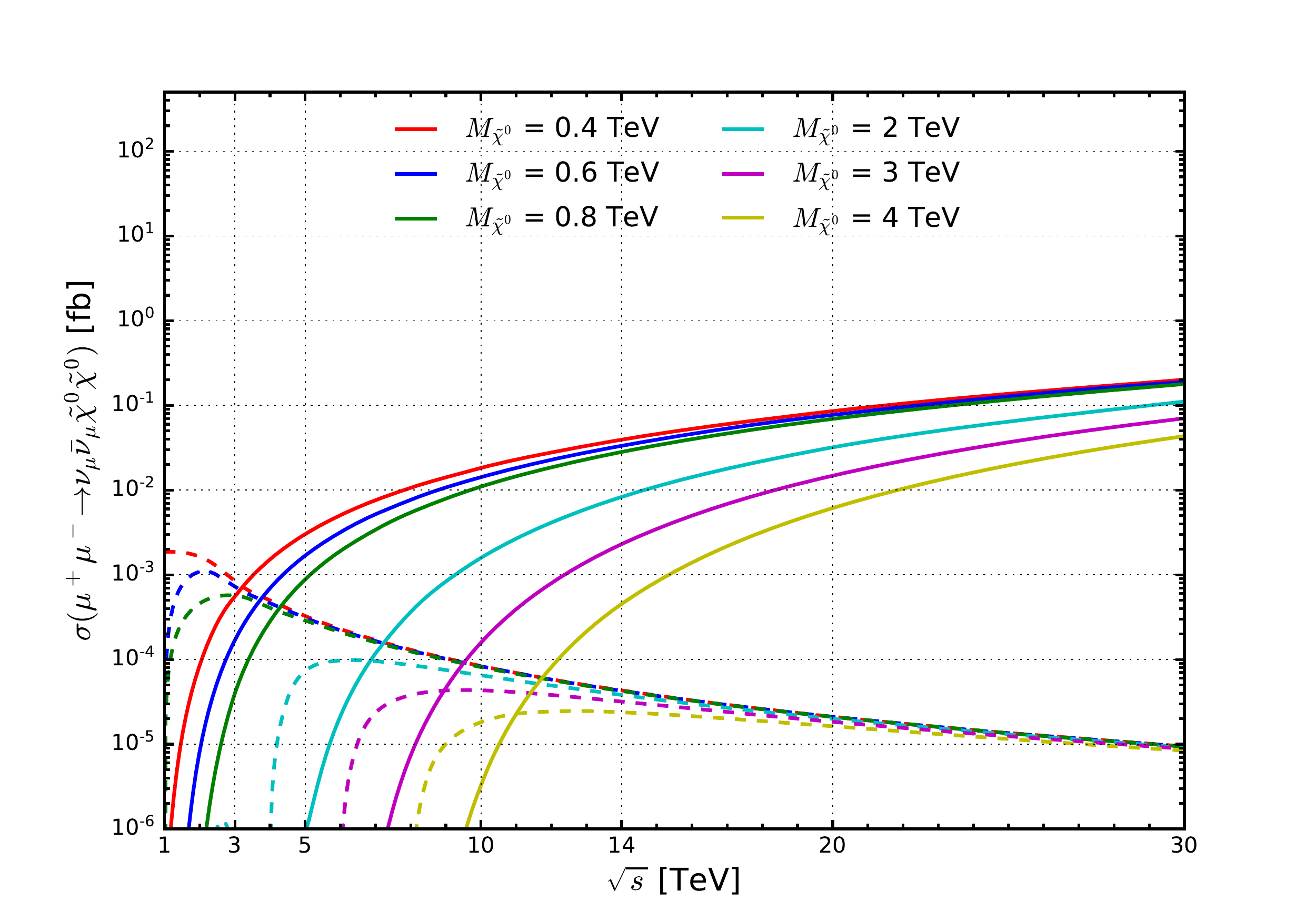}\label{fig:vbfschneucha_chi0}}
\subfigure[]{\includegraphics[width=.49\textwidth]{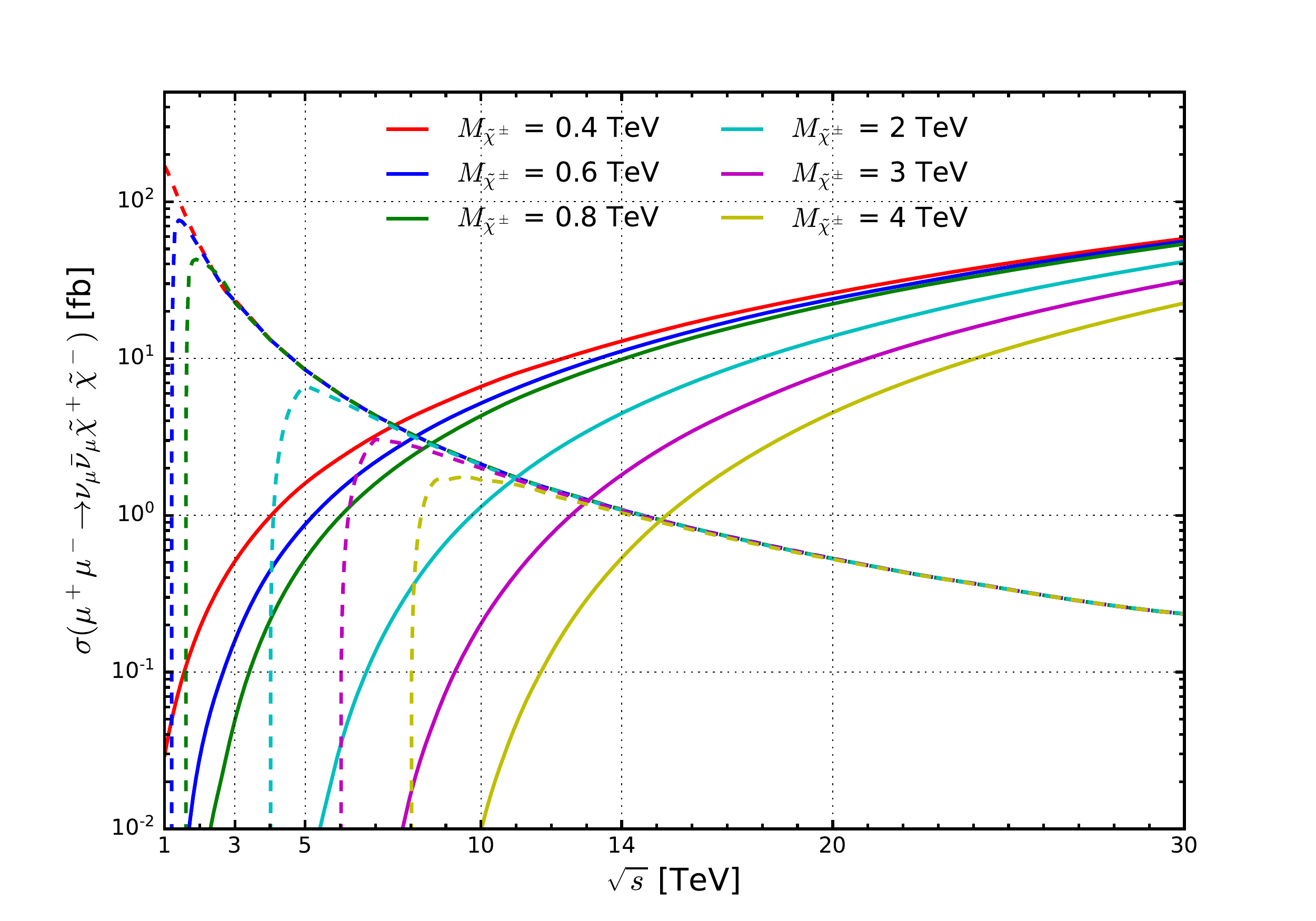}\label{fig:vbfschneucha_chip}}
\caption{
For representative input parameters and  as a function of muon collider energy [TeV],
the cross section [fb] via VBF (solid lines) and  $s$-channel annihilation (dashed lines) for:
(a) $S Z$ associated production in a singlet-scalar extension of the SM (section~\ref{sec:bsm_scalar});
(b) $H_2 Z$ associated production in the 2HDM (section~\ref{sec:bsm_2hdm});
(c) $\tilde{t}\overline{\tilde{t}}$ pair production in the MSSM (section~\ref{sec:bsm_mssm});
(d) $t'\overline{t'}$ pair production in a vector-like quark scenario (section~\ref{sec:bsm_vlq});
(e) $\tilde{\chi}^0\tilde{\chi}^0$ pair production in the MSSM;
and (f) $\tilde{\chi}^+\tilde{\chi}^-$ pair production in the MSSM.
}
\label{fig:bsm_vbf}
\end{figure}

As in the SM case studies of section~\ref{sec:sm}, there is not a definite energy beyond which $s$-channel, $\mu^+\mu^-$ annihilations are categorically subdominant. 
The situation is more nuanced. 
For example: as in our SM cases, the more final-state particles involved, the larger the $\sqrt{s}$ needed for $\sigma^{\rm VBF}$ to exceed $\sigma^{s-ch.}$. 
For the resonant production of BSM states, there is of course another important parameter that plays a role: the mass scale of new, final state particle(s). 
New mass scales complicates the na\"ive scaling for VBS in two ways.
First is the aforementioned propensity for collinear emission of transverse gauge bosons, which, more precisely, grows with the invariant mass of the VBF system.
Second is the possible enhancement of ``soft'' EW boson emissions at small momentum fractions. 
Third is the role of matrix elements featuring large longitudinal gauge boson couplings that nevertheless possess a relatively suppressed $V_0 V_0$ parton luminosity
(see section~\ref{sec:ppvsmuon_vbf}).

To explore how the mass scale of new particles impacts the threshold at which $\sigma^{\rm VBF}$ surpasses $\sigma^{s-ch.}$,
and working in the context of the BSM scenarios of section~\ref{sec:bsm},
we compare in figure~\ref{fig:bsm_vbf}
a variety of VBF and analogous $s$-channel, annihilation processes in multi-TeV $\mpmm$ collisions.
Assuming representative input parameters and as a function of muon collider energy [TeV],
we show the VBF (solid lines) and  $s$-channel (dashed lines) cross sections for:
\ref{fig:vbfschh2s_sing} $S Z$ associated production in a singlet-scalar extension of the SM (see section~\ref{sec:bsm_scalar});
\ref{fig:vbfschh2s_2hdm} $H_2 Z$ associated production in the 2HDM (see section~\ref{sec:bsm_2hdm});
\ref{fig:vbfschsttp_stop} $\tilde{t}\overline{\tilde{t}}$, 
\ref{fig:vbfschneucha_chi0} $\tilde{\chi}^0\tilde{\chi}^0$, 
and \ref{fig:vbfschneucha_chip} $\tilde{\chi}^+\tilde{\chi}^-$ pair production in the MSSM (see section~\ref{sec:bsm_mssm});
as well as  \ref{fig:vbfschsttp_tptp} $t'\overline{t'}$ pair production in a vector-like quark scenario (see section~\ref{sec:bsm_vlq}).
Estimated collider energies $\sqrt{s}$ at which the VBF rates surpass the $s$-channel rates are summarized in Table~\ref{tab:bsm_vbf}.

From this exercise we observe several trends.
We start by noting that the VBF production rates supersede $s$-channel rates at relatively lower collider energies 
for $SZ$, $H_2 Z$,  $\tilde{t}\overline{\tilde{t}}$, and $\tilde{\chi}^0\tilde{\chi}^0$ production than for $t'\overline{t'}$ and $\tilde{\chi}^+\tilde{\chi}^-$ pair production.
In particular, for $S Z$ and $H_2 Z$,  we report that $\sigma^{\rm VBF}$ becomes larger than $\sigma^{s-ch.}$ at around $\sqrt{s}\sim2-3\TeV$ for $M_S,~M_{H_2}=0.4-0.8\TeV$.
For heavier masses of $M_S,~M_{H_2}=2-4\TeV$, the transition  energies both span $\sqrt{s}\sim4-5.5\TeV$.
The same mass dependence can be found for $\tilde{t}\overline{\tilde{t}}$ and $\tilde{\chi}^0\tilde{\chi}^0$ production.
For the same ranges of lighter and heavier masses, the VBF cross sections become prevalent at $\sqrt{s}\sim3-4\TeV$ and $\sqrt{s}\sim7-13\TeV$.
The two sets of processes can further be linked by noting that the $M_S,~M_{H_2}=0.8~(2.0)~[4.0]\TeV$ benchmark masses  probe approximately  the same scales 
as the $M_{\tilde{t}},~M_{\tilde{\chi}^0}=0.4~(0.8)~[2.0]\TeV$ benchmarks, with reasonable consistency.
This trend suggests  some universal-like scaling behavior.

\begin{table}[!t]
\begin{center}
\resizebox{\textwidth}{!}{
\begin{tabular}{c | c c c c c c | c}
mass $(M_X)$ [TeV]	& $S Z$ (Singlet) & $H_2 Z$ (2HDM) & $t'\overline{t'}$ (VLQ) & $\tilde{t}\overline{\tilde{t}}$ (MSSM) & $\tilde{\chi}^0{\tilde{\chi}^0}$ (MSSM) & $\tilde{\chi}^+\tilde{\chi}^-$ (MSSM)  
& Scaling (Eq.~\ref{eq:bsm_vbf_scaling})\\	
\hline\hline
400	GeV &	2.1	TeV &	2.1	TeV &	11	TeV &	2.9	TeV &	3.2	TeV &	7.5	TeV 	&	1.0 (1.7) TeV\\
600	GeV &	2.5	TeV &	2.5	TeV &	16	TeV &	3.8	TeV &	3.8	TeV &	8.1	TeV 	&	1.3 (2.4) TeV \\
800	GeV &	2.8	TeV &	2.8	TeV &	22	TeV &	4.3	TeV &	4.3	TeV &	8.5	TeV 	&	1.7 (3.1) TeV \\
2.0	TeV &	4.0	TeV &	4.0	TeV &	>30	TeV &	7.8	TeV &	6.9	TeV &	11	TeV 	&	3.7 (6.8) TeV \\
3.0	TeV &	4.8	TeV &	4.8	TeV &	>30	TeV &	10	TeV &	9.0	TeV &	13	TeV 	&	5.3 (9.8) TeV \\
4.0	TeV &	5.5	TeV &	5.5	TeV &	>30	TeV &	13	TeV &	11	TeV &	15	TeV 	&	6.8 (13) TeV \\
\hline
\end{tabular}
} 
\end{center}		
\caption{
For representative processes and inputs, the required muon collider energy $\sqrt{s}$ [TeV]
at which the VBF production cross section surpasses the $s$-channel, annihilation cross section,
as shown in figure~\ref{fig:bsm_vbf}.
Also shown are the cross over energies as estimated from the scaling relationship in equation~(\ref{eq:bsm_vbf_scaling})
assuming a mass scale $M_X~(2M_X)$.
 } \label{tab:bsm_vbf}
\end{table}

For pair production of $t'\overline{t'}$ and $\tilde{\chi}^+\tilde{\chi}^-$, we find that the VBF channels become more important than $s$-channel production at much higher collider energies than the
previously discussed processes. 
More specifically, for $\tilde{\chi}^+\tilde{\chi}^-$, we find that collider energies must exceed $\sqrt{s}\sim7.5-8.5~(11-15)$ TeV for lighter~(heavier) mass scales.
For  $t'\overline{t'}$, the outlook is even worse. We find that VBF production only becomes important for $\sqrt{s}\sim11-22\TeV$ for relatively light masses of $M_{t'}=0.4-0.8\TeV$,
whereas for heavier masses of $M_{t'}=2-4\TeV$, one requires collider energies that exceed $\sqrt{s}=30\TeV$.

We attribute the qualitative differences between these two processes and the previous four processes to differences between subprocesses in the $s$-channel and VBF mechanisms.
In the first four cases, both $s$-channel and VBF proceed largely through the same EW gauge interactions.
In the latter two cases, the $s$-channel and VBF channels differ by additional $t$-channel exchanges that are governed not by gauge couplings but by mixing factors and Yukawa couplings.
The crossover, then, exhibits a stronger model dependency when VBF and $s$-channel diagrams adhere to different dynamics or interaction strengths.

As already stated, for the $S Z$, $H_2 Z$,  $\tilde{t}\overline{\tilde{t}}$, and $\tilde{\chi}^0\tilde{\chi}^0$ channels, we observe a suggestive, universal-like
behavior at which the VBF cross sections surpass their $s$-channel counter parts for a given final-state mass scale $M_X$.
This behavior can be roughly estimated by noting that the kinematic scaling for
$\mpmm \to X$, $s$-channel cross sections are of the form
\begin{equation}
\sigma^{s-ch.} \sim \frac{(s-M_X^2)}{(s-M_V^2)^2} \sim \frac{(s-M_X^2)}{s^2}.
\label{eq:bsm_sch_scaling}
\end{equation}
The denominator takes its form from the propagator of some intermediate state of mass $M_V\ll \sqrt{s}$ (it make no difference which state),
and the numerator from momentum conservation, which requires the cross section to vanish when the collider energy $\sqrt{s}$ dips to the mass threshold $M_X$ of the final-state.
Likewise, the differential rate for VBF processes that (importantly) proceed through the same interactions as $s$-channel process scale as
\begin{equation}
\frac{d\sigma^{\rm VBF}}{dz_1 dz_2} \sim f_V(z_1)f_{V'}(z_2)  \frac{(z_1 z_2 s-M_X^2)}{(z_1 z_2 s-M_V^2)^2}
\sim
   f_V(z_1)f_{V'}(z_2)  \frac{(z_1 z_2 s-M_X^2) \sigma^{s-ch.}}{(z_1 z_2)^2~(s-M_X^2)}. 
\label{eq:bsm_vbf_dSigma}
\end{equation}
Here we use the Effective $W$ Approximation (see section~\ref{sec:ppvsmuon_vbf}) to model the $VV'\to X$ hard process, which is mediated by EW bosons $VV'$ carrying energy fractions $z_1, z_2$.
(We make implicit a summation over all $VV'$ permutations that contribute to $VV'\to X$.)
In the final step, we assume that the invariant mass of the $VV'$-system is large, i.e., $M_{VV'}=z_1 z_2 s\gg M_V$, and express the $VV'\to X$ scaling in terms of equation~\ref{eq:bsm_sch_scaling}.
Now, as seen in equation~\ref{eq:ewa_VT}, the EWA PDFs contribute largest at small momentum fractions $(z_i\ll1)$, i.e., the limit where gauge radiation goes soft.
Moreover, as shown in figure~\ref{fig:p_vs_muon_VBF}, the $VV'$ luminosity is dominated by transverse polarizations.
Hence, in the small-$z_i$ limit (and setting the factorization scale $\mu_f=\sqrt{s}$),
the leading contribution to  equation~\ref{eq:bsm_vbf_dSigma} scales as
\begin{equation}
\frac{d\sigma^{\rm VBF}}{dz_1 dz_2} \sim \mathcal{S} \times \frac{g^2_W}{4\pi z_1}\log\frac{s}{M_V^2} \times \frac{g_W^2}{4\pi z_2} \log\frac{s}{M_{V'}^2} \times
 \frac{(z_1 z_2 s-M_X^2)}{(z_1 z_2)^2~(s-M_X^2)}\sigma^{s-ch.}. 
\end{equation}
Here we introduce explicitly a multiplicity factor $ \mathcal{S}=4$ to account for (sum) the four transverse polarization permutations that contribute to $V_TV'_T\to X$ production.

If we make the strong assumption that the  $VV'$-system's mass is also large in comparison to $M_X$, 
then the VBF scaling, in terms of the $s$-channel scaling, simplifies to
\begin{equation}
\frac{\sigma^{\rm VBF}}{\sigma^{s-ch.}}
 \sim  \mathcal{S} \left(\frac{g_W^2}{4\pi}\right)^2  \log^2\frac{s}{M_V^2}  \int \frac{dz_1 dz_2}{(z_1 z_2)^2} =  
\mathcal{S} \left(\frac{g_W^2}{4\pi}\right)^2 \log^2\frac{s}{M_V^2}  \int_{\tau_0}^1 d\tau \int_\tau^1 \frac{dz}{z} \frac{1}{\tau^2},
\end{equation}
where $\tau=z_1 z_2 = M_{VV'}^2/s$ is the dimensionless scale at which $VV'\to X$ proceeds,
and $\tau_0 = \min(\tau) = M_X^2/s$ is the smallest $\tau$ at which the hard process can kinematically occur.
In the first step, we group collinear logs under the stipulation that the $V-V'$ mass difference is negligible.
In the second, we made a change of variable to express the momentum integrals in terms of traditional collider variables.
After integrating and in terms of the $s$-channel scaling,
 the VBF dependence on collider energy for  $s\gg M_X^2$ scales as
\begin{eqnarray}
\sigma^{\rm VBF} 
&\sim& \sigma^{s-ch.}  \times \mathcal{S} \left(\frac{g_W^2}{4\pi}\right)^2   \log^2\frac{s}{M_V^2} \times \left[\frac{1}{\tau}-\frac{1}{\tau}\log\frac{1}{\tau}\right]_{\tau_0}^{1} 
\\
&\sim&  \sigma^{s-ch.}  \times \mathcal{S} \left(\frac{g_W^2}{4\pi}\right)^2    \left(\frac{s}{M_X^2}\right)  \log^2\frac{s}{M_V^2}\log\frac{s}{M_X^2}.
 \label{eq:bsm_vbf_soln}
\end{eqnarray}
We observe that the scaling behavior for VBF processes exhibits a double collinear logarithmic dependence on $s$, 
which stems from two collinear, EW PDFs, but remarkably only a single soft logarithm.
The double soft logarithm does not arise as the $VV'\to X$ hard process is power-suppressed by a relative factor of $1/(z_1z_2s) = 1/(\tau s)$.
This in turn manifests as a power-law factor that grows  linearly with $(s/M_X^2)$.

Altogether, this enables us to roughly estimate the collider energy $\sqrt{s}$ at which $\sigma^{\rm VBF}$ surpasses $\sigma^{s-ch.}$ for a given final-state mass $M_X$.
Essentially, one must solve for when
\begin{equation}
\frac{\sigma^{\rm VBF} }{\sigma^{s-ch.}} \sim 
\mathcal{S} \left(\frac{g_W^2}{4\pi}\right)^2   \left(\frac{s}{M_X^2}\right) \log^2\frac{s}{M_V^2}\log\frac{s}{M_X^2} > 1.
\label{eq:bsm_vbf_scaling}
\end{equation}
While the result is transcendental, the solution can easily be extracted numerically\footnote{Explicitly, we use the \texttt{Mathematica} function \texttt{NSolve} and report very quick runtime on a personal laptop.} for the representative $M_X$,
which we report in the rightmost  column of Table~\ref{tab:bsm_vbf} assuming a mass scale $M_X~(2M_X)$.
For sub-TeV masses, the scaling behavior systematically underestimates the true crossover by roughly a factor of two.
This is unsurprising as equation~\ref{eq:bsm_vbf_scaling} assumes a large hierarchy between relevant scales.
For TeV masses and above, however, we find good agreement between equation~\ref{eq:bsm_vbf_scaling} and explicit computation from Monte Carlo computations.
We report differences ranging from the percent level to the 20\% level.

\section{Conclusions}\label{sec:conclusions}

The next generation of particle accelerators needed to explore the energy frontier will offer tremendous challenges. 
Among these is a muon collider running at energies up to  several TeVs and luminosities in the tens of inverse attobarns, 
a dream machine both from the technology and physics points of view. 
Overcoming the challenges posed by producing, storing, and colliding high-intensity beams of high-energy muons will take years of further research and development.
Exploring the physics potential of such machines, on the other hand, is a relatively easy task that can be undertaken on a short time scale. 

In this paper, we have moved a small step forward in the latter direction by considering electroweak vector boson fusion/scattering  (VBF) processes at a future multi-TeV lepton collider in a rather systematic way. 
Our study is motivated by the simple observation that while $s$-channel production rates decrease with increasing collider energy as $1/s$, 
VBF rates grow as a power of $\log s$, 
and therefore, for any final state, VBF is consigned to eventually emerge as the leading production mechanism. 

In this context, we have investigated and show in section~\ref{sec:ppvsmuon} that, compared to hadron colliders, VBF is a much more relevant production mechanism at a high-energy lepton collider.  
We  continue  in section~\ref{sec:sm} and present
for a rather large set of SM final states involving EW vector bosons, Higgs bosons, and top quarks
the corresponding VBF cross sections and at what collider energy they surpass $s$-channel production modes.
We find  that VBF becomes the dominant production mechanism at relatively low collider energies,
starting  at just a few TeV  for low final-state  multiplicities and increases for higher multiplicities.

In order to further illustrate what could be attainable in terms of  new physics reach, we then moved in two directions, focusing mostly on luminosity scenarios envisaged for a muon collider.  
First, in section~\ref{sec:eft},  we considered prospects for precision measurements  of the Higgs's self-couplings and the top quark's EW couplings, 
and interpreted sensitivity in terms of Wilson coefficients within the SMEFT framework.
Second, in section~\ref{sec:bsm}, we explored a variety of simplified extensions of the SM and how large VBF luminosities can maximize the direct search for new physics.
In particular we find evidence that in several instances the reach of a multi-TeV muon collider is comparable or better than that attainable at a  100 TeV proton-proton collider.
A detailed comparison of VBF's utility over $s$-channel annihilations in BSM searches was then  summarized in  section~\ref{sec:bsm_vbf}.

The results presented in this work are meant to provide a first glimpse of what could be achieved at a multi-TeV muon collider in VBF channels,
and certainly motivate further and more refined investigations.
We close by stressing that while we focus on the specific prospects of a muon collider, our conclusions hold equally for other lepton colliders.

\section*{Acknowledgements}
FM, LM, and XZ would like to thank Mauro Chiesa, Roberto Franceschini, Barbara Mele and Fulvio Piccinini for many discussions on the physics of muon colliders. RR thanks Carlos Alishaan for helpful discussions.

This work has received funding from the European Union's Horizon 2020 research and innovation programme as part of the Marie Skłodowska-Curie Innovative Training Network MCnetITN3 (grant agreement no. 722104),
FNRS “Excellence of Science” EOS be.h Project No. 30820817.
The work of AC is supported by INFN research grant n. 20286/2018.
RR is supported under the UCLouvain fund “MOVE-IN Louvain” and  acknowledge the contribution of the VBSCan COST Action CA16108.

Computational resources have been provided by the supercomputing facilities of the Universit\'e catholique de Louvain (CISM/UCL) 
and the Consortium des \'Equipements de Calcul Intensif en F\'ed\'eration Wallonie Bruxelles (C\'ECI) funded by the Fond de la Recherche Scientifique de Belgique 
(F.R.S.-FNRS) under convention 2.5020.11 and by the Walloon Region.

\bibliography{madMuon_refs}

\providecommand{\href}[2]{#2}\begingroup\raggedright\begin{thebibliography}{100}

\bibitem{Strategy:2019vxc}
R.~K. Ellis and others ({European Strategy for Particle Physics Preparatory
  Group}), \textit{{Physics Briefing Book}: {Input for the European Strategy
  for Particle Physics Update 2020}},
  \href{https://arxiv.org/abs/1910.11775}{{\ttfamily 1910.11775}}.

\bibitem{EuropeanStrategyGroup:2020pow}
{\scshape European Strategy Group} collaboration, \textit{{2020 Update of the
  European Strategy for Particle Physics}}.
\newblock CERN Council, Geneva, 2020,
  \href{https://doi.org/10.17181/ESU2020}{10.17181/ESU2020}.

\bibitem{Gschwendtner:2015rni}
{\scshape AWAKE} collaboration, E.~Gschwendtner et~al., \textit{{AWAKE, The
  Advanced Proton Driven Plasma Wakefield Acceleration Experiment at CERN}},
  \href{https://doi.org/10.1016/j.nima.2016.02.026}{\textit{Nucl. Instrum.
  Meth.} {\bfseries A829} (2016) 76--82},
  [\href{https://arxiv.org/abs/1512.05498}{{\ttfamily 1512.05498}}].

\bibitem{Palmer:1996gs}
R.~Palmer et~al., \textit{{Muon collider design}},
  \href{https://doi.org/10.1016/0920-5632(96)00417-3}{\textit{Nucl. Phys. Proc.
  Suppl.} {\bfseries 51A} (1996) 61--84},
  [\href{https://arxiv.org/abs/acc-phys/9604001}{{\ttfamily
  acc-phys/9604001}}].

\bibitem{Ankenbrandt:1999cta}
C.~M. Ankenbrandt et~al., \textit{{Status of muon collider research and
  development and future plans}},
  \href{https://doi.org/10.1103/PhysRevSTAB.2.081001}{\textit{Phys. Rev. ST
  Accel. Beams} {\bfseries 2} (1999) 081001},
  [\href{https://arxiv.org/abs/physics/9901022}{{\ttfamily physics/9901022}}].

\bibitem{Delahaye:2019omf}
J.~P. Delahaye, M.~Diemoz, K.~Long, B.~Mansouli{\'e}, N.~Pastrone, L.~Rivkin
  et~al., \textit{{Muon Colliders}},
  \href{https://arxiv.org/abs/1901.06150}{{\ttfamily 1901.06150}}.

\bibitem{Palmer:2014nza}
R.~B. Palmer, \textit{{Muon Colliders}},
  \href{https://doi.org/10.1142/S1793626814300072}{\textit{Rev. Accel. Sci.
  Tech.} {\bfseries 7} (2014) 137--159}.

\bibitem{Antonelli:2013mmk}
M.~Antonelli and P.~Raimondi, \textit{{Snowmass Report: Ideas for Muon
  Production from Positron Beam Interaction on a Plasma Target}},  in
  \textit{{Proceedings, 2013 Community Summer Study on the Future of U.S.
  Particle Physics: Snowmass on the Mississippi (CSS2013): Minneapolis, MN,
  USA, July 29-August 6, 2013}}, 2013,
  \href{http://www.lnf.infn.it/sis/preprint/detail-new.php?id=5331}{http://www.lnf.infn.it/sis/preprint/detail-new.php?id=5331}.

\bibitem{Antonelli:2015nla}
M.~Antonelli, M.~Boscolo, R.~Di~Nardo and P.~Raimondi, \textit{{Novel proposal
  for a low emittance muon beam using positron beam on target}},
  \href{https://doi.org/10.1016/j.nima.2015.10.097}{\textit{Nucl. Instrum.
  Meth.} {\bfseries A807} (2016) 101--107},
  [\href{https://arxiv.org/abs/1509.04454}{{\ttfamily 1509.04454}}].

\bibitem{Bartosik:2019dzq}
N.~Bartosik et~al., \textit{{Preliminary Report on the Study of Beam-Induced
  Background Effects at a Muon Collider}},
  \href{https://arxiv.org/abs/1905.03725}{{\ttfamily 1905.03725}}.

\bibitem{Barger:1995hr}
V.~D. Barger, M.~S. Berger, J.~F. Gunion and T.~Han, \textit{{s channel Higgs
  boson production at a muon muon collider}},
  \href{https://doi.org/10.1103/PhysRevLett.75.1462}{\textit{Phys. Rev. Lett.}
  {\bfseries 75} (1995) 1462--1465},
  [\href{https://arxiv.org/abs/hep-ph/9504330}{{\ttfamily hep-ph/9504330}}].

\bibitem{Han:2012rb}
T.~Han and Z.~Liu, \textit{{Potential precision of a direct measurement of the
  Higgs boson total width at a muon collider}},
  \href{https://doi.org/10.1103/PhysRevD.87.033007}{\textit{Phys. Rev. D}
  {\bfseries 87} (2013) 033007},
  [\href{https://arxiv.org/abs/1210.7803}{{\ttfamily 1210.7803}}].

\bibitem{Chakrabarty:2014pja}
N.~Chakrabarty, T.~Han, Z.~Liu and B.~Mukhopadhyaya, \textit{{Radiative Return
  for Heavy Higgs Boson at a Muon Collider}},
  \href{https://doi.org/10.1103/PhysRevD.91.015008}{\textit{Phys. Rev. D}
  {\bfseries 91} (2015) 015008},
  [\href{https://arxiv.org/abs/1408.5912}{{\ttfamily 1408.5912}}].

\bibitem{Greco:2016izi}
M.~Greco, T.~Han and Z.~Liu, \textit{{ISR effects for resonant Higgs production
  at future lepton colliders}}, vol.~763, pp.~409--415.
\newblock 12, 2016.
\newblock \href{https://arxiv.org/abs/1607.03210}{{\ttfamily 1607.03210}}.
\newblock 10.1016/j.physletb.2016.10.078.

\bibitem{Buttazzo:2018qqp}
D.~Buttazzo, D.~Redigolo, F.~Sala and A.~Tesi, \textit{{Fusing Vectors into
  Scalars at High Energy Lepton Colliders}},
  \href{https://doi.org/10.1007/JHEP11(2018)144}{\textit{JHEP} {\bfseries 11}
  (2018) 144}, [\href{https://arxiv.org/abs/1807.04743}{{\ttfamily
  1807.04743}}].

\bibitem{Ruhdorfer:2019utl}
M.~Ruhdorfer, E.~Salvioni and A.~Weiler, \textit{{A Global View of the
  Off-Shell Higgs Portal}},
  \href{https://doi.org/10.21468/SciPostPhys.8.2.027}{\textit{SciPost Phys.}
  {\bfseries 8} (2020) 027},
  [\href{https://arxiv.org/abs/1910.04170}{{\ttfamily 1910.04170}}].

\bibitem{Dawson:1984ta}
S.~Dawson and J.~L. Rosner, \textit{{Capabilities of $e^+ e^-$ Collisions for
  Producing Very Heavy Higgs Bosons}},
  \href{https://doi.org/10.1016/0370-2693(84)90746-9}{\textit{Phys. Lett. B}
  {\bfseries 148} (1984) 497--501}.

\bibitem{Hikasa:1985ee}
K.-i. Hikasa, \textit{{Heavy Higgs Production in $e^+ e^-$ and $e^- e^-$
  Collisions}},
  \href{https://doi.org/10.1016/0370-2693(85)90346-6}{\textit{Phys. Lett. B}
  {\bfseries 164} (1985) 385}. [Erratum: Phys.Lett.B 195, 623 (1987)].

\bibitem{Altarelli:1987ue}
G.~Altarelli, B.~Mele and F.~Pitolli, \textit{{Heavy Higgs Production at Future
  Colliders}},
  \href{https://doi.org/10.1016/0550-3213(87)90103-9}{\textit{Nucl. Phys.}
  {\bfseries B287} (1987) 205--224}.

\bibitem{Kilian:1995tr}
W.~Kilian, M.~Kramer and P.~Zerwas, \textit{{Higgsstrahlung and W W fusion in
  e+ e- collisions}},
  \href{https://doi.org/10.1016/0370-2693(96)00100-1}{\textit{Phys. Lett. B}
  {\bfseries 373} (1996) 135--140},
  [\href{https://arxiv.org/abs/hep-ph/9512355}{{\ttfamily hep-ph/9512355}}].

\bibitem{Gunion:1998jc}
J.~Gunion, T.~Han and R.~Sobey, \textit{{Measuring the coupling of a Higgs
  boson to Z Z at linear colliders}},
  \href{https://doi.org/10.1016/S0370-2693(98)00450-X}{\textit{Phys. Lett. B}
  {\bfseries 429} (1998) 79--86},
  [\href{https://arxiv.org/abs/hep-ph/9801317}{{\ttfamily hep-ph/9801317}}].

\bibitem{Grzadkowski:2010es}
B.~Grzadkowski, M.~Iskrzynski, M.~Misiak and J.~Rosiek, \textit{{Dimension-Six
  Terms in the Standard Model Lagrangian}},
  \href{https://doi.org/10.1007/JHEP10(2010)085}{\textit{JHEP} {\bfseries 10}
  (2010) 085}, [\href{https://arxiv.org/abs/1008.4884}{{\ttfamily 1008.4884}}].

\bibitem{Aebischer:2017ugx}
J.~Aebischer et~al., \textit{{WCxf: an exchange format for Wilson coefficients
  beyond the Standard Model}},
  \href{https://doi.org/10.1016/j.cpc.2018.05.022}{\textit{Comput. Phys.
  Commun.} {\bfseries 232} (2018) 71--83},
  [\href{https://arxiv.org/abs/1712.05298}{{\ttfamily 1712.05298}}].

\bibitem{Brivio:2017btx}
I.~Brivio, Y.~Jiang and M.~Trott, \textit{{The SMEFTsim package, theory and
  tools}}, \href{https://doi.org/10.1007/JHEP12(2017)070}{\textit{JHEP}
  {\bfseries 12} (2017) 070},
  [\href{https://arxiv.org/abs/1709.06492}{{\ttfamily 1709.06492}}].

\bibitem{Chiesa:2020awd}
M.~Chiesa, F.~Maltoni, L.~Mantani, B.~Mele, F.~Piccinini and X.~Zhao,
  \textit{{Measuring the quartic Higgs self-coupling at a multi-TeV muon
  collider}},  \href{https://arxiv.org/abs/2003.13628}{{\ttfamily 2003.13628}}.

\bibitem{Alwall:2014hca}
J.~Alwall, R.~Frederix, S.~Frixione, V.~Hirschi, F.~Maltoni, O.~Mattelaer
  et~al., \textit{{The automated computation of tree-level and next-to-leading
  order differential cross sections, and their matching to parton shower
  simulations}}, \href{https://doi.org/10.1007/JHEP07(2014)079}{\textit{JHEP}
  {\bfseries 07} (2014) 079},
  [\href{https://arxiv.org/abs/1405.0301}{{\ttfamily 1405.0301}}].

\bibitem{Ball:2014uwa}
{\scshape NNPDF} collaboration, R.~D. Ball et~al., \textit{{Parton
  distributions for the LHC Run II}},
  \href{https://doi.org/10.1007/JHEP04(2015)040}{\textit{JHEP} {\bfseries 04}
  (2015) 040}, [\href{https://arxiv.org/abs/1410.8849}{{\ttfamily 1410.8849}}].

\bibitem{Buckley:2014ana}
A.~Buckley, J.~Ferrando, S.~Lloyd, K.~Nordstr{\"o}m, B.~Page, M.~R{\"u}fenacht
  et~al., \textit{{LHAPDF6: parton density access in the LHC precision era}},
  \href{https://doi.org/10.1140/epjc/s10052-015-3318-8}{\textit{Eur. Phys. J.}
  {\bfseries C75} (2015) 132},
  [\href{https://arxiv.org/abs/1412.7420}{{\ttfamily 1412.7420}}].

\bibitem{Kilian:2007gr}
W.~Kilian, T.~Ohl and J.~Reuter, \textit{{WHIZARD: Simulating Multi-Particle
  Processes at LHC and ILC}},
  \href{https://doi.org/10.1140/epjc/s10052-011-1742-y}{\textit{Eur. Phys. J.}
  {\bfseries C71} (2011) 1742},
  [\href{https://arxiv.org/abs/0708.4233}{{\ttfamily 0708.4233}}].

\bibitem{Denner:2017wsf}
A.~Denner, J.-N. Lang and S.~Uccirati, \textit{{Recola2: REcursive Computation
  of One-Loop Amplitudes 2}},
  \href{https://doi.org/10.1016/j.cpc.2017.11.013}{\textit{Comput. Phys.
  Commun.} {\bfseries 224} (2018) 346--361},
  [\href{https://arxiv.org/abs/1711.07388}{{\ttfamily 1711.07388}}].

\bibitem{Quigg:2009gg}
C.~Quigg, \textit{{LHC Physics Potential versus Energy}},
  \href{https://arxiv.org/abs/0908.3660}{{\ttfamily 0908.3660}}.

\bibitem{Dawson:1984gx}
S.~Dawson, \textit{{The Effective W Approximation}},
  \href{https://doi.org/10.1016/0550-3213(85)90038-0}{\textit{Nucl. Phys.}
  {\bfseries B249} (1985) 42--60}.

\bibitem{Kane:1984bb}
G.~L. Kane, W.~W. Repko and W.~B. Rolnick, \textit{{The Effective W+-, Z0
  Approximation for High-Energy Collisions}},
  \href{https://doi.org/10.1016/0370-2693(84)90105-9}{\textit{Phys. Lett.}
  {\bfseries 148B} (1984) 367--372}.

\bibitem{Collins:2011zzd}
J.~Collins, \textit{{Foundations of perturbative QCD}}, {\textit{Camb. Monogr.
  Part. Phys. Nucl. Phys. Cosmol.} {\bfseries 32} (2011) 1--624}.

\bibitem{Cahn:1984tx}
R.~N. Cahn, \textit{{Production of Heavy Higgs Bosons: Comparisons of Exact and
  Approximate Results}}, \href{https://doi.org/10.1016/0550-3213(85)90139-7,
  10.1016/0550-3213(85)90514-0}{\textit{Nucl. Phys.} {\bfseries B255} (1985)
  341}. [Erratum: Nucl. Phys.B262,744(1985)].

\bibitem{Willenbrock:1986cr}
S.~S.~D. Willenbrock and D.~A. Dicus, \textit{{Production of Heavy Quarks from
  W Gluon Fusion}},
  \href{https://doi.org/10.1103/PhysRevD.34.155}{\textit{Phys. Rev.} {\bfseries
  D34} (1986) 155}.

\bibitem{Dawson:1986tc}
S.~Dawson and S.~S.~D. Willenbrock, \textit{{Heavy Fermion Production in the
  Effective $W$ Approximation}},
  \href{https://doi.org/10.1016/0550-3213(87)90044-7}{\textit{Nucl. Phys.}
  {\bfseries B284} (1987) 449}.

\bibitem{Kunszt:1987tk}
Z.~Kunszt and D.~E. Soper, \textit{{On the Validity of the Effective $W$
  Approximation}},
  \href{https://doi.org/10.1016/0550-3213(88)90673-6}{\textit{Nucl. Phys.}
  {\bfseries B296} (1988) 253--289}.

\bibitem{Borel:2012by}
P.~Borel, R.~Franceschini, R.~Rattazzi and A.~Wulzer, \textit{{Probing the
  Scattering of Equivalent Electroweak Bosons}},
  \href{https://doi.org/10.1007/JHEP06(2012)122}{\textit{JHEP} {\bfseries 06}
  (2012) 122}, [\href{https://arxiv.org/abs/1202.1904}{{\ttfamily 1202.1904}}].

\bibitem{Wulzer:2013mza}
A.~Wulzer, \textit{{An Equivalent Gauge and the Equivalence Theorem}},
  \href{https://doi.org/10.1016/j.nuclphysb.2014.05.021}{\textit{Nucl. Phys.}
  {\bfseries B885} (2014) 97--126},
  [\href{https://arxiv.org/abs/1309.6055}{{\ttfamily 1309.6055}}].

\bibitem{Chen:2019dkx}
J.~Chen, \textit{{On the Feynman Rules of Massive Gauge Theory in Physical
  Gauges}},  \href{https://arxiv.org/abs/1902.06738}{{\ttfamily 1902.06738}}.

\bibitem{Chen:2016wkt}
J.~Chen, T.~Han and B.~Tweedie, \textit{{Electroweak Splitting Functions and
  High Energy Showering}},
  \href{https://doi.org/10.1007/JHEP11(2017)093}{\textit{JHEP} {\bfseries 11}
  (2017) 093}, [\href{https://arxiv.org/abs/1611.00788}{{\ttfamily
  1611.00788}}].

\bibitem{Chen:2017ekt}
J.~Chen, \textit{{Electroweak Splitting Functions and High Energy Showering}},
  Ph.D. thesis, U. Pittsburgh (main), 2017.

\bibitem{Cuomo:2019siu}
G.~Cuomo, L.~Vecchi and A.~Wulzer, \textit{{Goldstone Equivalence and High
  Energy Electroweak Physics}},
  \href{https://doi.org/10.21468/SciPostPhys.8.5.078}{\textit{SciPost Phys.}
  {\bfseries 8} (2020) 078},
  [\href{https://arxiv.org/abs/1911.12366}{{\ttfamily 1911.12366}}].

\bibitem{Denner:1999gp}
A.~Denner, S.~Dittmaier, M.~Roth and D.~Wackeroth, \textit{{Predictions for all
  processes e+ e- ---> 4 fermions + gamma}},
  \href{https://doi.org/10.1016/S0550-3213(99)00437-X}{\textit{Nucl. Phys.}
  {\bfseries B560} (1999) 33--65},
  [\href{https://arxiv.org/abs/hep-ph/9904472}{{\ttfamily hep-ph/9904472}}].

\bibitem{Denner:2005fg}
A.~Denner, S.~Dittmaier, M.~Roth and L.~H. Wieders, \textit{{Electroweak
  corrections to charged-current e+ e- ---> 4 fermion processes: Technical
  details and further results}},
  \href{https://doi.org/10.1016/j.nuclphysb.2011.09.001,
  10.1016/j.nuclphysb.2005.06.033}{\textit{Nucl. Phys.} {\bfseries B724} (2005)
  247--294}, [\href{https://arxiv.org/abs/hep-ph/0505042}{{\ttfamily
  hep-ph/0505042}}]. [Erratum: Nucl. Phys.B854,504(2012)].

\bibitem{Cai:2017mow}
Y.~Cai, T.~Han, T.~Li and R.~Ruiz, \textit{{Lepton Number Violation: Seesaw
  Models and Their Collider Tests}},
  \href{https://doi.org/10.3389/fphy.2018.00040}{\textit{Front.in Phys.}
  {\bfseries 6} (2018) 40}, [\href{https://arxiv.org/abs/1711.02180}{{\ttfamily
  1711.02180}}].

\bibitem{Kobach:2016ami}
A.~Kobach, \textit{{Baryon Number, Lepton Number, and Operator Dimension in the
  Standard Model}},
  \href{https://doi.org/10.1016/j.physletb.2016.05.050}{\textit{Phys. Lett.}
  {\bfseries B758} (2016) 455--457},
  [\href{https://arxiv.org/abs/1604.05726}{{\ttfamily 1604.05726}}].

\bibitem{Helset:2019eyc}
A.~Helset and A.~Kobach, \textit{{Baryon Number, Lepton Number, and Operator
  Dimension in the SMEFT with Flavor Symmetries}},
  \href{https://doi.org/10.1016/j.physletb.2019.135132}{\textit{Phys. Lett.}
  {\bfseries B800} (2020) 135132},
  [\href{https://arxiv.org/abs/1909.05853}{{\ttfamily 1909.05853}}].

\bibitem{Ellis:2018gqa}
J.~Ellis, C.~W. Murphy, V.~Sanz and T.~You, \textit{{Updated Global SMEFT Fit
  to Higgs, Diboson and Electroweak Data}},
  \href{https://doi.org/10.1007/JHEP06(2018)146}{\textit{JHEP} {\bfseries 06}
  (2018) 146}, [\href{https://arxiv.org/abs/1803.03252}{{\ttfamily
  1803.03252}}].

\bibitem{Hartland:2019bjb}
N.~P. Hartland, F.~Maltoni, E.~R. Nocera, J.~Rojo, E.~Slade, E.~Vryonidou
  et~al., \textit{{A Monte Carlo global analysis of the Standard Model
  Effective Field Theory: the top quark sector}},
  \href{https://doi.org/10.1007/JHEP04(2019)100}{\textit{JHEP} {\bfseries 04}
  (2019) 100}, [\href{https://arxiv.org/abs/1901.05965}{{\ttfamily
  1901.05965}}].

\bibitem{Buckley:2015lku}
A.~Buckley, C.~Englert, J.~Ferrando, D.~J. Miller, L.~Moore, M.~Russell et~al.,
  \textit{{Constraining top quark effective theory in the LHC Run II era}},
  \href{https://doi.org/10.1007/JHEP04(2016)015}{\textit{JHEP} {\bfseries 04}
  (2016) 015}, [\href{https://arxiv.org/abs/1512.03360}{{\ttfamily
  1512.03360}}].

\bibitem{Butter:2016cvz}
A.~Butter, O.~J.~P. {\'E}boli, J.~Gonzalez-Fraile, M.~C. Gonzalez-Garcia,
  T.~Plehn and M.~Rauch, \textit{{The Gauge-Higgs Legacy of the LHC Run I}},
  \href{https://doi.org/10.1007/JHEP07(2016)152}{\textit{JHEP} {\bfseries 07}
  (2016) 152}, [\href{https://arxiv.org/abs/1604.03105}{{\ttfamily
  1604.03105}}].

\bibitem{Chung:2012vg}
D.~J.~H. Chung, A.~J. Long and L.-T. Wang, \textit{{125 GeV Higgs boson and
  electroweak phase transition model classes}},
  \href{https://doi.org/10.1103/PhysRevD.87.023509}{\textit{Phys. Rev.}
  {\bfseries D87} (2013) 023509},
  [\href{https://arxiv.org/abs/1209.1819}{{\ttfamily 1209.1819}}].

\bibitem{Sirunyan:2017guj}
{\scshape CMS} collaboration, A.~M. Sirunyan et~al., \textit{{Search for
  resonant and nonresonant Higgs boson pair production in the $
  \mathrm{b}\overline{\mathrm{b}}\mathit{\ell \nu \ell \nu } $ final state in
  proton-proton collisions at $ \sqrt{s}=13 $ TeV}},
  \href{https://doi.org/10.1007/JHEP01(2018)054}{\textit{JHEP} {\bfseries 01}
  (2018) 054}, [\href{https://arxiv.org/abs/1708.04188}{{\ttfamily
  1708.04188}}].

\bibitem{CMS:2018rig}
{\scshape CMS} collaboration, C.~Collaboration, \textit{{Constraints on the
  Higgs boson self-coupling from ttH+tH, H to gamma gamma differential
  measurements at the HL-LHC}}, .

\bibitem{CMS:2018ccd}
{\scshape CMS} collaboration, C.~Collaboration, \textit{{Prospects for HH
  measurements at the HL-LHC}}, .

\bibitem{ATLAS:2018otd}
{\scshape ATLAS} collaboration, T.~A. collaboration, \textit{{Combination of
  searches for Higgs boson pairs in $pp$ collisions at 13 TeV with the ATLAS
  experiment.}}, .

\bibitem{Aaboud:2018zhh}
{\scshape ATLAS} collaboration, M.~Aaboud et~al., \textit{{Search for Higgs
  boson pair production in the $b\bar{b}WW^{*}$ decay mode at $\sqrt{s}=13$ TeV
  with the ATLAS detector}},
  \href{https://doi.org/10.1007/JHEP04(2019)092}{\textit{JHEP} {\bfseries 04}
  (2019) 092}, [\href{https://arxiv.org/abs/1811.04671}{{\ttfamily
  1811.04671}}].

\bibitem{Aad:2019uzh}
{\scshape ATLAS} collaboration, G.~Aad et~al., \textit{{Combination of searches
  for Higgs boson pairs in $pp$ collisions at $\sqrt{s} = $13 TeV with the
  ATLAS detector}},
  \href{https://doi.org/10.1016/j.physletb.2019.135103}{\textit{Phys. Lett.}
  {\bfseries B800} (2020) 135103},
  [\href{https://arxiv.org/abs/1906.02025}{{\ttfamily 1906.02025}}].

\bibitem{Aad:2019yxi}
{\scshape ATLAS} collaboration, G.~Aad et~al., \textit{{Search for non-resonant
  Higgs boson pair production in the $bb\ell\nu\ell\nu$ final state with the
  ATLAS detector in $pp$ collisions at $\sqrt{s} = 13$ TeV}},
  \href{https://doi.org/10.1016/j.physletb.2019.135145}{\textit{Phys. Lett.}
  {\bfseries B801} (2020) 135145},
  [\href{https://arxiv.org/abs/1908.06765}{{\ttfamily 1908.06765}}].

\bibitem{Sirunyan:2018iwt}
{\scshape CMS} collaboration, A.~M. Sirunyan et~al., \textit{{Search for Higgs
  boson pair production in the $\gamma\gamma\mathrm{b\overline{b}}$ final state
  in pp collisions at $\sqrt{s}=$ 13 TeV}},
  \href{https://doi.org/10.1016/j.physletb.2018.10.056}{\textit{Phys. Lett.}
  {\bfseries B788} (2019) 7--36},
  [\href{https://arxiv.org/abs/1806.00408}{{\ttfamily 1806.00408}}].

\bibitem{CMS:2018dvu}
{\scshape CMS} collaboration, C.~Collaboration, \textit{{Search for resonant
  double Higgs production with $bbZZ$ decays in the $b\bar{b} \ell\ell\nu\nu$
  final state}}, .

\bibitem{ATLAS:2019pbo}
{\scshape ATLAS} collaboration, T.~A. collaboration, \textit{{Constraints on
  the Higgs boson self-coupling from the combination of single-Higgs and
  double-Higgs production analyses performed with the ATLAS experiment}},
  (Geneva), CERN, CERN, 2019.

\bibitem{ATL-PHYS-PUB-2014-019}
T.~A. collaboration, \textit{{Prospects for measuring Higgs pair production in
  the channel $H(\rightarrow\gamma\gamma)H(\rightarrow b\overline{b}) $ using
  the ATLAS detector at the HL-LHC}}, .

\bibitem{ATL-PHYS-PUB-2017-001}
{\scshape ATLAS} collaboration, T.~A. collaboration, \textit{{Study of the
  double Higgs production channel $H(\rightarrow b\bar{b})H(\rightarrow
  \gamma\gamma)$ with the ATLAS experiment at the HL-LHC}},  2017.

\bibitem{Kim:2018cxf}
J.~H. Kim, K.~Kong, K.~T. Matchev and M.~Park, \textit{{Probing the Triple
  Higgs Self-Interaction at the Large Hadron Collider}},
  \href{https://doi.org/10.1103/PhysRevLett.122.091801}{\textit{Phys. Rev.
  Lett.} {\bfseries 122} (2019) 091801},
  [\href{https://arxiv.org/abs/1807.11498}{{\ttfamily 1807.11498}}].

\bibitem{Roloff:2018dqu}
{\scshape CLIC, CLICdp} collaboration, P.~Roloff, R.~Franceschini, U.~Schnoor
  and A.~Wulzer, \textit{{The Compact Linear e$^+$e$^-$ Collider (CLIC):
  Physics Potential}},  \href{https://arxiv.org/abs/1812.07986}{{\ttfamily
  1812.07986}}.

\bibitem{Vasquez:2019muw}
A.~Vasquez, C.~Degrande, A.~Tonero and R.~Rosenfeld, \textit{{New physics in
  double Higgs production at future e$^{+}$e$^{-}$ colliders}},
  \href{https://doi.org/10.1007/JHEP05(2019)020}{\textit{JHEP} {\bfseries 05}
  (2019) 020}, [\href{https://arxiv.org/abs/1901.05979}{{\ttfamily
  1901.05979}}].

\bibitem{Roloff:2019crr}
{\scshape CLICdp} collaboration, P.~Roloff, U.~Schnoor, R.~Simoniello and
  B.~Xu, \textit{{Double Higgs boson production and Higgs self-coupling
  extraction at CLIC}},  \href{https://arxiv.org/abs/1901.05897}{{\ttfamily
  1901.05897}}.

\bibitem{Liu:2018peg}
T.~Liu, K.-F. Lyu, J.~Ren and H.~X. Zhu, \textit{{Probing the quartic Higgs
  boson self-interaction}},
  \href{https://doi.org/10.1103/PhysRevD.98.093004}{\textit{Phys. Rev.}
  {\bfseries D98} (2018) 093004},
  [\href{https://arxiv.org/abs/1803.04359}{{\ttfamily 1803.04359}}].

\bibitem{Maltoni:2018ttu}
F.~Maltoni, D.~Pagani and X.~Zhao, \textit{{Constraining the Higgs
  self-couplings at e$^{+}$e$^{-}$ colliders}},
  \href{https://doi.org/10.1007/JHEP07(2018)087}{\textit{JHEP} {\bfseries 07}
  (2018) 087}, [\href{https://arxiv.org/abs/1802.07616}{{\ttfamily
  1802.07616}}].

\bibitem{deBlas:2019rxi}
J.~de~Blas et~al., \textit{{Higgs Boson Studies at Future Particle Colliders}},
  \href{https://doi.org/10.1007/JHEP01(2020)139}{\textit{JHEP} {\bfseries 01}
  (2020) 139}, [\href{https://arxiv.org/abs/1905.03764}{{\ttfamily
  1905.03764}}].

\bibitem{Giammanco:2017xyn}
A.~Giammanco and R.~Schwienhorst, \textit{{Single top-quark production at the
  Tevatron and the LHC}},
  \href{https://doi.org/10.1103/RevModPhys.90.035001}{\textit{Rev. Mod. Phys.}
  {\bfseries 90} (2018) 035001},
  [\href{https://arxiv.org/abs/1710.10699}{{\ttfamily 1710.10699}}].

\bibitem{Aad:2015eua}
{\scshape ATLAS} collaboration, G.~Aad et~al., \textit{{Measurement of the $
  t\overline{t}W $ and $ t\overline{t}Z $ production cross sections in pp
  collisions at $ \sqrt{s}=8 $ TeV with the ATLAS detector}},
  \href{https://doi.org/10.1007/JHEP11(2015)172}{\textit{JHEP} {\bfseries 11}
  (2015) 172}, [\href{https://arxiv.org/abs/1509.05276}{{\ttfamily
  1509.05276}}].

\bibitem{Khachatryan:2015sha}
{\scshape CMS} collaboration, V.~Khachatryan et~al., \textit{{Observation of
  top quark pairs produced in association with a vector boson in pp collisions
  at $ \sqrt{s}=8 $ TeV}},
  \href{https://doi.org/10.1007/JHEP01(2016)096}{\textit{JHEP} {\bfseries 01}
  (2016) 096}, [\href{https://arxiv.org/abs/1510.01131}{{\ttfamily
  1510.01131}}].

\bibitem{Aaboud:2017ylb}
{\scshape ATLAS} collaboration, M.~Aaboud et~al., \textit{{Measurement of the
  production cross-section of a single top quark in association with a Z boson
  in proton–proton collisions at 13 TeV with the ATLAS detector}},
  \href{https://doi.org/10.1016/j.physletb.2018.03.023}{\textit{Phys. Lett.}
  {\bfseries B780} (2018) 557--577},
  [\href{https://arxiv.org/abs/1710.03659}{{\ttfamily 1710.03659}}].

\bibitem{Sirunyan:2018zgs}
{\scshape CMS} collaboration, A.~M. Sirunyan et~al., \textit{{Observation of
  Single Top Quark Production in Association with a $Z$ Boson in Proton-Proton
  Collisions at $\sqrt {s}$ =13 TeV}},
  \href{https://doi.org/10.1103/PhysRevLett.122.132003}{\textit{Phys. Rev.
  Lett.} {\bfseries 122} (2019) 132003},
  [\href{https://arxiv.org/abs/1812.05900}{{\ttfamily 1812.05900}}].

\bibitem{CMS:2018rbc}
{\scshape CMS} collaboration, C.~Collaboration, \textit{{Measurement of the
  associated production of a Higgs boson and a pair of top-antitop quarks with
  the Higgs boson decaying to two photons in proton-proton collisions at
  $\sqrt{s}=13~\mathrm{TeV}$}}, .

\bibitem{Sirunyan:2018mvw}
{\scshape CMS} collaboration, A.~M. Sirunyan et~al., \textit{{Search for $
  \mathrm{t}\overline{\mathrm{t}}\mathrm{H} $ production in the $ \mathrm{H}\to
  \mathrm{b}\overline{\mathrm{b}} $ decay channel with leptonic $
  \mathrm{t}\overline{\mathrm{t}} $ decays in proton-proton collisions at $
  \sqrt{s}=13 $ TeV}},
  \href{https://doi.org/10.1007/JHEP03(2019)026}{\textit{JHEP} {\bfseries 03}
  (2019) 026}, [\href{https://arxiv.org/abs/1804.03682}{{\ttfamily
  1804.03682}}].

\bibitem{Sirunyan:2018shy}
{\scshape CMS} collaboration, A.~M. Sirunyan et~al., \textit{{Evidence for
  associated production of a Higgs boson with a top quark pair in final states
  with electrons, muons, and hadronically decaying $\tau$ leptons at $\sqrt{s}
  =$ 13 TeV}}, \href{https://doi.org/10.1007/JHEP08(2018)066}{\textit{JHEP}
  {\bfseries 08} (2018) 066},
  [\href{https://arxiv.org/abs/1803.05485}{{\ttfamily 1803.05485}}].

\bibitem{Aaboud:2017jvq}
{\scshape ATLAS} collaboration, M.~Aaboud et~al., \textit{{Evidence for the
  associated production of the Higgs boson and a top quark pair with the ATLAS
  detector}}, \href{https://doi.org/10.1103/PhysRevD.97.072003}{\textit{Phys.
  Rev.} {\bfseries D97} (2018) 072003},
  [\href{https://arxiv.org/abs/1712.08891}{{\ttfamily 1712.08891}}].

\bibitem{Aaboud:2017rss}
{\scshape ATLAS} collaboration, M.~Aaboud et~al., \textit{{Search for the
  standard model Higgs boson produced in association with top quarks and
  decaying into a $b\bar{b}$ pair in $pp$ collisions at $\sqrt{s}$ = 13 TeV
  with the ATLAS detector}},
  \href{https://doi.org/10.1103/PhysRevD.97.072016}{\textit{Phys. Rev.}
  {\bfseries D97} (2018) 072016},
  [\href{https://arxiv.org/abs/1712.08895}{{\ttfamily 1712.08895}}].

\bibitem{Maltoni:2019aot}
F.~Maltoni, L.~Mantani and K.~Mimasu, \textit{{Top-quark electroweak
  interactions at high energy}},
  \href{https://doi.org/10.1007/JHEP10(2019)004}{\textit{JHEP} {\bfseries 10}
  (2019) 004}, [\href{https://arxiv.org/abs/1904.05637}{{\ttfamily
  1904.05637}}].

\bibitem{Henning:2018kys}
B.~Henning, D.~Lombardo, M.~Riembau and F.~Riva, \textit{{Measuring Higgs
  Couplings without Higgs Bosons}},
  \href{https://doi.org/10.1103/PhysRevLett.123.181801}{\textit{Phys. Rev.
  Lett.} {\bfseries 123} (2019) 181801},
  [\href{https://arxiv.org/abs/1812.09299}{{\ttfamily 1812.09299}}].

\bibitem{Bauer:2016kkv}
C.~W. Bauer and N.~Ferland, \textit{{Resummation of electroweak Sudakov
  logarithms for real radiation}},
  \href{https://doi.org/10.1007/JHEP09(2016)025}{\textit{JHEP} {\bfseries 09}
  (2016) 025}, [\href{https://arxiv.org/abs/1601.07190}{{\ttfamily
  1601.07190}}].

\bibitem{Bauer:2017isx}
C.~W. Bauer, N.~Ferland and B.~R. Webber, \textit{{Standard Model Parton
  Distributions at Very High Energies}},
  \href{https://doi.org/10.1007/JHEP08(2017)036}{\textit{JHEP} {\bfseries 08}
  (2017) 036}, [\href{https://arxiv.org/abs/1703.08562}{{\ttfamily
  1703.08562}}].

\bibitem{Manohar:2018kfx}
A.~V. Manohar and W.~J. Waalewijn, \textit{{Electroweak Logarithms in Inclusive
  Cross Sections}},
  \href{https://doi.org/10.1007/JHEP08(2018)137}{\textit{JHEP} {\bfseries 08}
  (2018) 137}, [\href{https://arxiv.org/abs/1802.08687}{{\ttfamily
  1802.08687}}].

\bibitem{Han:2020uid}
T.~Han, Y.~Ma and K.~Xie, \textit{{High Energy Leptonic Collisions and
  Electroweak Parton Distribution Functions}},
  \href{https://arxiv.org/abs/2007.14300}{{\ttfamily 2007.14300}}.

\bibitem{ATLAS:2018doi}
{\scshape ATLAS} collaboration, T.~A. collaboration, \textit{{Combined
  measurements of Higgs boson production and decay using up to 80 fb$^{-1}$ of
  proton--proton collision data at $\sqrt{s}=$ 13 TeV collected with the ATLAS
  experiment}}, .

\bibitem{Sirunyan:2018koj}
{\scshape CMS} collaboration, A.~M. Sirunyan et~al., \textit{{Combined
  measurements of Higgs boson couplings in proton–proton collisions at
  $\sqrt{s}=13\,\text {Te}\text {V} $}},
  \href{https://doi.org/10.1140/epjc/s10052-019-6909-y}{\textit{Eur. Phys. J.}
  {\bfseries C79} (2019) 421},
  [\href{https://arxiv.org/abs/1809.10733}{{\ttfamily 1809.10733}}].

\bibitem{Gunion:1989we}
J.~F. Gunion, H.~E. Haber, G.~L. Kane and S.~Dawson, \textit{{The Higgs
  Hunter's Guide}}, {\textit{Front. Phys.} {\bfseries 80} (2000) 1--404}.

\bibitem{Branco:2011iw}
G.~C. Branco, P.~M. Ferreira, L.~Lavoura, M.~N. Rebelo, M.~Sher and J.~P.
  Silva, \textit{{Theory and phenomenology of two-Higgs-doublet models}},
  \href{https://doi.org/10.1016/j.physrep.2012.02.002}{\textit{Phys. Rept.}
  {\bfseries 516} (2012) 1--102},
  [\href{https://arxiv.org/abs/1106.0034}{{\ttfamily 1106.0034}}].

\bibitem{Morrissey:2012db}
D.~E. Morrissey and M.~J. Ramsey-Musolf, \textit{{Electroweak baryogenesis}},
  \href{https://doi.org/10.1088/1367-2630/14/12/125003}{\textit{New J. Phys.}
  {\bfseries 14} (2012) 125003},
  [\href{https://arxiv.org/abs/1206.2942}{{\ttfamily 1206.2942}}].

\bibitem{Ivanov:2017dad}
I.~P. Ivanov, \textit{{Building and testing models with extended Higgs
  sectors}}, \href{https://doi.org/10.1016/j.ppnp.2017.03.001}{\textit{Prog.
  Part. Nucl. Phys.} {\bfseries 95} (2017) 160--208},
  [\href{https://arxiv.org/abs/1702.03776}{{\ttfamily 1702.03776}}].

\bibitem{Khan:2015ipa}
N.~Khan and S.~Rakshit, \textit{{Constraints on inert dark matter from the
  metastability of the electroweak vacuum}},
  \href{https://doi.org/10.1103/PhysRevD.92.055006}{\textit{Phys. Rev.}
  {\bfseries D92} (2015) 055006},
  [\href{https://arxiv.org/abs/1503.03085}{{\ttfamily 1503.03085}}].

\bibitem{Kanemura:2016sos}
S.~Kanemura, M.~Kikuchi and K.~Sakurai, \textit{{Testing the dark matter
  scenario in the inert doublet model by future precision measurements of the
  Higgs boson couplings}},
  \href{https://doi.org/10.1103/PhysRevD.94.115011}{\textit{Phys. Rev.}
  {\bfseries D94} (2016) 115011},
  [\href{https://arxiv.org/abs/1605.08520}{{\ttfamily 1605.08520}}].

\bibitem{Ilnicka:2018def}
A.~Ilnicka, T.~Robens and T.~Stefaniak, \textit{{Constraining Extended Scalar
  Sectors at the LHC and beyond}},
  \href{https://doi.org/10.1142/S0217732318300070}{\textit{Mod. Phys. Lett.}
  {\bfseries A33} (2018) 1830007},
  [\href{https://arxiv.org/abs/1803.03594}{{\ttfamily 1803.03594}}].

\bibitem{Craig:2013xia}
N.~Craig, C.~Englert and M.~McCullough, \textit{{New Probe of Naturalness}},
  \href{https://doi.org/10.1103/PhysRevLett.111.121803}{\textit{Phys. Rev.
  Lett.} {\bfseries 111} (2013) 121803},
  [\href{https://arxiv.org/abs/1305.5251}{{\ttfamily 1305.5251}}].

\bibitem{vanderBij:2006ne}
J.~J. van~der Bij, \textit{{The Minimal non-minimal standard model}},
  \href{https://doi.org/10.1016/j.physletb.2006.03.018}{\textit{Phys. Lett.}
  {\bfseries B636} (2006) 56--59},
  [\href{https://arxiv.org/abs/hep-ph/0603082}{{\ttfamily hep-ph/0603082}}].

\bibitem{Abe:2012eu}
T.~Abe, R.~Kitano, Y.~Konishi, K.-y. Oda, J.~Sato and S.~Sugiyama,
  \textit{{Minimal Dilaton Model}},
  \href{https://doi.org/10.1103/PhysRevD.86.115016}{\textit{Phys. Rev.}
  {\bfseries D86} (2012) 115016},
  [\href{https://arxiv.org/abs/1209.4544}{{\ttfamily 1209.4544}}].

\bibitem{Crivellin:2013wna}
A.~Crivellin, A.~Kokulu and C.~Greub, \textit{{Flavor-phenomenology of
  two-Higgs-doublet models with generic Yukawa structure}},
  \href{https://doi.org/10.1103/PhysRevD.87.094031}{\textit{Phys. Rev.}
  {\bfseries D87} (2013) 094031},
  [\href{https://arxiv.org/abs/1303.5877}{{\ttfamily 1303.5877}}].

\bibitem{Craig:2013hca}
N.~Craig, J.~Galloway and S.~Thomas, \textit{{Searching for Signs of the Second
  Higgs Doublet}},  \href{https://arxiv.org/abs/1305.2424}{{\ttfamily
  1305.2424}}.

\bibitem{Degrande:2014vpa}
C.~Degrande, \textit{{Automatic evaluation of UV and R2 terms for beyond the
  Standard Model Lagrangians: a proof-of-principle}},
  \href{https://doi.org/10.1016/j.cpc.2015.08.015}{\textit{Comput. Phys.
  Commun.} {\bfseries 197} (2015) 239--262},
  [\href{https://arxiv.org/abs/1406.3030}{{\ttfamily 1406.3030}}].

\bibitem{Georgi:1985nv}
H.~Georgi and M.~Machacek, \textit{{DOUBLY CHARGED HIGGS BOSONS}},
  \href{https://doi.org/10.1016/0550-3213(85)90325-6}{\textit{Nucl. Phys.}
  {\bfseries B262} (1985) 463--477}.

\bibitem{Konetschny:1977bn}
W.~Konetschny and W.~Kummer, \textit{{Nonconservation of Total Lepton Number
  with Scalar Bosons}},
  \href{https://doi.org/10.1016/0370-2693(77)90407-5}{\textit{Phys. Lett.}
  {\bfseries 70B} (1977) 433--435}.

\bibitem{Schechter:1980gr}
J.~Schechter and J.~W.~F. Valle, \textit{{Neutrino Masses in SU(2) x U(1)
  Theories}}, \href{https://doi.org/10.1103/PhysRevD.22.2227}{\textit{Phys.
  Rev.} {\bfseries D22} (1980) 2227}.

\bibitem{Cheng:1980qt}
T.~P. Cheng and L.-F. Li, \textit{{Neutrino Masses, Mixings and Oscillations in
  SU(2) x U(1) Models of Electroweak Interactions}},
  \href{https://doi.org/10.1103/PhysRevD.22.2860}{\textit{Phys. Rev.}
  {\bfseries D22} (1980) 2860}.

\bibitem{Lazarides:1980nt}
G.~Lazarides, Q.~Shafi and C.~Wetterich, \textit{{Proton Lifetime and Fermion
  Masses in an SO(10) Model}},
  \href{https://doi.org/10.1016/0550-3213(81)90354-0}{\textit{Nucl. Phys.}
  {\bfseries B181} (1981) 287--300}.

\bibitem{Mohapatra:1980yp}
R.~N. Mohapatra and G.~Senjanovic, \textit{{Neutrino Masses and Mixings in
  Gauge Models with Spontaneous Parity Violation}},
  \href{https://doi.org/10.1103/PhysRevD.23.165}{\textit{Phys. Rev.} {\bfseries
  D23} (1981) 165}.

\bibitem{Fuks:2019clu}
B.~Fuks, M.~Nemev{\v s}ek and R.~Ruiz, \textit{{Doubly Charged Higgs Boson
  Production at Hadron Colliders}},
  \href{https://doi.org/10.1103/PhysRevD.101.075022}{\textit{Phys. Rev.}
  {\bfseries D101} (2020) 075022},
  [\href{https://arxiv.org/abs/1912.08975}{{\ttfamily 1912.08975}}].

\bibitem{Chen:2005jx}
M.-C. Chen, S.~Dawson and T.~Krupovnickas, \textit{{Constraining new models
  with precision electroweak data}},
  \href{https://doi.org/10.1142/S0217751X0603388X}{\textit{Int. J. Mod. Phys.}
  {\bfseries A21} (2006) 4045--4070},
  [\href{https://arxiv.org/abs/hep-ph/0504286}{{\ttfamily hep-ph/0504286}}].

\bibitem{Han:2005nk}
T.~Han, H.~E. Logan, B.~Mukhopadhyaya and R.~Srikanth, \textit{{Neutrino masses
  and lepton-number violation in the littlest Higgs scenario}},
  \href{https://doi.org/10.1103/PhysRevD.72.053007}{\textit{Phys. Rev.}
  {\bfseries D72} (2005) 053007},
  [\href{https://arxiv.org/abs/hep-ph/0505260}{{\ttfamily hep-ph/0505260}}].

\bibitem{Chen:2008jg}
M.-C. Chen, S.~Dawson and C.~B. Jackson, \textit{{Higgs Triplets, Decoupling,
  and Precision Measurements}},
  \href{https://doi.org/10.1103/PhysRevD.78.093001}{\textit{Phys. Rev.}
  {\bfseries D78} (2008) 093001},
  [\href{https://arxiv.org/abs/0809.4185}{{\ttfamily 0809.4185}}].

\bibitem{Perez:2008ha}
P.~Fileviez~Perez, T.~Han, G.-y. Huang, T.~Li and K.~Wang, \textit{{Neutrino
  Masses and the CERN LHC: Testing Type II Seesaw}},
  \href{https://doi.org/10.1103/PhysRevD.78.015018}{\textit{Phys. Rev.}
  {\bfseries D78} (2008) 015018},
  [\href{https://arxiv.org/abs/0805.3536}{{\ttfamily 0805.3536}}].

\bibitem{Kanemura:2012rs}
S.~Kanemura and K.~Yagyu, \textit{{Radiative corrections to electroweak
  parameters in the Higgs triplet model and implication with the recent Higgs
  boson searches}},
  \href{https://doi.org/10.1103/PhysRevD.85.115009}{\textit{Phys. Rev.}
  {\bfseries D85} (2012) 115009},
  [\href{https://arxiv.org/abs/1201.6287}{{\ttfamily 1201.6287}}].

\bibitem{Das:2016bir}
D.~Das and A.~Santamaria, \textit{{Updated scalar sector constraints in the
  Higgs triplet model}},
  \href{https://doi.org/10.1103/PhysRevD.94.015015}{\textit{Phys. Rev.}
  {\bfseries D94} (2016) 015015},
  [\href{https://arxiv.org/abs/1604.08099}{{\ttfamily 1604.08099}}].

\bibitem{Ismail:2020zoz}
A.~Ismail, H.~E. Logan and Y.~Wu, \textit{{Updated constraints on the
  Georgi-Machacek model from LHC Run 2}},
  \href{https://arxiv.org/abs/2003.02272}{{\ttfamily 2003.02272}}.

\bibitem{Hartling:2014zca}
K.~Hartling, K.~Kumar and H.~E. Logan, \textit{{The decoupling limit in the
  Georgi-Machacek model}},
  \href{https://doi.org/10.1103/PhysRevD.90.015007}{\textit{Phys. Rev.}
  {\bfseries D90} (2014) 015007},
  [\href{https://arxiv.org/abs/1404.2640}{{\ttfamily 1404.2640}}].

\bibitem{Degrande:2015xnm}
C.~Degrande, K.~Hartling, H.~E. Logan, A.~D. Peterson and M.~Zaro,
  \textit{{Automatic predictions in the Georgi-Machacek model at
  next-to-leading order accuracy}},
  \href{https://doi.org/10.1103/PhysRevD.93.035004}{\textit{Phys. Rev.}
  {\bfseries D93} (2016) 035004},
  [\href{https://arxiv.org/abs/1512.01243}{{\ttfamily 1512.01243}}].

\bibitem{Nilles:1983ge}
H.~P. Nilles, \textit{{Supersymmetry, Supergravity and Particle Physics}},
  \href{https://doi.org/10.1016/0370-1573(84)90008-5}{\textit{Phys. Rept.}
  {\bfseries 110} (1984) 1--162}.

\bibitem{Haber:1984rc}
H.~E. Haber and G.~L. Kane, \textit{{The Search for Supersymmetry: Probing
  Physics Beyond the Standard Model}},
  \href{https://doi.org/10.1016/0370-1573(85)90051-1}{\textit{Phys. Rept.}
  {\bfseries 117} (1985) 75--263}.

\bibitem{Martin:1997ns}
S.~P. Martin, \textit{{A Supersymmetry primer}},
  \href{https://arxiv.org/abs/hep-ph/9709356}{{\ttfamily hep-ph/9709356}}.
  [Adv. Ser. Direct. High Energy Phys.21,1(2010); Adv. Ser. Direct. High Energy
  Phys.18,1(1998)].

\bibitem{Baer:2006rs}
H.~Baer and X.~Tata, \textit{{Weak scale supersymmetry: From superfields to
  scattering events}}.
\newblock Cambridge University Press, 2006.

\bibitem{Tanabashi:2018oca}
{\scshape Particle Data Group} collaboration, M.~Tanabashi et~al.,
  \textit{{Review of Particle Physics}},
  \href{https://doi.org/10.1103/PhysRevD.98.030001}{\textit{Phys. Rev.}
  {\bfseries D98} (2018) 030001}.

\bibitem{Aad:2019pfy}
{\scshape ATLAS} collaboration, G.~Aad et~al., \textit{{Search for
  bottom-squark pair production with the ATLAS detector in final states
  containing Higgs bosons, $b$-jets and missing transverse momentum}},
  \href{https://doi.org/10.1007/JHEP12(2019)060}{\textit{JHEP} {\bfseries 12}
  (2019) 060}, [\href{https://arxiv.org/abs/1908.03122}{{\ttfamily
  1908.03122}}].

\bibitem{Sirunyan:2019ctn}
{\scshape CMS} collaboration, A.~M. Sirunyan et~al., \textit{{Search for
  supersymmetry in proton-proton collisions at 13 TeV in final states with jets
  and missing transverse momentum}},
  \href{https://doi.org/10.1007/JHEP10(2019)244}{\textit{JHEP} {\bfseries 10}
  (2019) 244}, [\href{https://arxiv.org/abs/1908.04722}{{\ttfamily
  1908.04722}}].

\bibitem{Aad:2019ftg}
{\scshape ATLAS} collaboration, G.~Aad et~al., \textit{{Search for squarks and
  gluinos in final states with same-sign leptons and jets using 139 fb$^{-1}$
  of data collected with the ATLAS detector}},
  \href{https://arxiv.org/abs/1909.08457}{{\ttfamily 1909.08457}}.

\bibitem{CMS:2019tlp}
{\scshape CMS} collaboration, A.~M. Sirunyan et~al., \textit{{Search for
  supersymmetry in pp collisions at $\sqrt{s}=$ 13 TeV with 137 fb$^{-1}$ in
  final states with a single lepton using the sum of masses of large-radius
  jets}}, \href{https://doi.org/10.1103/PhysRevD.101.052010}{\textit{Phys.
  Rev.} {\bfseries D101} (2020) 052010},
  [\href{https://arxiv.org/abs/1911.07558}{{\ttfamily 1911.07558}}].

\bibitem{Aad:2019qnd}
{\scshape ATLAS} collaboration, G.~Aad et~al., \textit{{Searches for
  electroweak production of supersymmetric particles with compressed mass
  spectra in $\sqrt{s}=$ 13 TeV $pp$ collisions with the ATLAS detector}},
  \href{https://doi.org/10.1103/PhysRevD.101.052005}{\textit{Phys. Rev.}
  {\bfseries D101} (2020) 052005},
  [\href{https://arxiv.org/abs/1911.12606}{{\ttfamily 1911.12606}}].

\bibitem{Aad:2019vvi}
{\scshape ATLAS} collaboration, G.~Aad et~al., \textit{{Search for
  chargino-neutralino production with mass splittings near the electroweak
  scale in three-lepton final states in $\sqrt {s}$=13 TeV $pp$ collisions with
  the ATLAS detector}},
  \href{https://doi.org/10.1103/PhysRevD.101.072001}{\textit{Phys. Rev.}
  {\bfseries D101} (2020) 072001},
  [\href{https://arxiv.org/abs/1912.08479}{{\ttfamily 1912.08479}}].

\bibitem{Sirunyan:2019glc}
{\scshape CMS} collaboration, A.~M. Sirunyan et~al., \textit{{Search for direct
  top squark pair production in events with one lepton, jets, and missing
  transverse momentum at 13 TeV with the CMS experiment}},
  \href{https://doi.org/10.1007/JHEP05(2020)032}{\textit{JHEP} {\bfseries 05}
  (2020) 032}, [\href{https://arxiv.org/abs/1912.08887}{{\ttfamily
  1912.08887}}].

\bibitem{Sirunyan:2020ztc}
{\scshape CMS} collaboration, A.~M. Sirunyan et~al., \textit{{Search for
  physics beyond the standard model in events with jets and two same-sign or at
  least three charged leptons in proton-proton collisions at $\sqrt{s}=$ 13
  TeV}},  \href{https://arxiv.org/abs/2001.10086}{{\ttfamily 2001.10086}}.

\bibitem{Duhr:2011se}
C.~Duhr and B.~Fuks, \textit{{A superspace module for the FeynRules package}},
  \href{https://doi.org/10.1016/j.cpc.2011.06.009}{\textit{Comput. Phys.
  Commun.} {\bfseries 182} (2011) 2404--2426},
  [\href{https://arxiv.org/abs/1102.4191}{{\ttfamily 1102.4191}}].

\bibitem{Pati:1974yy}
J.~C. Pati and A.~Salam, \textit{{Lepton Number as the Fourth Color}},
  \href{https://doi.org/10.1103/PhysRevD.10.275,
  10.1103/PhysRevD.11.703.2}{\textit{Phys. Rev.} {\bfseries D10} (1974)
  275--289}. [Erratum: Phys. Rev.D11,703(1975)].

\bibitem{Georgi:1974sy}
H.~Georgi and S.~L. Glashow, \textit{{Unity of All Elementary Particle
  Forces}}, \href{https://doi.org/10.1103/PhysRevLett.32.438}{\textit{Phys.
  Rev. Lett.} {\bfseries 32} (1974) 438--441}.

\bibitem{Fritzsch:1974nn}
H.~Fritzsch and P.~Minkowski, \textit{{Unified Interactions of Leptons and
  Hadrons}}, \href{https://doi.org/10.1016/0003-4916(75)90211-0}{\textit{Annals
  Phys.} {\bfseries 93} (1975) 193--266}.

\bibitem{Dimopoulos:1979es}
S.~Dimopoulos and L.~Susskind, \textit{{Mass Without Scalars}},
  \href{https://doi.org/10.1016/0550-3213(79)90364-X}{\textit{Nucl. Phys.}
  {\bfseries B155} (1979) 237--252}. [2,930(1979)].

\bibitem{Senjanovic:1982ex}
G.~Senjanovic and A.~Sokorac, \textit{{Light Leptoquarks in SO(10)}},
  \href{https://doi.org/10.1007/BF01574858}{\textit{Z. Phys.} {\bfseries C20}
  (1983) 255}.

\bibitem{Schrempp:1984nj}
B.~Schrempp and F.~Schrempp, \textit{{LIGHT LEPTOQUARKS}},
  \href{https://doi.org/10.1016/0370-2693(85)91450-9}{\textit{Phys. Lett.}
  {\bfseries 153B} (1985) 101--107}.

\bibitem{Hewett:1988xc}
J.~L. Hewett and T.~G. Rizzo, \textit{{Low-Energy Phenomenology of Superstring
  Inspired E(6) Models}},
  \href{https://doi.org/10.1016/0370-1573(89)90071-9}{\textit{Phys. Rept.}
  {\bfseries 183} (1989) 193}.

\bibitem{Frampton:1989fu}
P.~H. Frampton and B.-H. Lee, \textit{{SU(15) GRAND UNIFICATION}},
  \href{https://doi.org/10.1103/PhysRevLett.64.619}{\textit{Phys. Rev. Lett.}
  {\bfseries 64} (1990) 619}.

\bibitem{Dorsner:2016wpm}
I.~Dor{\v s}ner, S.~Fajfer, A.~Greljo, J.~F. Kamenik and N.~Ko{\v s}nik,
  \textit{{Physics of leptoquarks in precision experiments and at particle
  colliders}},
  \href{https://doi.org/10.1016/j.physrep.2016.06.001}{\textit{Phys. Rept.}
  {\bfseries 641} (2016) 1--68},
  [\href{https://arxiv.org/abs/1603.04993}{{\ttfamily 1603.04993}}].

\bibitem{Lees:2013uzd}
{\scshape BaBar} collaboration, J.~P. Lees et~al., \textit{{Measurement of an
  Excess of $\bar{B} \to D^{(*)}\tau^- \bar{\nu}_\tau$ Decays and Implications
  for Charged Higgs Bosons}},
  \href{https://doi.org/10.1103/PhysRevD.88.072012}{\textit{Phys. Rev.}
  {\bfseries D88} (2013) 072012},
  [\href{https://arxiv.org/abs/1303.0571}{{\ttfamily 1303.0571}}].

\bibitem{Aaij:2014ora}
{\scshape LHCb} collaboration, R.~Aaij et~al., \textit{{Test of lepton
  universality using $B^{+}\rightarrow K^{+}\ell^{+}\ell^{-}$ decays}},
  \href{https://doi.org/10.1103/PhysRevLett.113.151601}{\textit{Phys. Rev.
  Lett.} {\bfseries 113} (2014) 151601},
  [\href{https://arxiv.org/abs/1406.6482}{{\ttfamily 1406.6482}}].

\bibitem{Aaij:2015yra}
{\scshape LHCb} collaboration, R.~Aaij et~al., \textit{{Measurement of the
  ratio of branching fractions $\mathcal{B}(\bar{B}^0 \to
  D^{*+}\tau^{-}\bar{\nu}_{\tau})/\mathcal{B}(\bar{B}^0 \to
  D^{*+}\mu^{-}\bar{\nu}_{\mu})$}},
  \href{https://doi.org/10.1103/PhysRevLett.115.159901,
  10.1103/PhysRevLett.115.111803}{\textit{Phys. Rev. Lett.} {\bfseries 115}
  (2015) 111803}, [\href{https://arxiv.org/abs/1506.08614}{{\ttfamily
  1506.08614}}]. [Erratum: Phys. Rev. Lett.115,no.15,159901(2015)].

\bibitem{Hirose:2016wfn}
{\scshape Belle} collaboration, S.~Hirose et~al., \textit{{Measurement of the
  $\tau$ lepton polarization and $R(D^*)$ in the decay $\bar{B} \to D^* \tau^-
  \bar{\nu}_\tau$}},
  \href{https://doi.org/10.1103/PhysRevLett.118.211801}{\textit{Phys. Rev.
  Lett.} {\bfseries 118} (2017) 211801},
  [\href{https://arxiv.org/abs/1612.00529}{{\ttfamily 1612.00529}}].

\bibitem{Aaij:2017deq}
{\scshape LHCb} collaboration, R.~Aaij et~al., \textit{{Test of Lepton Flavor
  Universality by the measurement of the $B^0 \to D^{*-} \tau^+ \nu_{\tau}$
  branching fraction using three-prong $\tau$ decays}},
  \href{https://doi.org/10.1103/PhysRevD.97.072013}{\textit{Phys. Rev.}
  {\bfseries D97} (2018) 072013},
  [\href{https://arxiv.org/abs/1711.02505}{{\ttfamily 1711.02505}}].

\bibitem{Aaij:2017vbb}
{\scshape LHCb} collaboration, R.~Aaij et~al., \textit{{Test of lepton
  universality with $B^{0} \rightarrow K^{*0}\ell^{+}\ell^{-}$ decays}},
  \href{https://doi.org/10.1007/JHEP08(2017)055}{\textit{JHEP} {\bfseries 08}
  (2017) 055}, [\href{https://arxiv.org/abs/1705.05802}{{\ttfamily
  1705.05802}}].

\bibitem{Cerri:2018ypt}
A.~Cerri et~al., \textit{{Report from Working Group 4}},
  \href{https://doi.org/10.23731/CYRM-2019-007.867}{\textit{CERN Yellow Rep.
  Monogr.} {\bfseries 7} (2019) 867--1158},
  [\href{https://arxiv.org/abs/1812.07638}{{\ttfamily 1812.07638}}].

\bibitem{Barbieri:2015yvd}
R.~Barbieri, G.~Isidori, A.~Pattori and F.~Senia, \textit{{Anomalies in
  $B$-decays and $U(2)$ flavour symmetry}},
  \href{https://doi.org/10.1140/epjc/s10052-016-3905-3}{\textit{Eur. Phys. J.}
  {\bfseries C76} (2016) 67},
  [\href{https://arxiv.org/abs/1512.01560}{{\ttfamily 1512.01560}}].

\bibitem{Buttazzo:2016kid}
D.~Buttazzo, A.~Greljo, G.~Isidori and D.~Marzocca, \textit{{Toward a coherent
  solution of diphoton and flavor anomalies}},
  \href{https://doi.org/10.1007/JHEP08(2016)035}{\textit{JHEP} {\bfseries 08}
  (2016) 035}, [\href{https://arxiv.org/abs/1604.03940}{{\ttfamily
  1604.03940}}].

\bibitem{Barbieri:2016las}
R.~Barbieri, C.~W. Murphy and F.~Senia, \textit{{B-decay Anomalies in a
  Composite Leptoquark Model}},
  \href{https://doi.org/10.1140/epjc/s10052-016-4578-7}{\textit{Eur. Phys. J.}
  {\bfseries C77} (2017) 8},
  [\href{https://arxiv.org/abs/1611.04930}{{\ttfamily 1611.04930}}].

\bibitem{DiLuzio:2017vat}
L.~Di~Luzio, A.~Greljo and M.~Nardecchia, \textit{{Gauge leptoquark as the
  origin of B-physics anomalies}},
  \href{https://doi.org/10.1103/PhysRevD.96.115011}{\textit{Phys. Rev.}
  {\bfseries D96} (2017) 115011},
  [\href{https://arxiv.org/abs/1708.08450}{{\ttfamily 1708.08450}}].

\bibitem{Baker:2019sli}
M.~J. Baker, J.~Fuentes-Mart{\'i}n, G.~Isidori and M.~K{\"o}nig, \textit{{High-
  $p_T$ signatures in vector–leptoquark models}},
  \href{https://doi.org/10.1140/epjc/s10052-019-6853-x}{\textit{Eur. Phys. J.}
  {\bfseries C79} (2019) 334},
  [\href{https://arxiv.org/abs/1901.10480}{{\ttfamily 1901.10480}}].

\bibitem{Ahmad:2002jz}
{\scshape SNO} collaboration, Q.~R. Ahmad et~al., \textit{{Direct evidence for
  neutrino flavor transformation from neutral current interactions in the
  Sudbury Neutrino Observatory}},
  \href{https://doi.org/10.1103/PhysRevLett.89.011301}{\textit{Phys. Rev.
  Lett.} {\bfseries 89} (2002) 011301},
  [\href{https://arxiv.org/abs/nucl-ex/0204008}{{\ttfamily nucl-ex/0204008}}].

\bibitem{Ashie:2005ik}
{\scshape Super-Kamiokande} collaboration, Y.~Ashie et~al., \textit{{A
  Measurement of atmospheric neutrino oscillation parameters by
  SUPER-KAMIOKANDE I}},
  \href{https://doi.org/10.1103/PhysRevD.71.112005}{\textit{Phys. Rev.}
  {\bfseries D71} (2005) 112005},
  [\href{https://arxiv.org/abs/hep-ex/0501064}{{\ttfamily hep-ex/0501064}}].

\bibitem{Aker:2019uuj}
{\scshape KATRIN} collaboration, M.~Aker et~al., \textit{{Improved Upper Limit
  on the Neutrino Mass from a Direct Kinematic Method by KATRIN}},
  \href{https://doi.org/10.1103/PhysRevLett.123.221802}{\textit{Phys. Rev.
  Lett.} {\bfseries 123} (2019) 221802},
  [\href{https://arxiv.org/abs/1909.06048}{{\ttfamily 1909.06048}}].

\bibitem{Schechter:1981bd}
J.~Schechter and J.~W.~F. Valle, \textit{{Neutrinoless Double beta Decay in
  SU(2) x U(1) Theories}},
  \href{https://doi.org/10.1103/PhysRevD.25.2951}{\textit{Phys. Rev.}
  {\bfseries D25} (1982) 2951}.

\bibitem{Hirsch:2006yk}
M.~Hirsch, S.~Kovalenko and I.~Schmidt, \textit{{Extended black box theorem for
  lepton number and flavor violating processes}},
  \href{https://doi.org/10.1016/j.physletb.2006.09.012}{\textit{Phys. Lett.}
  {\bfseries B642} (2006) 106--110},
  [\href{https://arxiv.org/abs/hep-ph/0608207}{{\ttfamily hep-ph/0608207}}].

\bibitem{Duerr:2011zd}
M.~Duerr, M.~Lindner and A.~Merle, \textit{{On the Quantitative Impact of the
  Schechter-Valle Theorem}},
  \href{https://doi.org/10.1007/JHEP06(2011)091}{\textit{JHEP} {\bfseries 06}
  (2011) 091}, [\href{https://arxiv.org/abs/1105.0901}{{\ttfamily 1105.0901}}].

\bibitem{Moffat:2017feq}
K.~Moffat, S.~Pascoli and C.~Weiland, \textit{{Equivalence between massless
  neutrinos and lepton number conservation in fermionic singlet extensions of
  the Standard Model}},  \href{https://arxiv.org/abs/1712.07611}{{\ttfamily
  1712.07611}}.

\bibitem{Ma:1998dn}
E.~Ma, \textit{{Pathways to naturally small neutrino masses}},
  \href{https://doi.org/10.1103/PhysRevLett.81.1171}{\textit{Phys. Rev. Lett.}
  {\bfseries 81} (1998) 1171--1174},
  [\href{https://arxiv.org/abs/hep-ph/9805219}{{\ttfamily hep-ph/9805219}}].

\bibitem{Cai:2017jrq}
Y.~Cai, J.~Herrero-Garc{\'i}a, M.~A. Schmidt, A.~Vicente and R.~R. Volkas,
  \textit{{From the trees to the forest: a review of radiative neutrino mass
  models}}, \href{https://doi.org/10.3389/fphy.2017.00063}{\textit{Front.in
  Phys.} {\bfseries 5} (2017) 63},
  [\href{https://arxiv.org/abs/1706.08524}{{\ttfamily 1706.08524}}].

\bibitem{Pilaftsis:1991ug}
A.~Pilaftsis, \textit{{Radiatively induced neutrino masses and large Higgs
  neutrino couplings in the standard model with Majorana fields}},
  \href{https://doi.org/10.1007/BF01482590}{\textit{Z. Phys.} {\bfseries C55}
  (1992) 275--282}, [\href{https://arxiv.org/abs/hep-ph/9901206}{{\ttfamily
  hep-ph/9901206}}].

\bibitem{Kersten:2007vk}
J.~Kersten and A.~{\relax Yu}. Smirnov, \textit{{Right-Handed Neutrinos at CERN
  LHC and the Mechanism of Neutrino Mass Generation}},
  \href{https://doi.org/10.1103/PhysRevD.76.073005}{\textit{Phys. Rev.}
  {\bfseries D76} (2007) 073005},
  [\href{https://arxiv.org/abs/0705.3221}{{\ttfamily 0705.3221}}].

\bibitem{delAguila:2008cj}
F.~del Aguila and J.~A. Aguilar-Saavedra, \textit{{Distinguishing seesaw models
  at LHC with multi-lepton signals}},
  \href{https://doi.org/10.1016/j.nuclphysb.2008.12.029}{\textit{Nucl. Phys.}
  {\bfseries B813} (2009) 22--90},
  [\href{https://arxiv.org/abs/0808.2468}{{\ttfamily 0808.2468}}].

\bibitem{Atre:2009rg}
A.~Atre, T.~Han, S.~Pascoli and B.~Zhang, \textit{{The Search for Heavy
  Majorana Neutrinos}},
  \href{https://doi.org/10.1088/1126-6708/2009/05/030}{\textit{JHEP} {\bfseries
  05} (2009) 030}, [\href{https://arxiv.org/abs/0901.3589}{{\ttfamily
  0901.3589}}].

\bibitem{Pascoli:2018heg}
S.~Pascoli, R.~Ruiz and C.~Weiland, \textit{{Heavy neutrinos with dynamic jet
  vetoes: multilepton searches at $ \sqrt{s}=14 $ , 27, and 100 TeV}},
  \href{https://doi.org/10.1007/JHEP06(2019)049}{\textit{JHEP} {\bfseries 06}
  (2019) 049}, [\href{https://arxiv.org/abs/1812.08750}{{\ttfamily
  1812.08750}}].

\bibitem{Alva:2014gxa}
D.~Alva, T.~Han and R.~Ruiz, \textit{{Heavy Majorana neutrinos from $W\gamma$
  fusion at hadron colliders}},
  \href{https://doi.org/10.1007/JHEP02(2015)072}{\textit{JHEP} {\bfseries 02}
  (2015) 072}, [\href{https://arxiv.org/abs/1411.7305}{{\ttfamily 1411.7305}}].

\bibitem{Degrande:2016aje}
C.~Degrande, O.~Mattelaer, R.~Ruiz and J.~Turner, \textit{{Fully-Automated
  Precision Predictions for Heavy Neutrino Production Mechanisms at Hadron
  Colliders}}, \href{https://doi.org/10.1103/PhysRevD.94.053002}{\textit{Phys.
  Rev.} {\bfseries D94} (2016) 053002},
  [\href{https://arxiv.org/abs/1602.06957}{{\ttfamily 1602.06957}}].

\bibitem{Kobayashi:1973fv}
M.~Kobayashi and T.~Maskawa, \textit{{CP Violation in the Renormalizable Theory
  of Weak Interaction}},
  \href{https://doi.org/10.1143/PTP.49.652}{\textit{Prog. Theor. Phys.}
  {\bfseries 49} (1973) 652--657}.

\bibitem{Eberhardt:2010bm}
O.~Eberhardt, A.~Lenz and J.~Rohrwild, \textit{{Less space for a new family of
  fermions}}, \href{https://doi.org/10.1103/PhysRevD.82.095006}{\textit{Phys.
  Rev.} {\bfseries D82} (2010) 095006},
  [\href{https://arxiv.org/abs/1005.3505}{{\ttfamily 1005.3505}}].

\bibitem{Djouadi:2012ae}
A.~Djouadi and A.~Lenz, \textit{{Sealing the fate of a fourth generation of
  fermions}},
  \href{https://doi.org/10.1016/j.physletb.2012.07.060}{\textit{Phys. Lett.}
  {\bfseries B715} (2012) 310--314},
  [\href{https://arxiv.org/abs/1204.1252}{{\ttfamily 1204.1252}}].

\bibitem{Eberhardt:2012gv}
O.~Eberhardt, G.~Herbert, H.~Lacker, A.~Lenz, A.~Menzel, U.~Nierste et~al.,
  \textit{{Impact of a Higgs boson at a mass of 126 GeV on the standard model
  with three and four fermion generations}},
  \href{https://doi.org/10.1103/PhysRevLett.109.241802}{\textit{Phys. Rev.
  Lett.} {\bfseries 109} (2012) 241802},
  [\href{https://arxiv.org/abs/1209.1101}{{\ttfamily 1209.1101}}].

\bibitem{ArkaniHamed:2002qy}
N.~Arkani-Hamed, A.~G. Cohen, E.~Katz and A.~E. Nelson, \textit{{The Littlest
  Higgs}}, \href{https://doi.org/10.1088/1126-6708/2002/07/034}{\textit{JHEP}
  {\bfseries 07} (2002) 034},
  [\href{https://arxiv.org/abs/hep-ph/0206021}{{\ttfamily hep-ph/0206021}}].

\bibitem{Schmaltz:2005ky}
M.~Schmaltz and D.~Tucker-Smith, \textit{{Little Higgs review}},
  \href{https://doi.org/10.1146/annurev.nucl.55.090704.151502}{\textit{Ann.
  Rev. Nucl. Part. Sci.} {\bfseries 55} (2005) 229--270},
  [\href{https://arxiv.org/abs/hep-ph/0502182}{{\ttfamily hep-ph/0502182}}].

\bibitem{Martin:2009bg}
S.~P. Martin, \textit{{Extra vector-like matter and the lightest Higgs scalar
  boson mass in low-energy supersymmetry}},
  \href{https://doi.org/10.1103/PhysRevD.81.035004}{\textit{Phys. Rev.}
  {\bfseries D81} (2010) 035004},
  [\href{https://arxiv.org/abs/0910.2732}{{\ttfamily 0910.2732}}].

\bibitem{Han:2003wu}
T.~Han, H.~E. Logan, B.~McElrath and L.-T. Wang, \textit{{Phenomenology of the
  little Higgs model}},
  \href{https://doi.org/10.1103/PhysRevD.67.095004}{\textit{Phys. Rev.}
  {\bfseries D67} (2003) 095004},
  [\href{https://arxiv.org/abs/hep-ph/0301040}{{\ttfamily hep-ph/0301040}}].

\bibitem{Hubisz:2004ft}
J.~Hubisz and P.~Meade, \textit{{Phenomenology of the littlest Higgs with
  T-parity}}, \href{https://doi.org/10.1103/PhysRevD.71.035016}{\textit{Phys.
  Rev.} {\bfseries D71} (2005) 035016},
  [\href{https://arxiv.org/abs/hep-ph/0411264}{{\ttfamily hep-ph/0411264}}].

\bibitem{Perelstein:2005ka}
M.~Perelstein, \textit{{Little Higgs models and their phenomenology}},
  \href{https://doi.org/10.1016/j.ppnp.2006.04.001}{\textit{Prog. Part. Nucl.
  Phys.} {\bfseries 58} (2007) 247--291},
  [\href{https://arxiv.org/abs/hep-ph/0512128}{{\ttfamily hep-ph/0512128}}].

\bibitem{Aguilar-Saavedra:2013qpa}
J.~A. Aguilar-Saavedra, R.~Benbrik, S.~Heinemeyer and M.~P{\'e}rez-Victoria,
  \textit{{Handbook of vectorlike quarks: Mixing and single production}},
  \href{https://doi.org/10.1103/PhysRevD.88.094010}{\textit{Phys. Rev.}
  {\bfseries D88} (2013) 094010},
  [\href{https://arxiv.org/abs/1306.0572}{{\ttfamily 1306.0572}}].

\bibitem{Buchkremer:2013bha}
M.~Buchkremer, G.~Cacciapaglia, A.~Deandrea and L.~Panizzi, \textit{{Model
  Independent Framework for Searches of Top Partners}},
  \href{https://doi.org/10.1016/j.nuclphysb.2013.08.010}{\textit{Nucl. Phys.}
  {\bfseries B876} (2013) 376--417},
  [\href{https://arxiv.org/abs/1305.4172}{{\ttfamily 1305.4172}}].

\bibitem{Aaboud:2018xpj}
{\scshape ATLAS} collaboration, M.~Aaboud et~al., \textit{{Search for new
  phenomena in events with same-charge leptons and $b$-jets in $pp$ collisions
  at $\sqrt{s}= 13$ TeV with the ATLAS detector}},
  \href{https://doi.org/10.1007/JHEP12(2018)039}{\textit{JHEP} {\bfseries 12}
  (2018) 039}, [\href{https://arxiv.org/abs/1807.11883}{{\ttfamily
  1807.11883}}].

\bibitem{Aaboud:2018pii}
{\scshape ATLAS} collaboration, M.~Aaboud et~al., \textit{{Combination of the
  searches for pair-produced vector-like partners of the third-generation
  quarks at $\sqrt{s} =$ 13 TeV with the ATLAS detector}},
  \href{https://doi.org/10.1103/PhysRevLett.121.211801}{\textit{Phys. Rev.
  Lett.} {\bfseries 121} (2018) 211801},
  [\href{https://arxiv.org/abs/1808.02343}{{\ttfamily 1808.02343}}].

\bibitem{Aaboud:2018wxv}
{\scshape ATLAS} collaboration, M.~Aaboud et~al., \textit{{Search for pair
  production of heavy vector-like quarks decaying into hadronic final states in
  $pp$ collisions at $\sqrt{s} = 13$ TeV with the ATLAS detector}},
  \href{https://doi.org/10.1103/PhysRevD.98.092005}{\textit{Phys. Rev.}
  {\bfseries D98} (2018) 092005},
  [\href{https://arxiv.org/abs/1808.01771}{{\ttfamily 1808.01771}}].

\bibitem{Sirunyan:2018qau}
{\scshape CMS} collaboration, A.~M. Sirunyan et~al., \textit{{Search for
  vector-like quarks in events with two oppositely charged leptons and jets in
  proton-proton collisions at $\sqrt{s} =$ 13 TeV}},
  \href{https://doi.org/10.1140/epjc/s10052-019-6855-8}{\textit{Eur. Phys. J.}
  {\bfseries C79} (2019) 364},
  [\href{https://arxiv.org/abs/1812.09768}{{\ttfamily 1812.09768}}].

\bibitem{Sirunyan:2019sza}
{\scshape CMS} collaboration, A.~M. Sirunyan et~al., \textit{{Search for pair
  production of vectorlike quarks in the fully hadronic final state}},
  \href{https://doi.org/10.1103/PhysRevD.100.072001}{\textit{Phys. Rev.}
  {\bfseries D100} (2019) 072001},
  [\href{https://arxiv.org/abs/1906.11903}{{\ttfamily 1906.11903}}].

\bibitem{Sirunyan:2019xeh}
{\scshape CMS} collaboration, A.~M. Sirunyan et~al., \textit{{Search for
  electroweak production of a vector-like T quark using fully hadronic final
  states}}, \href{https://doi.org/10.1007/JHEP01(2020)036}{\textit{JHEP}
  {\bfseries 01} (2020) 036},
  [\href{https://arxiv.org/abs/1909.04721}{{\ttfamily 1909.04721}}].

\end{thebibliography}\endgroup

\end{document}